\documentclass[11pt]{article}

\usepackage{array}	
\newcolumntype{P}[1]{>{\centering\arraybackslash}p{#1}}		
\newcolumntype{M}[1]{>{\centering\arraybackslash}m{#1}}

\usepackage{tikz}
\usetikzlibrary{patterns}
\usepackage{bm}
\usepackage{empheq}
\usepackage{cases}
\usepackage{multirow}
\usepackage{booktabs}
\usepackage{soul}
\usepackage{xcolor}

\usetikzlibrary{decorations.pathreplacing,angles,quotes}

\usepackage{listings}
\usepackage[colorlinks=false, backref]{hyperref}

\usepackage{amsmath,amsthm,amsfonts,amssymb,amscd}
\usepackage{float}
\usepackage{graphicx,subfigure}
\graphicspath {{figures/}}
\usepackage[font=footnotesize,labelfont=bf]{caption}

\usepackage[mathscr]{euscript}
\usepackage{stmaryrd}
\usepackage{microtype}
\usepackage{booktabs}
\usepackage{cleveref}
\usepackage{bookmark}
\usepackage{mathrsfs}

\usepackage{emptypage}
\usepackage{tikz}
\usepackage{subfigure}
\usepackage[utf8]{inputenc}
\usepackage[T1]{fontenc}
\usepackage{mathtools}
\usepackage{colonequals}
\usepackage{bbm}
\usepackage{mathrsfs}

\DeclareMathAlphabet{\mathpzc}{OT1}{pzc}{m}{it}

\theoremstyle{plain}

\theoremstyle{definition}

\newtheorem{remark}{\scshape Remark}

\def\p{\partial}

\def\divv{{\operatorname{div}}}
\def\u{ \bm{u} }

\usepackage{fancyhdr}
\title{{\bf A space-time smooth artificial viscosity method with wavelet noise indicator and shock collision scheme,  Part 2: the 2-$D$ case}}
\author{
 {\small {\bf Raaghav Ramani}}
 \vspace{-.05 in}
\\{\footnotesize Department of Mathematics}
\vspace{-.05 in}
\\{\footnotesize University of California}
\vspace{-.05 in}
\\{\footnotesize Davis, CA 95616 USA}
\vspace{-.05 in}
\\{\footnotesize  {\it rramani@math.ucdavis.edu}}\and
 {\small {\bf Jon Reisner}}
 \vspace{-.05 in}
\\{\footnotesize Los Alamos National Lab}
\vspace{-.05 in}
\\{\footnotesize XCP-4 MSF605}
\vspace{-.05 in}
\\{\footnotesize Los Alamos, NM 87544}
\vspace{-.05 in}
\\{\footnotesize  {\it reisner@lanl.gov}}
\and
 {\small {\bf Steve Shkoller}}
 \vspace{-.05 in}
\\{\footnotesize Department of Mathematics}
\vspace{-.05 in}
\\{\footnotesize University of California}
\vspace{-.05 in}
\\{\footnotesize Davis, CA 95616 USA}
\vspace{-.05 in}
\\{\footnotesize  {\it shkoller@math.ucdavis.edu}}
}
\date{\today}

\begin{document}

\maketitle

\begin{abstract}
This is the second part to our companion paper \cite{RaReSh2018a}.  Herein, 
we generalize  to two space dimensions the $C$-method developed in \cite{ReSeSh2012,RaReSh2018a} for adding localized, space-time smooth artificial viscosity to nonlinear systems 
of conservation laws that propagate shock waves, rarefaction waves, and contact discontinuities. For gas dynamics, the $C$-method 
couples the Euler equations to 
scalar reaction-diffusion equations, which we call $C$-equations, whose solutions serve as space-time smooth artificial viscosity indicators for shocks
and contacts.

We develop a high-order numerical algorithm for gas dynamics in 2-$D$ which can
accurately simulate  the Rayleigh-Taylor (RT)  instability with Kelvin-Helmholtz (KH) roll-up of the contact discontinuity, as well as shock collision and
bounce-back.   Solutions to our $C$-equations not only indicate the location of the shocks and contacts, but also  track the geometry of
the evolving fronts.  This allows us to implement both directionally isotropic and anisotropic artificial viscosity schemes, the latter adding diffusion
only in directions tangential to the evolving front. We additionally  produce a novel
shock collision indicator function, which naturally activates during shock
collision, and then smoothly deactivates.   Moreover,  we implement a  high-frequency 2-$D$ wavelet-based noise detector together with an efficient and localized noise removal algorithm.    

 To test the 
methodology, we use 
a highly simplified WENO-based discretization scheme.  We provide numerical results for some classical
2-$D$ test problems, including the RT problem, the Noh problem, a circular explosion problem from the
 Liska \& Wendroff \cite{LiWe2003} review paper, the Sedov blast wave problem, 
the double Mach 10 reflection test, and a shock-wall collision problem. In particular, we show that our 
artificial viscosity method can eliminate the wall-heating phenomenon for the Noh problem, and thereby 
produce an accurate, non-oscillatory solution, even though our simplified WENO-type scheme fails to run 
for this problem.
\end{abstract}

\newpage

{\small
\tableofcontents}

\section{Introduction}\label{sec:intro}
This is the second in a two-part series of papers, in which we develop a high-order numerical algorithm, the $C$-method,  to simulate compressible fluid flow
with shock waves and contact discontinuities, as well as shock-wall collision and bounce-back.  In  the first part \cite{RaReSh2018a}, 
we developed our scheme in one space dimension.  This second part is devoted to the more geometric problem of simulating
 shocks, contacts, and collisions in two
space dimensions.

Compressible fluid flow is an example of a system of nonlinear conservation laws, which in two space dimensions,  can be written as  
\begin{align*}
\partial_t {\bm U}(\bm{x},t)+ \p_x {\bm F}({\bm U}(\bm{x},t)) + \p_y {\bm G}({\bm U}(\bm{x},t)) & = \bm{0}\,, \\
{\bm U}(\bm{x},t=0)  &= {\bm U}_0(\bm{x})   \,,   
\end{align*}
where $\bm{x} = (x,y)$ denotes coordinates on the plane, $t$ denotes time, 
 $\bm{U}(\bm{x},t)$ is a vector of conserved quantities, and  $\bm{F}(\bm{U})$ and $\bm{G}(\bm{U})$ denote the horizontal and vertical
 flux functions, respectively.

Even with smooth initial conditions, multi-dimensional conservation laws such as the compressible Euler equations develop singularities in finite-time
\cite{Si1985, CoSh2014} and, in general, solutions consist of propagating shock waves and contact discontinuities.   In two space dimensions,
shocks and contacts produce {\it curves of discontinuities}, also known as {\it fronts}, which 
propagate according to the \emph{Rankine-Hugoniot}
conditions (see \S\ref{sec:euler-2d-description}).   The objective of a high-order numerical scheme is to produce a simulation which keeps the
fronts sharp, while simultaneously providing high-order accuracy in smooth regions away from the front.

\subsection{Numerical discretization}
In \S 1.1 of   \cite{RaReSh2018a}, we described the tools necessary for designing high-order accurate numerical schemes in 1-$D$.
In multi-$D$, similar tools are required to obtain non-oscillatory numerical schemes, but the multi-dimensional analogues  are generally limited by mesh considerations. For structured grids 
(such as products of uniform 1-$D$ grids), dimensional splitting is commonly used, decomposing the problem into a sequence of 1-$D$ 
problems. This technique is quite successful, but stringent mesh requirements prohibits its use on complex domains. Moreover, applications to
 PDE outside of variants of the Euler equations may be somewhat limited. For further discussion of the limitations of dimensional splitting, we 
 refer the reader to Crandall \& Majda \cite{Crandall80}, and  Jiang \& Tadmor \cite{Jiang98}. For unstructured grids, dimensional splitting is not available and alternative approaches  must be employed, 
necessitated by the lack of multi-$D$ Riemann solvers. WENO schemes on unstructured  triangular grids have been developed in Hu \& Shu \cite{Hu199997}, but using simplified methods, which employ reduced characteristic decompositions,  can lead   to a loss of   monotonicity and stability.    

Algorithms that explicitly introduce diffusion provide a simple way to stabilize higher-order numerical schemes and subsequently remove non-physical oscillations near shocks.  We refer the reader to the introductory sections in  \cite{ReSeSh2012,RaReSh2018a} for a review of the classical 
artificial viscosity method \cite{vNRi1950}.  

In this paper, we develop a stable high-order 2-$D$ numerical scheme that does not use approximate Riemann solvers or characteristic decompositions, but instead relies upon a 2-$D$ generalization of the 
1-$D$ $C$-method \mbox{\cite{ReSeSh2012,RaReSh2018a}}. The extensive error analysis and 
convergence tests performed for the 1-$D$ $C$-method in {\cite{RaReSh2018a}} indicate 
that the scheme is high order accurate. We expect that the 2-$D$ implementation of the method presented 
in this paper is similarly high-order accurate, and we perform a number of numerical experiments to 
qualitatively support this claim. We also present error analysis and convergence studies for the Sedov 
problem to provide quantitative evidence for the claim of high order accuracy; further extensive 
convergence tests will be presented in {\cite{RaReSh2019}}. 

\subsection{Localization to shocks and contacts and capturing the geometry of the front}

As we noted above, part one \cite{RaReSh2018a}  provides a self-contained development of the 1-$D$ $C$-method; for  problems in one space dimension, the
function $C(x,t)$ is the solution to a forced scalar reaction-diffusion equation, and serves as a highly localized space-time smooth indicator for the shock location.   In 2-$D$, we again solve a scalar reaction-diffusion equation for $C( \bm{x}, t)$ on the plane; in this
instance, the function $C( \bm{x}, t)$ not only serves as a localized indicator for both shocks and contacts, but is also able to accurately 
represent the geometry of the evolving fronts.   Having this geometry allows us to define time-dependent normal and tangent vectors to the
fronts, which, in turn, enables the  construction of artificial viscosity methods that add diffusion only in certain directions rather than uniformly across the
mesh.   As we shall explain below, this type of anisotropic artificial viscosity scheme is essential in accurately capturing Kelvin-Helmholtz roll-up without overly diffusing the mixing regions in such flows.

\subsection{Shock collision and wavelet-based noise removal}  

In part one \cite{RaReSh2018a}, we consider the difficult problem of shock-wall collision and bounce-back
in 1-$D$. In particular, when a shock wave collides with and reflects off of 
a fixed boundary, spurious oscillations develop behind the 
reflected shock. In addition to these post-collision oscillations, 
most schemes produce solutions that exhibit the phenomenon of 
anomalous \emph{wall heating} \cite{Rider2000}.
A novel modification to the $C$-method in 1-$D$ \cite{RaReSh2018a}, wherein the time-dependent artificial 
viscosity parameter $\mathcal{B}(t)$ naturally increases during shock-wall collision and bounce-back, 
allows for the 
addition of extra ``wall viscosity'' to the shock front. This suppresses
post-collision noise and the wall heating error, and ensures that the solution retains high-order accuracy away 
from the shock and prior to collision. In \S\ref{sec:C-wall-collision} of this paper, we 
present a  generalization of the 1-$D$ shock collision
scheme to the 2-$D$ setting, and apply the method to a circular explosion problem in \S\ref{subsec:sod-collision}. 

The occurrence of high frequency noise in numerical solutions to PDE is a well-known issue. 
The first part of this work \cite{RaReSh2018a} provides a description of an efficient  wavelet-based noise detection and 
heat equation-based noise removal algorithm  for 1-$D$ gas dynamics simulations.
Error analysis and convergence tests in \cite{RaReSh2018a}
show that the noise detection and removal algorithm decreases the errors of numerical solutions and
improves the rate of convergence.  In \S\ref{sec:noise-ind} of this paper, we present the natural extension 
of this 1-$D$ algorithm to 
the 2-$D$ case, with applications to the RT problem in \S\ref{sec:RT} and the Noh problem in \S\ref{sec:Noh}.

\subsection{Outline of the paper}
In \S\ref{sec:euler-2d-description}, we introduce the 2-$D$ compressible Euler equations and the 
Rankine-Hugoniot jump conditions relating the speed of propagation of curves of discontinuity with the jump
discontinuities in the conservative variables. In \S\ref{sec:extension-2d}, we present the \emph{isotropic} 
$C$-method, and in \S\ref{sec:anisotropic-C-method}, 
the \emph{anisotropic} $C$-method, the latter designed for the long-time 
evolution of contact discontinuities. A new 2-$D$ shock-wall collision scheme is introduced in 
\S\ref{sec:C-wall-collision}, and a wavelet-based noise detection and {\it localized}
heat equation-based noise removal procedure is developed in 
\S\ref{sec:noise-ind}. Details about the numerical methods implemented are 
provided in \S\ref{sec:numerical-implementation} and Appendix \ref{appendix:WENO}. 
We then apply the methods to a number of test problems. In particular, the $C$-method produces
an accurate, non-oscillatory solution for the difficult Noh problem, whereas our simplified WENO-type 
algorithm (as well as more sophisticated advection schemes)  fails to run for this problem.

\section{The compressible Euler equations in 2-$D$}\label{sec:euler-2d-description}

The compressible Euler equations in a 2-$D$ rectangular domain 
$\Omega = (x_1,x_M) \times (y_1,y_N) \subset \mathbb{R}^2$ and a time interval $[0,T]$ are given in conservation law form as
 \begin{subequations}
\label{Euler-2d}
\begin{alignat}{2}
\partial_t {\bm U}(\bm{x},t)+ \p_x {\bm F}({\bm U}(\bm{x},t)) + \p_y {\bm G}({\bm U}(\bm{x},t)) = \bm{0},& && \ \ \  \bm{x} \in \Omega \,,   t > 0,  \label{Euler-2d-motion} \\
{\bm U}(\bm{x},0)  = {\bm U}_0(\bm{x}),& && \ \ \ \bm{x} \in {\Omega} \,,   t = 0,
\end{alignat}
\end{subequations}
where the 4-vector $\bm{U}(\bm{x},t)$ and the flux functions $\bm{F}(\bm{U})$ and $\bm{G}(\bm{U})$ 
are defined as
$$
{\bm U} = \left ( \begin{array}{c} \rho \\ \rho u \\ \rho v \\ E \end{array} \right ) \quad \text{ and } \quad {\bm F}({\bm U}) = \begin{pmatrix}
		\rho u \\
		\rho u^2+p \\
		\rho u v \\
		u(E+p)
		\end{pmatrix}\, \quad \text{ and } \quad 
\bm{G}(\bm{U}) = \begin{pmatrix}
		\rho v \\
		\rho uv \\
		\rho v^2 + p \\
		v(E+p)
		\end{pmatrix}\,. 
$$
Here, the vector $\bm{x} = (x,y)$ denotes the two Cartesian coordinates, $\rho$ and $E$ 
are the density and energy fields, respectively, $u$ and $v$ are the velocities in 
the $x$-direction and $y$-direction, respectively, and we use the notation
$\bm{u}$ to denote the velocity vector field $\bm{u} = (u,v)$. 
The pressure function $p$ is defined by the equation of state
\begin{equation}\label{eqn-of-state}
p = (\gamma - 1) \left( E - \frac{1}{2} \rho | \bm{u} | ^2 \right), 
\end{equation}
where $| \bm{u} | = \sqrt{ u^2 + v^2 }$ and $\gamma>1$ is the adiabatic constant.  
The system \eqref{Euler-2d} represents the conservation of mass, momentum, and energy:
$$
 \partial_t\rho + \operatorname{div} (\rho \u)=0\,, \ \ \partial_t (\rho \u) + \operatorname{div} ( \rho \u \otimes \u)+ \nabla p =0 \,,  \ \ 
  \partial_t E + \operatorname{div} \left((E+p)\u \right)=0 \,,
$$
and defines a hyperbolic system, in the sense that both $ \nabla \bm{F}$ and $ \nabla \bm{G}$ have real eigenvalues; in particular,  the
four eigenvalues of $ \nabla \bm{F}$ are
$$
\lambda_1 = u - c\,, \ \ \lambda_2 = \lambda_3 = u\,, \ \ \lambda_4 = u + c\,,
$$
and the four eigenvalues of $ \nabla \bm{G}$ are $$
\lambda_5 = v - c\,, \ \ \lambda_6 = \lambda_7 = v\,, \ \ \lambda_8 = v + c\,.
$$
The eigenvalues $\lambda _1, \lambda_5, \lambda _4$ and $\lambda _8$ correspond to sound waves, while the  repeated eigenvalues
$ \lambda _2,\lambda _3,\lambda _6$, and $ \lambda _7$ correspond to vorticity and entropy waves.
We define the \emph{maximum wave speed}  $S(\bm{u})$ as
\begin{equation}\label{S-wave-speed}
S(\bm{u}) = [S(\bm{u})](t) = \max_{i=1,\ldots,8} \max_{\Omega} \left\{ |\lambda_i (x,y,t) | \right\}\,.
\end{equation}

We focus on solutions $\bm{U}$ to the compressible Euler equations \eqref{Euler-2d} that have a jump 
discontinuity across a time-dependent curve 
\begin{equation}\label{Gamma_curve}
\Gamma(t) = \cup_{i=1}^p \Gamma_i(t)\,,
\end{equation} 
where for each index $i=1,...,p$,
$\Gamma_i(t)$ represents either a shock front or a contact discontinuity.  In the case that $\Gamma_i(t)$ is a closed curve, we define
$ \bm{U}^{+}(x,t)$ to be the value of $ \bm{U}(x,t)$ inside of $\Gamma_i(t)$ and $ \bm{U}^{-}(x,t)$ to be the value of outside of $\Gamma_i(t)$.
We then set
$$
[\bm{U}]^{+}_{-} = \bm{U}^{+} - \bm{U}^{-}\, \text{ at } \Gamma(t). 
$$
Let $\vec{n}=\vec{n}(x,y,t)$ and $\vec{\tau}=\vec{\tau}(x,y,t)$ be the unit normal and tangent 
vectors to $\Gamma(t)$, respectively. 
The Rankine-Hugoniot conditions relate the speed of propagation $\dot{\sigma}$ of the curve of 
discontinuity $\Gamma(t)$ with the jump discontinuity in the variables $\bm{U}$ via the relation
$$
\dot{\sigma} [\bm{U}]^{+}_{-} = [\bm{F} \cdot \vec{n}\, ]^{+}_{-}\,.
$$
If $\Gamma_i(t)$ is a shock, 
$$
\left[ \bm{u} \cdot \vec{n} \right]^{+}_{-} \ne 0 \text{ and } \left[ \bm{u} \cdot \vec{\tau} \right]^{+}_{-} = 0 \,,
$$
while if $\Gamma_i(t)$ is a contact discontinuity, then
$$
\left[ \bm{u} \cdot \vec{n} \right]^{+}_{-} = 0 \text{ and } \left[ \bm{u} \cdot \vec{\tau} \right]^{+}_{-} \ne 0.  
$$
For the problems we consider in this paper, $\left[ \rho \right]^{+}_{-} \neq 0$ for both shocks and contacts.

\section{The $C$-method in the 2-$D$ setting}\label{sec:extension-2d}

\subsection{Smoothly localizing the curves of discontinuity and tracking the geometry}


As noted above, the jump discontinuities of the density function describe the location of both the  shock and contact fronts; as such, it is
natural to use $| \nabla \rho(\bm{x},t)|$ as an indicator for these time-dependent curves.
Consequently, we track discontinuities by considering the quantity 
\begin{equation}\label{Grho}
G_\rho \coloneqq \frac{ |\nabla \rho | }{ \max_{\Omega} | \nabla \rho |}\,.
\end{equation} 

We note that in the 1-$D$ $C$-method formulation in {\cite{RaReSh2018a}}, we use $|\p_x u|$ instead of $|\p_x \rho|$  as the forcing
function for the $C$-equation.  In 1-$D$, using either  $|\p_x u|$  or $|\p_x \rho|$ as the forcing, together with the compression switch, produces identical
results. In 2-$D$, the velocity is a vector quantity, whereas the density is a scalar function, so that 
the latter provides a simpler approach to tracking discontinuities.

\begin{figure}[H]
\centering
\subfigure[]{\label{fig:G-forcing}\includegraphics[width=60mm]{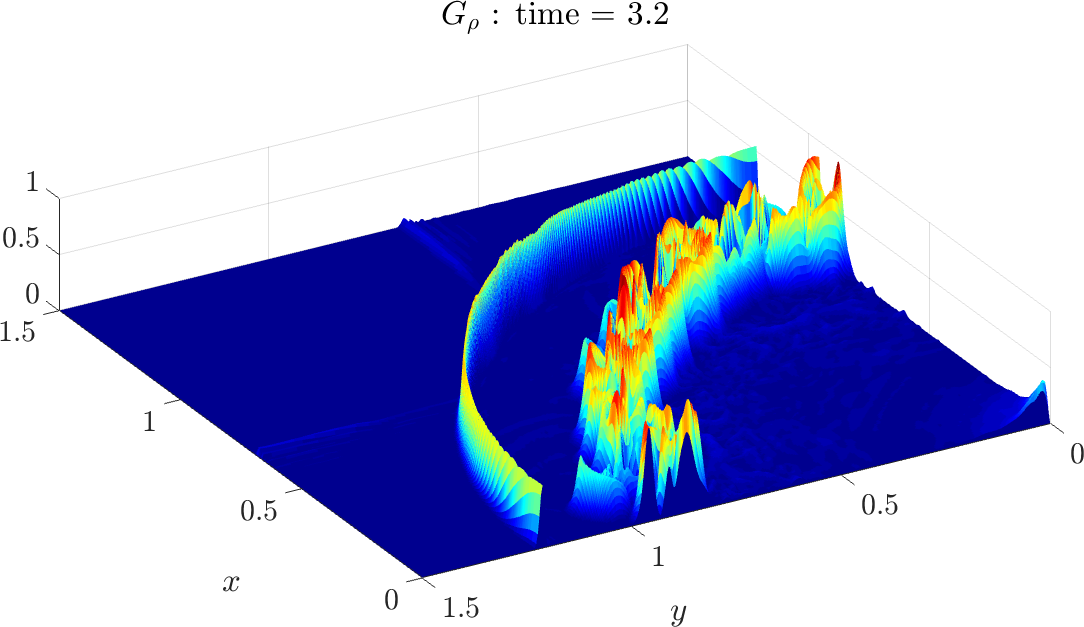}}
\hspace{2em}
\subfigure[]{\label{fig:G-forcing2}\includegraphics[width=40mm]{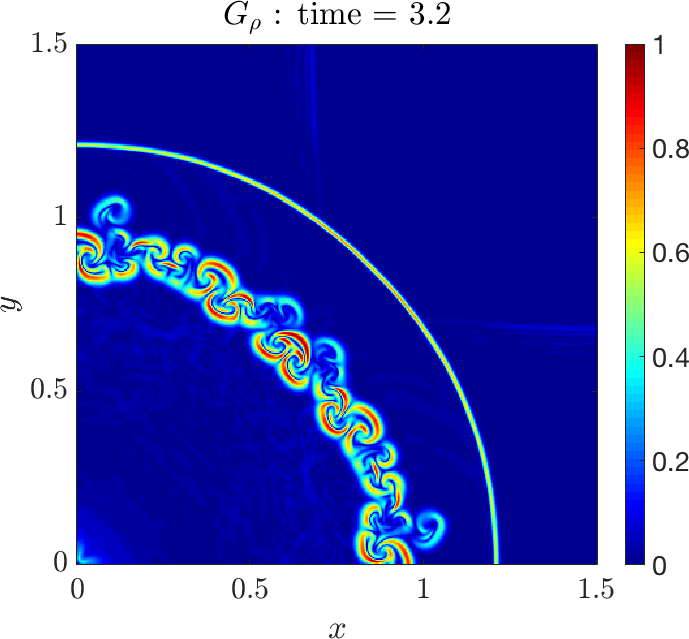}}
\caption{The function $G_\rho$ for the Sod explosion problem at time $t=3.2$.}
\label{fig:track-contour-Sod}
\end{figure}

In 2-$D$, we are interested in tracking both shock fronts and contact curves, and the 
density gradient provides a natural method for tracking both of these discontinuities.
In Fig.\ref{fig:track-contour-Sod}, we see that $G_\rho$ captures both 
of the discontinuities, namely the shock front and the contact discontinuity, 
present in the solution.

Although the function $G_\rho$ is able to localize artificial viscosity to a curve of discontinuity, its use
in artificial viscosity operators often serves to produce an oscillatory solution \cite{ReSeSh2012,RaReSh2018a}. This is due to the rough nature of the localizing function $G_\rho$.  
Consequently, we first produce a space-time smooth variant of $G_\rho$ through the use of the $C$-method. 
We describe 
the natural 2-$D$ generalization of the 1-$D$ $C$-equation as follows: we first define the operator 
$\mathscr{L}$ as
\begin{equation}\label{C-2d-operator}
\mathscr{L}[C\,; \varepsilon, \kappa] \coloneqq - \frac{S(\bm{u})}{\varepsilon | \delta \bm{x} |} C + \kappa S(\bm{u}) | \delta \bm{x} | \, \Delta C\,,
\end{equation}
where $\delta \bm{x} = (\delta x\,,\delta y)$ with $\delta x$, $\delta y$ the grid spacings in the 
$x$ and $y$ directions, respectively, $S(\bm{u})$ is the maximum wave-speed \eqref{S-wave-speed}, 
and $\Delta= \p_{xx} + \p_{yy}$ denotes the 2-$D$ Laplace operator. 
The space-time smooth version of $G_\rho$, which we denote by $C = C(\bm{x},t)$, is the solution to the 
scalar linear
parabolic PDE
\begin{equation}\label{C-eqn-2D}
\p_t C(\bm{x},t) - \mathscr{L}[C(\bm{x},t);\varepsilon,\kappa] = \mathbbm{1}_{\divv \bm{u} < 0}\, \frac{S(\bm{u})}{\varepsilon | \delta \bm{x} |} G_\rho(\bm{x},t)\,, 
\end{equation}
where the function $\mathbbm{1}_{\divv \bm{u} < 0}$ is a \emph{compression switch}\footnote{$\mathbbm{1}_{\divv \bm{u} < 0}$ 
 is equal to $1$ on the set ${\divv \bm{u} < 0}$ and is equal to zero on ${\divv \bm{u} \ge 0}$.}
that 
ensures that $C$ vanishes
in regions of expansion, where there are no discontinuities in the solution. The parameters $\varepsilon$ and $\kappa$ in \eqref{C-2d-operator} control the support and smoothness
of the solution $C$, respectively \cite{ReSeSh2012,RaReSh2018a}. 

\begin{figure}[H]
\centering
\subfigure[]{\label{fig:G-C-compare1}\includegraphics[width=45mm]{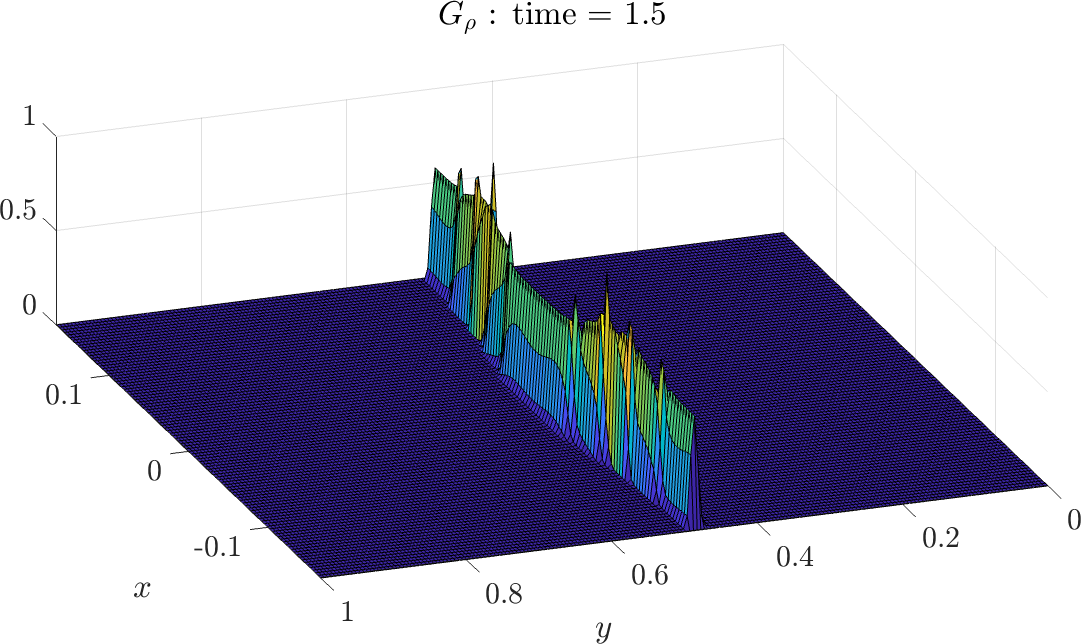}}
\hspace{0.5em}
\subfigure[]{\label{fig:G-C-compare2}\includegraphics[width=45mm]{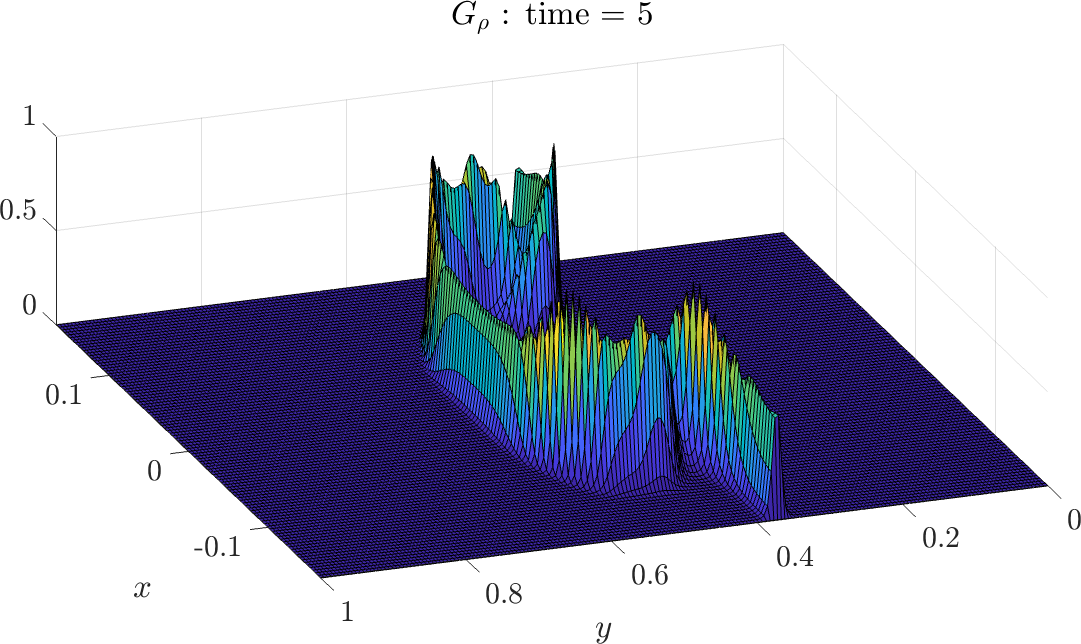}}
\hspace{0.5em}
\subfigure[]{\label{fig:G-C-compare3}\includegraphics[width=45mm]{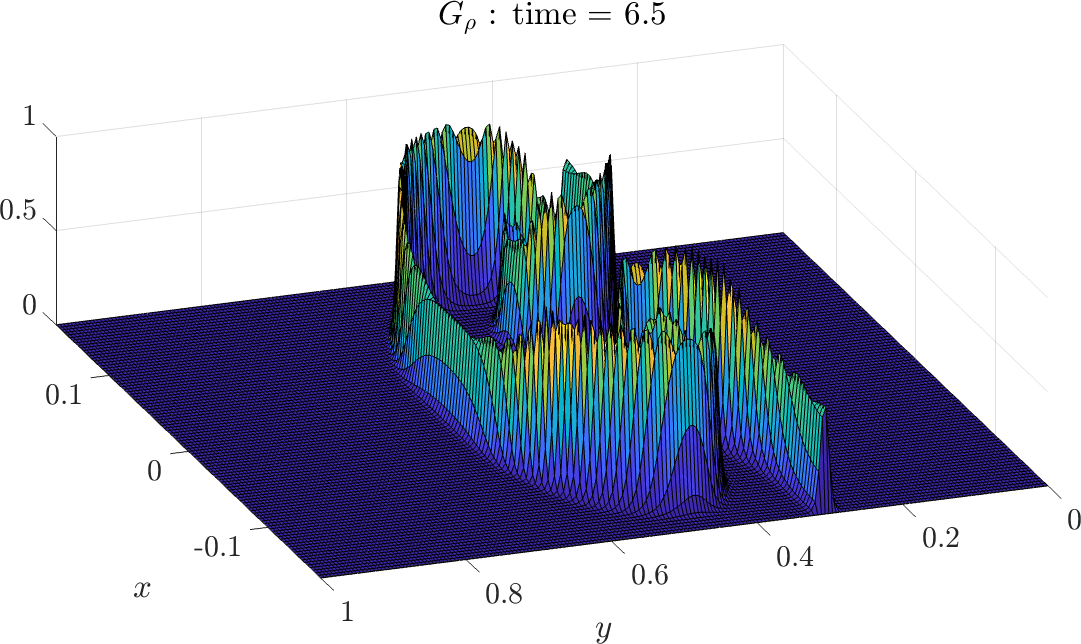}}
\subfigure[]{\label{fig:G-C-compare4}\includegraphics[width=45mm]{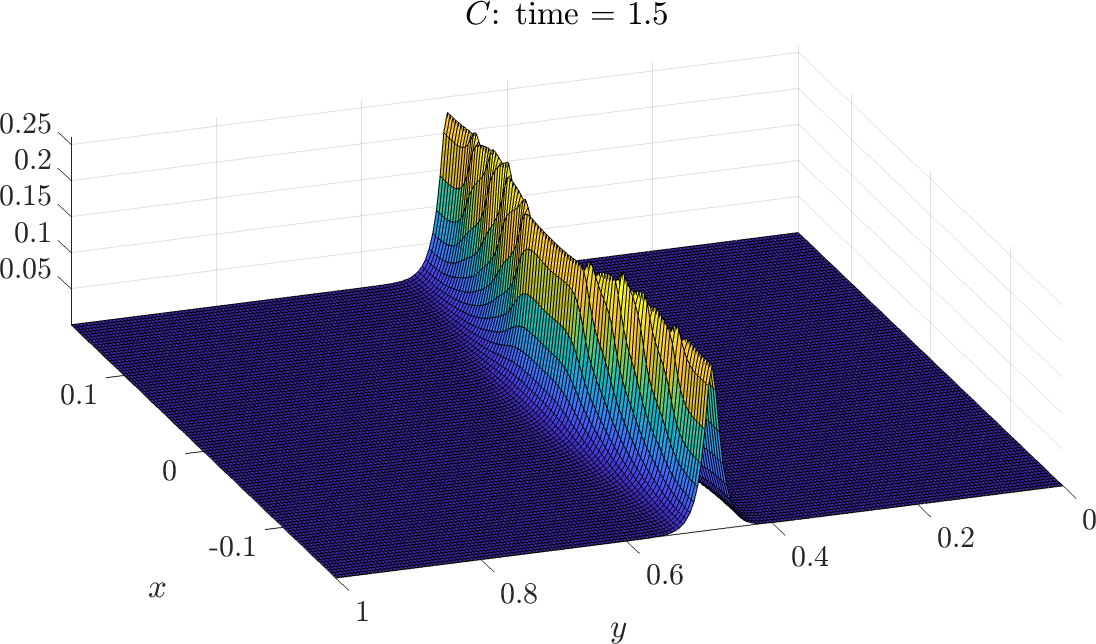}}
\hspace{0.5em}
\subfigure[]{\label{fig:G-C-compare5}\includegraphics[width=45mm]{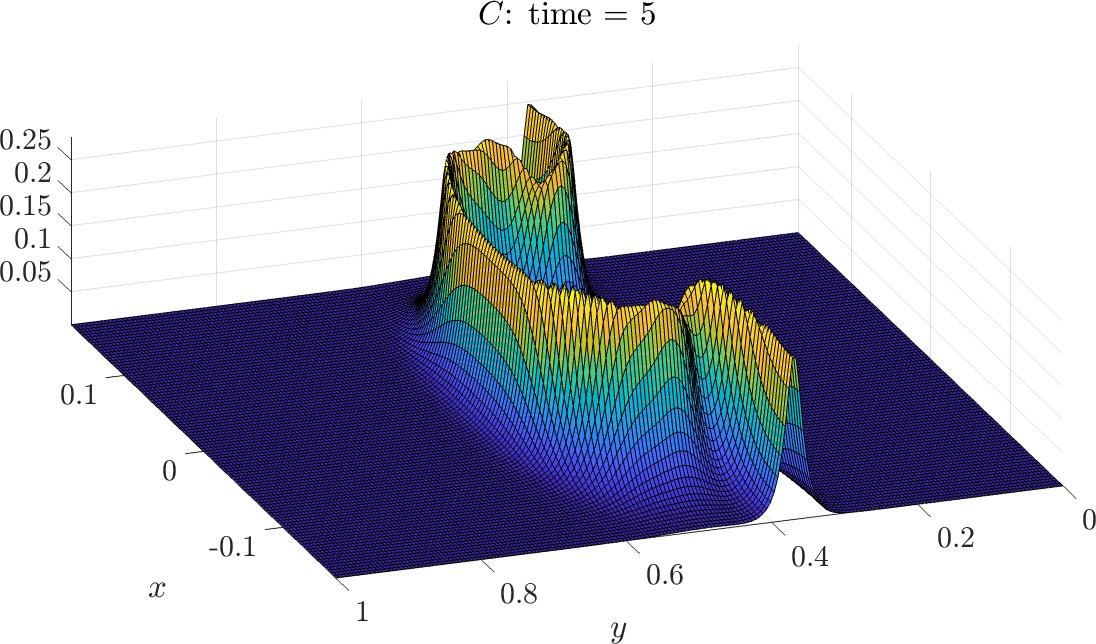}}
\hspace{0.5em}
\subfigure[]{\label{fig:G-C-compare6}\includegraphics[width=45mm]{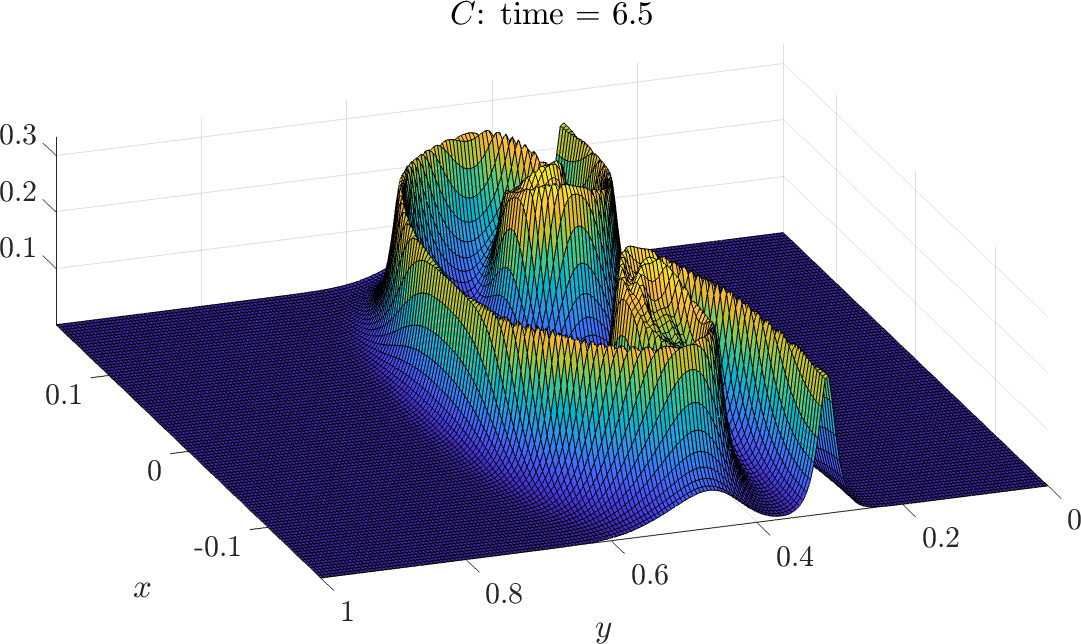}}
\caption{Comparison of $G_{\rho}$ and $C$ for the Rayleigh-Taylor instability at various times. The top row 
(a,b,c) are surface plots of $G_{\rho}$, and the bottom row (d,e,f) are surface plots of the corresponding $C$.}
\label{fig:G-C-compare}
\end{figure}

The function $C(\bm{x},t)$ provides not only the location of the shock and contact fronts, but also a good approximation to
the  geometry of the
front.   Specifically, the $C$ function is sufficiently localized so as to provide the shape of the evolving front.
 In 
Fig.\ref{fig:G-C-compare}, we show results of the evolution of a contact discontinuity associated to the Rayleigh-Taylor instability (which
is discussed in \S\ref{sec:RT}).   As can be seen,  $| \nabla \rho|$ and hence the function $G_\rho$ track
 the material interface accurately but the function  $G_\rho$ is very rough, particularly in directions tangential to the contact front.  The function $C$, by contrast, is smooth and exhibits no oscillatory behavior, but is still
  highly localized to the contact discontinuity.

\subsection{The isotropic $C$-method and the Euler-$C$ system}\label{sec:isotropic}

We begin with a very natural generalization of the 1-$D$ $C$-method to the 2-$D$ setting.  In particular, we 
 first consider the
2-$D$ Euler-$C$ system with isotropic space-time smooth artificial viscosity:
\begin{subequations}\label{EulerC-2D}
\begin{align}
\partial_t \rho + \divv (\rho \bm{u}) &=0, \label{EulerC-density}\\ 
\partial_t (\rho u) + \partial_x (\rho u^2 + p) + \p_y (\rho uv) &=  \divv \left( \tilde{\beta}^u \, \rho \, C \, \nabla u \right), \label{EulerC-momentum-u}\\
\partial_t (\rho v) + \p_x(\rho uv) + \partial_y (\rho u^2 + p) &=  \divv \left( \tilde{\beta}^u \, \rho \, C \, \nabla v \right), \label{EulerC-momentum-v}\\
\partial_t E + \divv (\bm{u}(E + p)) &=  \divv \left( \tilde{\beta}^E \, \rho \, C \, \nabla(E/\rho) \right) \,, \label{EulerC-energy}  \\
\p_t C - \mathscr{L}[C\,; \varepsilon,\kappa] &
 = \mathbbm{1}_{\divv \bm{u} < 0} \, \frac{S(\bm{u})}{\varepsilon | \delta \bm{x} |} G_\rho  \,,  \label{C-Sod}
\end{align}
\end{subequations}
where the pressure $p$ is given by the equation of state \eqref{eqn-of-state}, 
$\mathscr{L}= \mathscr{L}[C\,; \varepsilon, \kappa] $ is the operator defined in \eqref{C-2d-operator}, $G_\rho$ is the forcing function defined in \eqref{Grho}, and 
the artificial viscosity parameters are given by
\begin{equation}\label{C-artificial-viscosity}
\tilde{\beta}^{(\cdot)} \coloneqq \frac{|\delta \bm{x}|^2}{\max_{\Omega} C} \, \beta^{(\cdot)} \,, 
\end{equation}
with $\beta^{(\cdot)}$ a constant.

We refer to the system \eqref{EulerC-2D} as the {\it isotropic} $C$-method, because artificial viscosity  is added uniformly in all directions (although clearly still localized to the fronts via the use of $C$).
We remark that the particular form of artificial viscosity used  in the momentum equations 
\eqref{EulerC-momentum-u} and \eqref{EulerC-momentum-v} as well as the energy equation 
\eqref{EulerC-energy} ensure that total energy remains conserved, and that $E$ continues to evolve as the total energy function. For simplicity, suppose that  periodic boundary conditions are enforced.
On the one hand, integrating the energy equation
\eqref{EulerC-energy} yields $ \frac{\mathrm{d}}{\mathrm{d}t}\int_{\Omega} E \,\mathrm{d}\bm{x} = 0$. 
On the other hand, multiplying \eqref{EulerC-momentum-u} and \eqref{EulerC-momentum-v} by $u$ and $v$, respectively, integrating over the domain $\Omega$, summing the resulting quantities and utilizing 
the conservation of mass equation
\eqref{EulerC-density}, the energy equation \eqref{EulerC-energy}, and the 
equation of state \eqref{eqn-of-state} yields
$$
\frac{\mathrm{d}}{\mathrm{d}t}\int_{\Omega} \frac{1}{2} \rho |\bm{u}|^2 + \frac{p}{\gamma -1} \,\mathrm{d}\bm{x} = 0\,. 
$$
This shows that the velocity $\bm{u}$ and the pressure $p$ adjust accordingly to maintain the relation  
\eqref{eqn-of-state}, and that the Euler-$C$ system conserves the total energy. 

The space-time smooth localizing function $C$ ensures that viscosity is added only at discontinuities, thereby
ensuring that the solution retains high-order accuracy away from discontinuities, and, moreover, given that $C(\bm{x},t)$ is a solution
to a reaction-diffusion equation, it is smooth in both space and time.   See \cite{ReSeSh2012} for the analysis of the solutions $C$.

\section{The anisotropic $C$-method for contact discontinuities}\label{sec:anisotropic-C-method}
We next consider the {\it anisotropic} $C$-method, specifically designed for the 
long-time evolution of contact discontinuities in
the presence of Rayleigh-Taylor (RT) instabilities which lead to Kelvin-Helmholtz (KH) roll-up.    We are particularly interested in the case that
$\left[ \rho \right]^{+}_{-} \neq 0$ across the contact discontinuity\footnote{KH instabilities can occur with a constant density
profile, but we consider problems for which $\left[ \rho \right]^{+}_{-} \neq 0$. The method described here requires the condition $\left[ \rho \right]^{+}_{-} \neq 0$ to calculate the normal and tangent 
vectors to the evolving front. However, we remark that 
our algorithm can be adapted for problems for which $\left[ \rho \right]^{+}_{-} = 0$, and details are 
provided in the paper {\cite{RaReSh2019}} under preparation.}, for which we also have that 
$\left[ \bm{u} \cdot \vec{n} \right]^{+}_{-} = 0$ and $ \left[ \bm{u} \cdot \vec{\tau} \right]^{+}_{-} \neq 0$.

Conservation of mass can be written as
$$
\partial_t \rho + {\bm u} \cdot \nabla \rho = - \rho \operatorname{div} {\bm u} \,,
$$
and near an evolving front with tangent and normal vectors given by $\vec{\tau}$ and $\vec{n}$, respectively, the divergence of the velocity 
vector field ${\bm u}$ is given by
$$
\operatorname{div} {\bm u} = \partial_{\vec{\tau}} {\bm u} \cdot \vec{\tau} + \partial_{\vec{n}} {\bm u} \cdot \vec{n}  \,.
$$
Across a contact curve, $\partial_{\vec{n}} {\bm u} \cdot \vec{n}$ remains smooth, while $\partial_{\vec{\tau}} {\bm u} \cdot \vec{\tau} $
can become extremely oscillatory due to a combination of the discontinuity of ${\bm u} \cdot \vec{\tau}$ across the contact curve together with
interpolation error of the contact curve onto a fixed grid (particularly in specifying the initial data).    For simulations that require a great deal of
time steps, it is important to add artificial viscosity in the tangential directions, but not in the normal directions.

In classical RT problems (and particularly for low Mach-number flows), 
instabilities that are generated by a small perturbation of the equilibrium interface position require a large number of time-steps to fully
develop the KH roll-up of the contact curve, and it is most often the case that fluid mixing (due to  numerical dissipation) is present in this roll-up
region, so that the amplitude of the density and the gradient of the density can be significantly smaller than their maximum values.   

The objective of our anisotropic $C$-method is to add diffusion only in directions that are tangent to the contact discontinuity, while adding no 
artificial diffusion in directions that are normal to the contact curve.  In doing so, we can maintain a very sharp interface, and prevent over-diffusion of the slight ``hills-and-valleys'' which arise in the KH mixing zones.    Moreover, when generating the RT instability of a small 
interface perturbation on a uniform rectangular grid  (rather than using a velocity perturbation as done by ATHENA\cite{Athena}), 
spurious {\it tangential spikes} can form in the velocity fields $u$ and $v$ along the contact curve; these spikes, in turn, generate small-scale
numerical KH structures which contaminate the solution (this is discussed further in
 \S\ref{subsec:oscillations}). Consequently, it is necessary to remove these spikes while maintaining a sharp
 interface; this may be accomplished through the use of anisotropic diffusion. 


\subsection{Calculating the normal and tangential directions to $\Gamma(t)$}
\begin{figure}[H]
\centering
\subfigure[Vector plot of $\vec{\tau}$ at $t=4.0$]{\label{fig:tau-vectors1}\includegraphics[width=60mm]{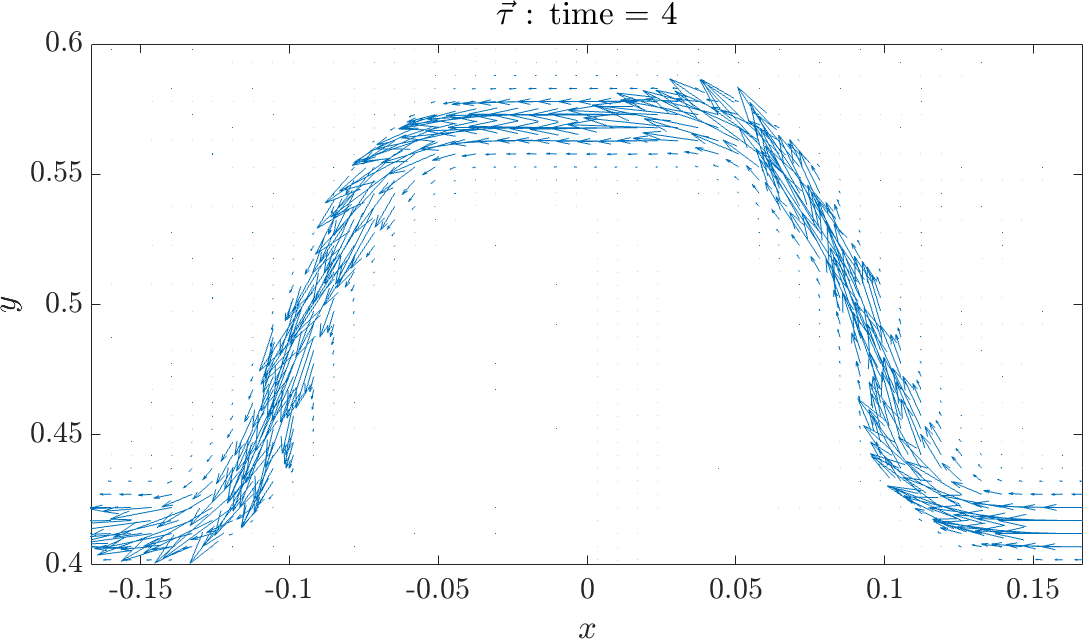}}
\hspace{2em}
\subfigure[Vector plot of $\vec{\tau}$ at $t=8.5$]{\label{fig:tau-vectors2}\includegraphics[width=60mm]{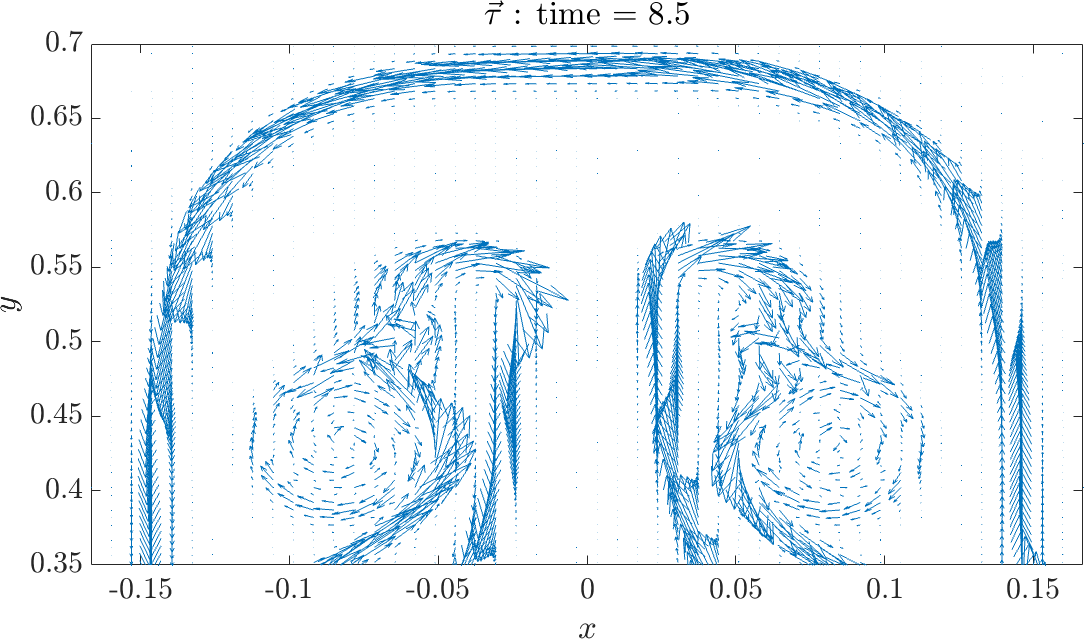}}
\par\bigskip
\subfigure[Surface plot of $\tau_1$ at $t=4.0$]{\label{fig:tau-vectors3}\includegraphics[width=60mm]{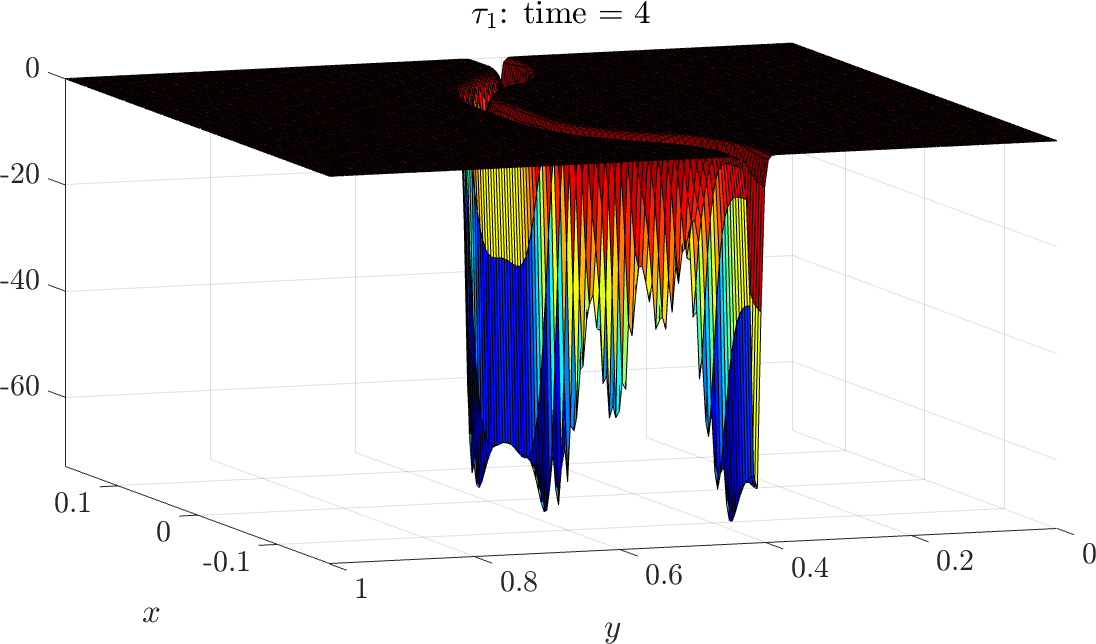}}
\hspace{2em}
\subfigure[Surface plot of $\tau_2$ at $t=4.0$]{\label{fig:tau-vectors4}\includegraphics[width=60mm]{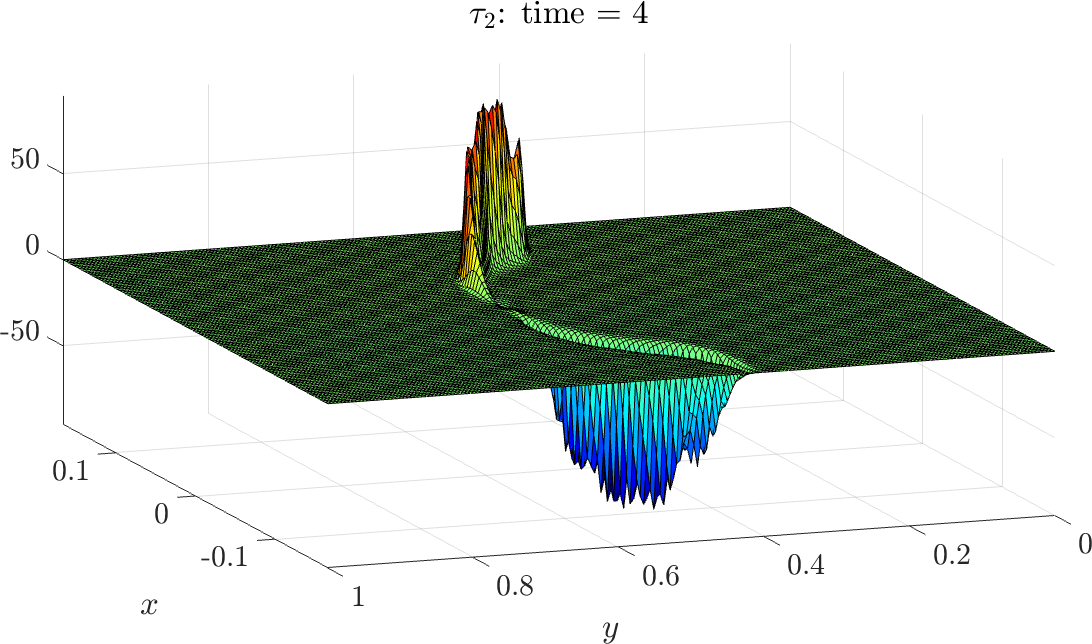}}
\caption{Calculation of the tangent vector $\vec{\tau}$ for the RT instability.}
\label{fig:tau-vectors}
\end{figure}

The first task is to accurately compute a good approximation to the tangent vectors $\vec{\tau}$ 
to any curve of discontinuity $\Gamma(t)$, defined in \eqref{Gamma_curve}. For
the problems we consider here, this may be accomplished by setting
\begin{subequations}\label{tau-vectors}
\begin{align}
\tau_1(x,y,t) \coloneqq \vec{\tau} \cdot \bm{e}_1 &= - \p_y \rho\,, \\
\tau_2(x,y,t) \coloneqq \vec{\tau} \cdot \bm{e}_2 &= \p_x \rho\,,
\end{align}
\end{subequations}
where $\bm{e}_1$ and $\bm{e}_2$ denote the unit vectors in the $x$ and $y$ directions, respectively. 
In Fig.\ref{fig:tau-vectors}, we provide vector plots of $\vec{\tau}$ calculated using 
\eqref{tau-vectors}, 
as well as surface plots of each component of $\vec{\tau}$, for the specific case of the RT instability\footnote{
We note that the vector plots in Figs.{\ref{fig:tau-vectors1}}  and {\ref{fig:tau-vectors2}} appear 
asymmetric only because of the Matlab plotting routine we use.  In fact, for the RT problem we consider in this 
paper, the interface $\Gamma(t)$ is symmetric across the  line $x=0$. Since we fix an orientation for 
$\Gamma(t)$, the horizontal component of the tangent  vector $\tau_1$ is symmetric across $x=0$, while the 
vertical component $\tau_2$ is anti-symmetric, as can  be seen in Figs.{\ref{fig:tau-vectors3}} 
and {\ref{fig:tau-vectors4}}. These facts, combined with the particular Matlab plotting routine we employ, cause
the perceived asymmetry in the vector plots.}. 

As shown in Fig.\ref{fig:tau-vectors}, the functions $\tau_1$ and $\tau_2$ suffer from the same issue affecting 
the localizing function $G_\rho$; namely, a lack of smoothness in both space and time. 
Consequently, we utilize the space-time smoothing 
mechanism provided by the $C$-method to produce regularized versions of $\tau_1$ and $\tau_2$, which we 
denote $C^{\tau_1}$ and $C^{\tau_2}$, respectively:
\begin{subequations}\label{Ctau-vectors}
\begin{align}
\p_t C - \mathscr{L} \left[ C^{\tau_1} \,;\varepsilon,\kappa\right] = \frac{S(\bm{u})}{\varepsilon |\delta \bm{x} |} \, \tau_1 \,, \\[0.5em]
\p_t C - \mathscr{L} \left[ C^{\tau_2} \,;\varepsilon,\kappa \right] = \frac{S(\bm{u})}{\varepsilon |\delta \bm{x} |} \, \tau_2 \,,
\end{align}
\end{subequations}
where $\mathscr{L}$ is the operator defined in \eqref{C-2d-operator}. We also define the function
\begin{equation}\label{mu-normalize}
\mu = \mu(t) \coloneqq \max_{\Omega} \left\{  \max \left\{  |C^{\tau_1}|\,, |C^{\tau_2}|  \right\} \right\} \,,
\end{equation}
which can be used to produce a ``normalized'' tangent vector 
$\frac{1}{\mu} \vec{C^{\tau}} = \frac{1}{\mu}C^{\tau_1} \bm{e}_1  + \frac{1}{\mu} C^{\tau_2} \bm{e}_2$. 

\begin{figure}[H]
\centering
\subfigure[Vector plot of $\vec{C^\tau}$ at $t=4.0$]{\label{fig:Ctau-vectors1}\includegraphics[width=60mm]{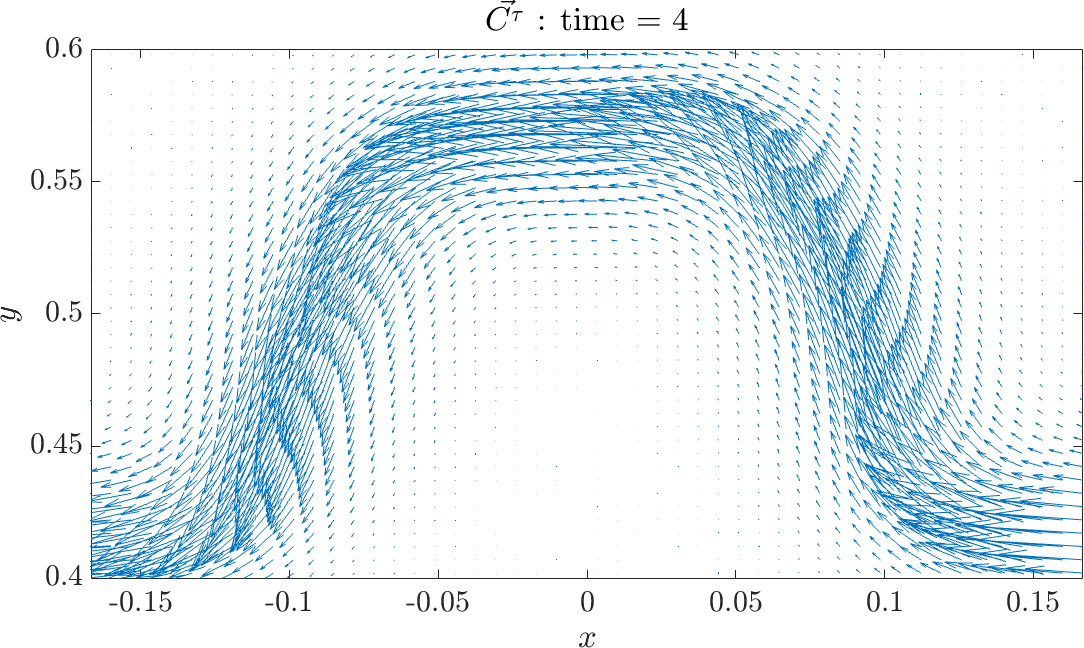}}
\hspace{2em}
\subfigure[Vector plot of $\vec{C^\tau}$ at $t=8.5$]{\label{fig:Ctau-vectors2}\includegraphics[width=60mm]{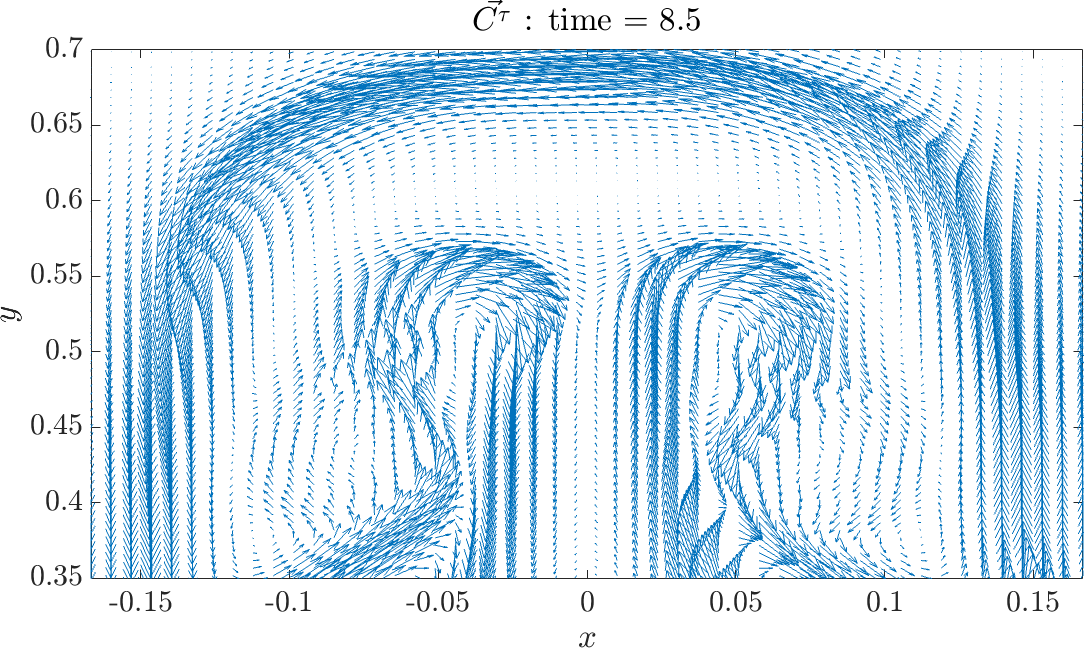}}
\par\bigskip
\subfigure[Surface plot of $\frac{1}{\mu} C^{\tau_1}$ at $t=4.0$]{\label{fig:Ctau-vectors3}\includegraphics[width=60mm]{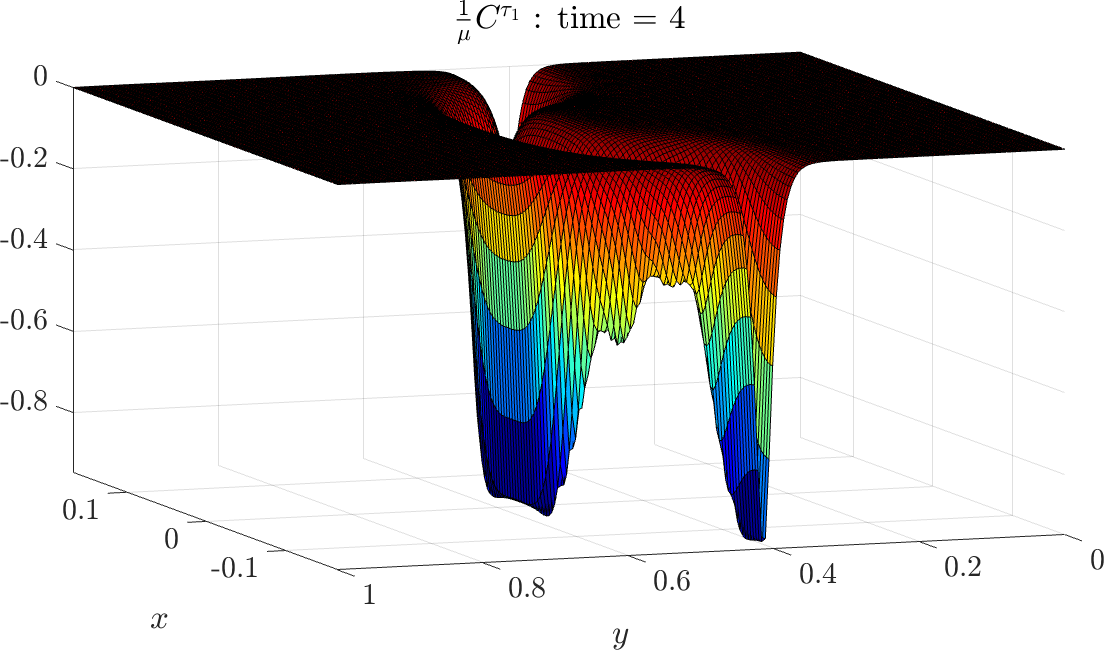}}
\hspace{2em}
\subfigure[Surface plot of $\frac{1}{\mu} C^{\tau_2}$ at $t=4.0$]{\label{fig:Ctau-vectors4}\includegraphics[width=60mm]{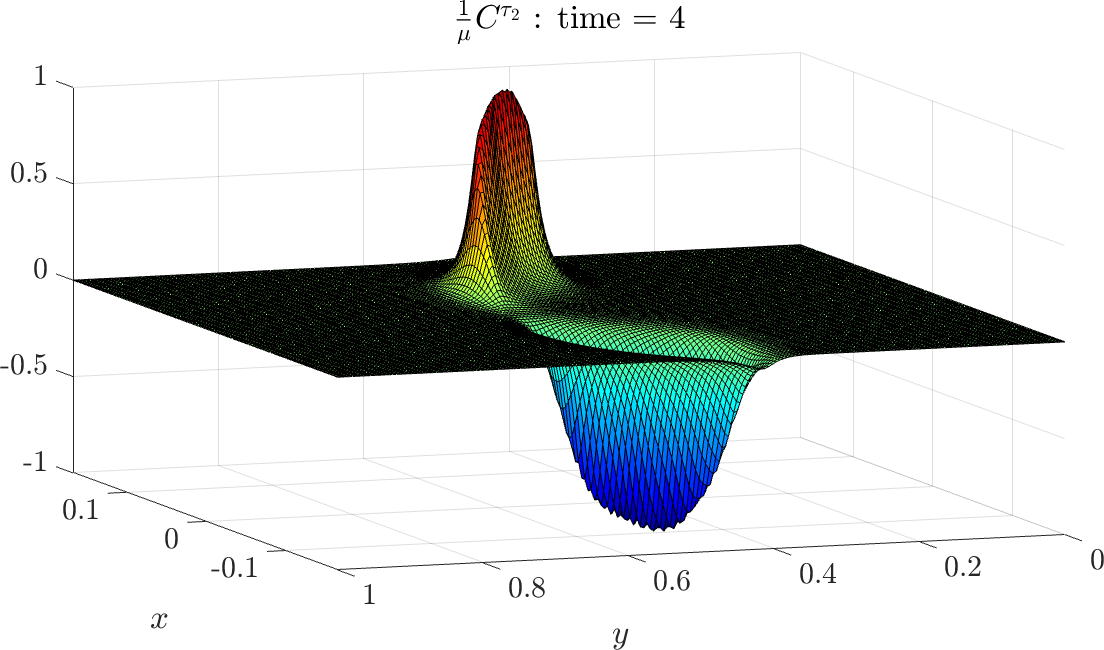}}
\caption{Calculation of the smoothed tangent vector $\vec{C^\tau}$ for the RT instability.}
\label{fig:Ctau-vectors}
\end{figure}

In Fig.\ref{fig:Ctau-vectors}, we 
provide vector plots of the vector $\vec{C^{\tau}}$, as well
as surface plots of the components $\frac{1}{\mu}C^{\tau_1}$ and $\frac{1}{\mu}C^{\tau_2}$. These should be contrasted with the 
corresponding figures in Fig.\ref{fig:tau-vectors}; we see that $\vec{C^{\tau}}$ is much smoother than 
$\vec{\tau}$, while remaining localized to the discontinuity $\Gamma(t)$. 

\subsection{Directional artificial viscosity and the Euler-$C^{\tau}$ system}

We now consider the following Euler-$C^{\tau}$ system for contact discontinuity evolution:
\begin{subequations}\label{EulerCtau-2D}
\begin{align}
\partial_t \rho + \divv (\rho \bm{u}) &=0 \,, \label{EulerCtau-density}\\[0.5em] 
\partial_t (\rho u) + \partial_x (\rho u^2 + p) + \p_y (\rho uv) &=  \p_i \left( \tilde{\beta} \, \rho \, C \, C^{\tau_i} \, C^{\tau_j} \, \p_j u \right)\, , \label{EulerCtau-momentum-u}\\[0.5em]
\partial_t (\rho v) + \p_x(\rho uv) + \partial_y (\rho v^2 + p) &=  \p_i \left( \tilde{\beta} \, \rho \, C \, C^{\tau_i} \, C^{\tau_j} \, \p_j v \right) \,, \label{EulerCtau-momentum-v}\\[0.5em]
\partial_t E + \divv (\bm{u}(E + p)) &=  0 \,, \label{EulerCtau-energy}
\end{align}
\begin{gather}
\p_t C - \mathscr{L}[C\,; \varepsilon,\kappa] 
 = \frac{S(\bm{u})}{\varepsilon | \delta \bm{x} |} G_\rho  \,, \quad  
 \p_t C^{\tau_i} - \mathscr{L}[C^{\tau_i} \,; \varepsilon,\kappa] 
 = \frac{S(\bm{u})}{\varepsilon | \delta \bm{x} |} \tau_i  \text{ for } i=1,2\, , \label{C-RT}
\end{gather}
\end{subequations}
where $G_{\rho}$ is given by \eqref{Grho}, $\tau_1$ and $\tau_2$ by \eqref{tau-vectors}, $\mathscr{L}$ by 
\eqref{C-2d-operator}, and 
we utilize the Einstein summation convention, where a repeated free index in the 
same term implies summation over all values of that index. 
We note that $\p_1 \equiv \p_x$ and $\p_2 \equiv \p_y$, and we use the two notations 
interchangeably. The artificial viscosity parameter is defined by 
\begin{equation}\label{artificial-viscosity-Ctau}
\tilde{\beta} = \frac{ |\delta \bm{x}|^2 }{\mu^2 \, \max_{\Omega} C} \, \beta\,,
\end{equation}
with $\mu = \max_{\Omega} \left\{ \max \left\{ |C^{\tau_1}| \,, |C^{\tau_2}| \right\} \right\}$.  

We note that the artificial viscosity operator $- \p_i \left( \tilde{\beta} \, \rho \, C \, C^{\tau_i} \, C^{\tau_j} \, \p_j  \right)$ is  a positive semi-definite
operator.  The proof is as follows:  taking the scalar product of  $- \p_i \left( \tilde{\beta} \, \rho \, C \, C^{\tau_i} \, C^{\tau_j} \, \p_j \bm{u} \right)$ with $\bm{u}$ and integrating
over $\Omega$ yields
$$
\int_{\Omega}  \tilde{\beta} \, \rho \, C \, C^{\tau_j} \, \p_j \bm{u} \cdot  \, C^{\tau_i} \p_i \bm{u} \,\mathrm{d}\bm{x}= 
\int_{\Omega}  \tilde{\beta} \, \rho \, C \,  |\p_{\vec{C^\tau}} \bm{u}|^2 \,  \mathrm{d}{\bm x} \ge \tilde \lambda \int_{\Omega}   |\p_{\vec{C^\tau}} \bm{u}|^2  \, \mathrm{d}{\bm x} \,,
$$
for some $\tilde \lambda \ge 0$.  Here, 
$\p_{\vec{C^\tau}} = \vec{C^{\tau}} \cdot \nabla $ denotes the smoothed tangential derivative operator.  Thus, the anisotropic artificial viscosity operator is
obtained as the Euler-Lagrange extremum associated to the function $\int  \tilde{\beta} \, \rho \, C \,  |\p_{\vec{C^\tau}} \bm{u}|^2 \,  \mathrm{d}{\bm x}$.
Just as in the case of isotropic artificial viscosity, our solutions to the Euler-$C^\tau$ system preserve total energy $E(t)$ (see \S \ref{sec:isotropic}).

\section{The $C$-method for shock-wall collision}\label{sec:C-wall-collision}

We now present a simple extension of the 1-$D$ shock-wall collision scheme (see \S3 \cite{RaReSh2018a}). Recall that the main novelty of the 1-$D$ shock
wall collision scheme is the use of a \emph{wall function} $\overline{C}(t)$ which naturally activates during
shock-wall collision and bounce-back. This allows for the addition of extra ``wall viscosity'' during 
shock-collision, which results in the suppression of post-collision noise while maintaining high-order accuracy 
prior to collision. 

The natural generalization to the two-dimensional setting results in the 2-$D$ Euler-$C$-$W$ scheme:
\begin{subequations}\label{EulerCW-2D}
\begin{align}
\partial_t \rho + \divv (\rho \bm{u}) &=0, \label{EulerCW-density}\\ 
\partial_t (\rho u) + \partial_x (\rho u^2 + p) + \p_y (\rho uv) &=  \divv \left( \mathcal{B}^u \, \rho \, C \, \nabla u \right), \label{EulerCW-momentum-u}\\
\partial_t (\rho v) + \p_x(\rho uv) + \partial_y (\rho u^2 + p) &=  \divv \left( \mathcal{B}^u \, \rho \, C \, \nabla v \right), \label{EulerCW-momentum-v}\\
\partial_t E + \divv (\bm{u}(E + p)) &=  \divv \left( \mathcal{B}^E \, \rho \, C \, \nabla(E/\rho) \right) \,, \label{EulerCW-energy}  \\
\p_t C - \mathscr{L}[C\,; \varepsilon,\kappa] &
 = \mathbbm{1}_{\divv \bm{u} < 0} \, \frac{S(\bm{u})}{\varepsilon | \delta \bm{x} |} G_\rho  \,,  \label{CW-Sod} \\
 \p_t C_w - \mathscr{L}[C_w\,; \varepsilon_w,\kappa_w] &
 = \mathbbm{1}_{\divv \bm{u} < 0} \, \frac{S(\bm{u})}{\varepsilon_w | \delta \bm{x} |} G_\rho  \,,  \label{CW-Sod2}
\end{align}
\end{subequations}
where $\mathscr{L}$ is the operator defined in \eqref{C-2d-operator}, $G_\rho$ is the forcing 
function defined in \eqref{Grho}, and $\mathbbm{1}_{\divv \bm{u} < 0}$ is a compression switch.  
The artificial viscosity parameters $\mathcal{B}$ are given by
\begin{equation}\label{CW-artificial-viscosity}
\mathcal{B}^{(\cdot)} = \frac{|\delta \bm{x}|^2}{\max_{\Omega} C} \left( \beta^{(\cdot)} + \beta^{(\cdot)}_w \overline{C}(x,t) \right) \,,
\end{equation}
with the wall function $\overline{C}(x,t)$ defined by 
$$
\overline{C}(x,t) = \frac{C_w(x,y_l,t)}{\max_{\Omega} C_w(x,y,t)}\,.
$$
Here, we  assume that the shock-wall collision occurs at the bottom boundary $y=y_l$. A Neumann 
boundary condition for $C_w$ is enforced at the bottom boundary $y=y_l$. As in the 1-$D$ case, this results  
in the smooth growth in time of the amplitude of $\overline{C}(x,t)$ as the shock approaches the wall, and 
allows for the addition of ``wall viscosity'' during shock-wall collision and bounce-back. 

\section{A wavelet-based 2-$D$ noise detection and removal algorithm}\label{sec:noise-ind}
In this section, we extend the noise indicator algorithm presented in \cite{RaReSh2018a} to the two-dimensional setting, 
using the same ideas as in the 1-$D$ case. Again, we construct a family of wavelets $\left\{ \psi_{i,j} \right\}$
and obtain a set of wavelet coefficients $\left\{ \mathcal{C}_{i,j} \right\}$, 
found by calculating the inner product of the 
highest frequency wavelets with the noisy function. These wavelet coefficients will indicate 
the location of the noise, and we employ a localized heat equation-based solver to remove this noise. 

\subsection{Construction of the highest frequency wavelets}\label{sec:2d-wavelet-construction}
\begin{figure}[H]
\centering\scalebox{.6}{
\begin{tikzpicture}
\filldraw [lightgray] (-5.5,7.5) --(5.5,7.5) -- (5.5,-0.5) -- (-5.5,-0.5);
\draw[very thick]  (-5.5,7.5) --(5.5,7.5) -- (5.5,-0.5) -- (-5.5,-0.5) -- cycle;
\draw[thick,dashed] (-5,7) -- (-5,0);
\draw[ thick,dashed] (5,7) -- (5,0);
\draw[ thick,dashed] (-5,7) -- (5,7);
\draw[ thick,dashed] (-5,0) -- (5,0);
\node[text width=1.75cm,align=center] at (-6.5,3.5) {left boundary};
\node[text width=1.75cm,align=center] at (6.5,3.5) {right boundary};
\node[text width=1.75cm,align=center] at (0,8.0) {top boundary};
\node[text width=1.75cm,align=center] at (0,-1) {bottom boundary};
\draw[very thick] (-5.5,7.5) -- (-2.5,7.5) -- (-2.5,4.5) -- (-5.5,4.5) -- cycle;
\draw[very thick] (-5.5,6.5)--(-2.5,6.5);
\draw[very thick] (-5.5,5.5)--(-2.5,5.5);
\draw[very thick] (-4.5,7.5)--(-4.5,4.5);
\draw[very thick] (-3.5,7.5)--(-3.5,4.5);
\draw[very thick] (-2.5,7.5) -- (0.5,7.5) -- (0.5,4.5) -- (-2.5,4.5) -- cycle;
\draw[very thick] (-2.5,6.5)--(0.5,6.5);
\draw[very thick] (-2.5,5.5)--(0.5,5.5);
\draw[very thick] (-1.5,7.5)--(-1.5,4.5);
\draw[very thick] (-0.5,7.5)--(-0.5,4.5);
\draw[thick,color=red] (-5,7)--(-3,7)--(-3,5)--(-5,5)-- cycle;
\draw[thick,color=red] (-3,7)--(-1,7)--(-1,5)--(-3,5)-- cycle;
\filldraw [black] (-5,7) circle (2pt);
\filldraw [black] (-4,7) circle (2pt);
\filldraw [black] (-3,7) circle (2pt);
\filldraw [black] (-5,6) circle (2pt);
\filldraw [black] (-4,6) circle (2pt);
\filldraw [black] (-3,6) circle (2pt);
\filldraw [black] (-5,5) circle (2pt);
\filldraw [black] (-4,5) circle (2pt);
\filldraw [black] (-3,5) circle (2pt);
\filldraw [black] (-2,5) circle (2pt);
\filldraw [black] (-1,5) circle (2pt);
\filldraw [black] (-2,6) circle (2pt);
\filldraw [black] (-1,6) circle (2pt);
\filldraw [black] (-2,7) circle (2pt);
\filldraw [black] (-1,7) circle (2pt);
\filldraw [black] (0,7) circle (2pt);
\filldraw [black] (0,6) circle (2pt);
\filldraw [black] (0,5) circle (2pt);
\filldraw [black] (1,6) circle (0.5pt);
\filldraw [black] (1.25,6) circle (0.5pt);
\filldraw [black] (1.5,6) circle (0.5pt);
\filldraw [black] (-4,4) circle (0.5pt);
\filldraw [black] (-4,3.75) circle (0.5pt);
\filldraw [black] (-4,3.5) circle (0.5pt);
\filldraw [black] (-1,4) circle (0.5pt);
\filldraw [black] (-1,3.75) circle (0.5pt);
\filldraw [black] (-1,3.5) circle (0.5pt);
\filldraw [black] (0.75,4) circle (0.5pt);
\filldraw [black] (1.0,3.75) circle (0.5pt);
\filldraw [black] (1.25,3.5) circle (0.5pt);
\end{tikzpicture}}
\caption{Grid setup for the construction of the wavelets in 2-$D$. The dashed black curve 
denotes the boundary $\partial \Omega$, while the solid black lines indicate cell edges.
The black dots indicate cell centers, and the domains bounded by 
red lines indicate the support of each of the 
highest frequency wavelets.}
\label{fig:wavelet-construction-2d}
\end{figure}
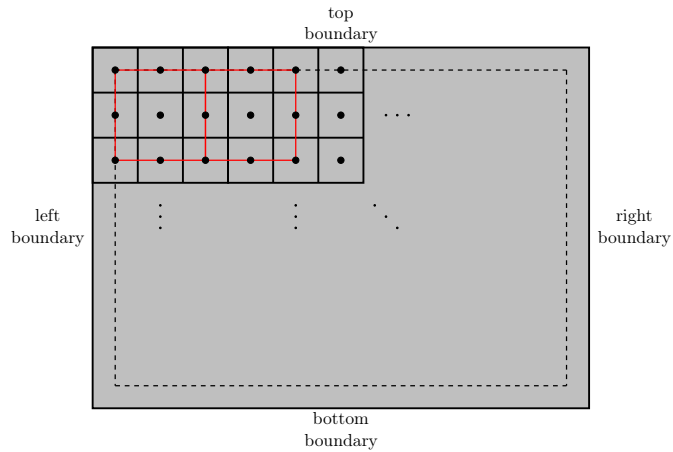

We first discretize our grid by assuming we have $M$ cells in the 
$x$-direction, and $N$ cells in the $y$-direction. Label the cell centers by $(x_i,y_j)$, 
where $x_i$ and $y_j$ are given by
\begin{align*}
x_i &= x_l + (i-1) \cdot \delta x \quad \text{for } i=1,\ldots,M, \\
y_j &= y_l + (j-1) \cdot \delta y \quad \text{for } j=1,\ldots,N,
\end{align*}
with $\delta x = (x_r-x_l)/(M-1)$ and $\delta y = (y_r - y_l)/(N-1)$. 
We group the cells into 3x3 blocks of 9 cells each, and 
then define the highest frequency wavelet with support over the domain spanned by the cell 
centers in each of these blocks, as shown in Fig.\ref{fig:wavelet-construction-2d}.

This yields a set of $\frac{(M-1)}{2} \times \frac{(N-1)}{2}$ highest frequency wavelets. We denote 
these wavelets by $\psi_{i,j} = \psi_{i,j}(x,y)$, for $i = 1,\ldots,(M-1)/2$ and $j = 1,\ldots,(N-1)/2$, 
with each $\psi_{i,j}$ is supported in the rectangular 
domain $\mathcal{I}_{i,j} = [x_{2i-1},x_{2i+1}] \times [y_{2j-1},y_{2j+1}]$. 

We now have to fix a form for the 2-$D$ wavelet.
The two key properties that are required of the  wavelet family are:
\begin{enumerate}
\item Zero mean: 
\[
\int_\Omega \psi_{i,j}(\bm{x}) \,\mathrm{d}\bm{x} = 0. 
\]  
\item ``Quasi-orthogonality'' of the form:
\[
\int_\Omega \psi_{i,j}(\bm{x}) \cdot \psi_{r,s}(\bm{x}) \,\mathrm{d}\bm{x} = \delta_{ir} \delta_{js}, \text{ for } i, r=1,\frac{M-1}{2} \text{ and } j, s =1,\ldots,\frac{N-1}{2}, 
\]
so that each of the highest frequency wavelets is orthogonal to every other wavelet of the same frequency.
\end{enumerate}

\begin{figure}[H]
\centering
\subfigure[the highest frequency wavelet $\psi$]{\label{fig:2d-wavelet1}\includegraphics[width=75mm]{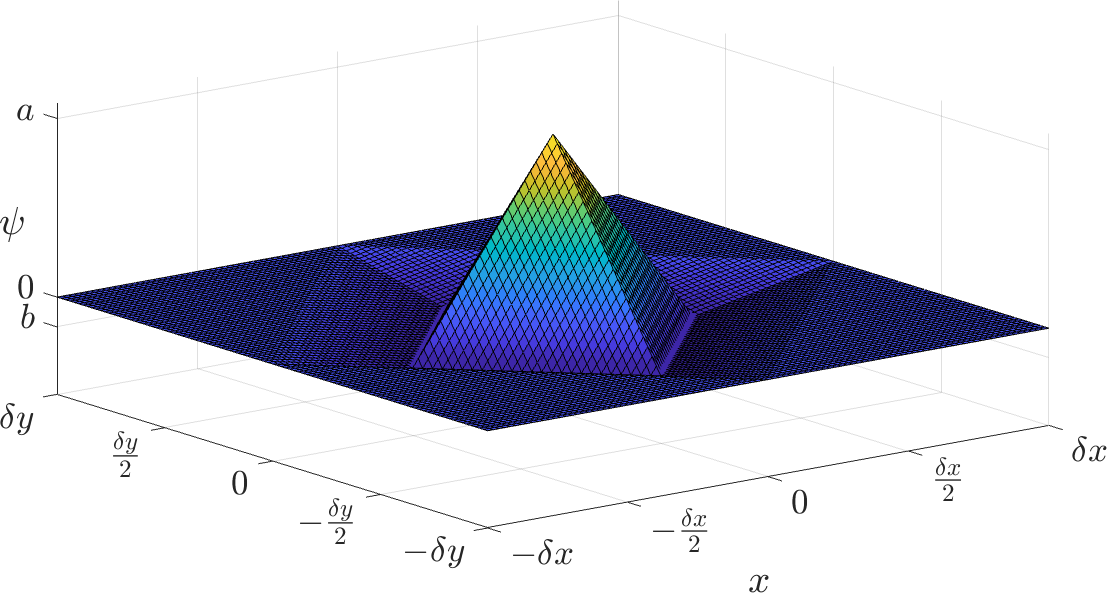}}
\subfigure[cross sectional view of $\psi$ along $y=0$]{\label{fig:2d-wavelet2}\includegraphics[width=75mm]{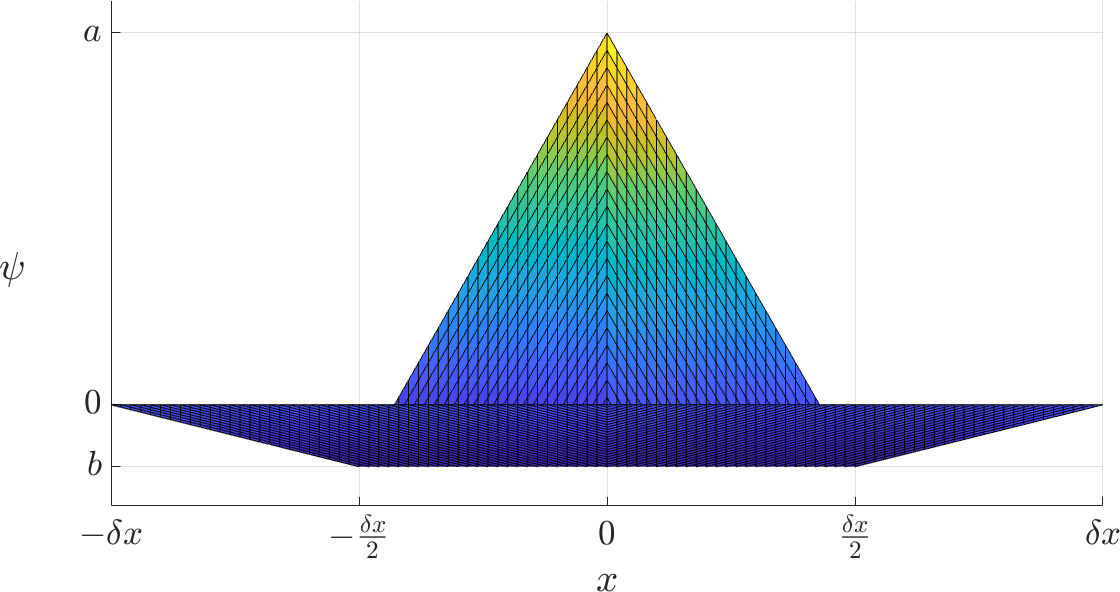}}
\caption{The highest frequency wavelet $\psi = \psi(x,y)$ in 2-$D$. We assume that the support of $\psi$ is 
$\mathcal{I} = [-\delta x,\delta x] \times [-\delta y, \delta y]$.}
\label{fig:2d-wavelet}
\end{figure}

We recall from \cite{RaReSh2018a} that each member of the 1-$D$ wavelet family is oscillatory  and
 orthogonal to all linear functions. Consequently, 
we design the 2-$D$ wavelet family to also have these two properties.  
For simplicity, assume that a highest frequency wavelet $\psi$
 is supported in the domain 
$[-\delta x,\delta x] \times [-\delta y, \delta y]$. The highest frequency wavelets $\psi_{i,j}$ are then obtained
by translation of $\psi$ to the domain $\mathcal{I}_{i,j}$. Our wavelets take the form shown in 
Fig.\ref{fig:2d-wavelet}. 

The exact formula for the highest frequency wavelet $\psi$ that is supported in the domain 
$\mathcal{I} = [-\delta x,\delta x] \times [-\delta y, \delta y]$ is
$$
\psi(x,y) = \begin{cases}
			\psi^{(1)}(x,y), &\text{if } 0 \leq \lvert \frac{x}{\delta x} \rvert + \lvert \frac{y}{\delta y} \rvert \leq \frac{1}{2}, \\[1em]
			\psi^{(2)}(x,y), &\text{if } \frac{1}{2} \leq \lvert \frac{x}{\delta x} \rvert + \lvert \frac{y}{\delta y} \rvert \leq 1,
			\end{cases}
$$
where $\psi^{(1)}$ and $\psi^{(2)}$ are given by
\begin{align*}
\psi^{(1)}(x,y) &= \begin{cases}
				a+{2(b-a)} \left( \frac{x}{\delta x} + \frac{y}{\delta y} \right), &\text{if } x>0, y>0, \\[1em]
				a+{2(b-a)} \left( -\frac{x}{\delta x} + \frac{y}{\delta y} \right), &\text{if } x<0, y>0, \\[1em]
				a+{2(b-a)} \left( -\frac{x}{\delta x} - \frac{y}{\delta y} \right), &\text{if } x<0, y<0, \\[1em]
				a+{2(b-a)} \left( \frac{x}{\delta x} - \frac{y}{\delta y} \right), &\text{if } x>0, y<0,
				\end{cases} \\[2em]
\psi^{(2)}(x,y) &= \begin{cases}
				2b-{2b} \left(\frac{x}{\delta x} + \frac{y}{\delta y}\right), &\text{if } x>0, y>0, \\[1em]
				2b-{2b} \left(-\frac{x}{\delta x} + \frac{y}{\delta y}\right), &\text{if } x<0, y>0, \\[1em]
				2b-{2b} \left(-\frac{x}{\delta x} - \frac{y}{\delta y}\right), &\text{if } x<0, y<0, \\[1em]
				2b-{2b} \left(\frac{x}{\delta x} - \frac{y}{\delta y}\right), &\text{if } x>0, y<0. 
				\end{cases}
\end{align*}
Here, the values of $a$ and $b$ are chosen so that $\psi$ satisfies 
$\lvert \lvert \psi \rvert \rvert_{L^2(\Omega)} = 1$ and the zero mean condition 
$\langle \psi\,,1 \rangle_{L^2(\Omega)} = 0$, and can be calculated as
$$
a = -6b \ \ \text{ and } \ \ b = -\sqrt{ \frac{3}{8} \cdot \frac{1}{\delta x \, \delta y}}. 
$$
The highest frequency wavelets $\psi_{i,j}$ are then obtained by translation of $\psi$ to the domain 
$\mathcal{I}_{i,j} = [x_{2i-1},x_{2i+1}] \times [y_{2j-1},y_{2j+1}]$. 

Now, given a function $f(x,y)$ defined at the cell centers $(x_i,y_j)$, we wish to calculate the inner 
product of $f$ with each of the highest frequency wavelets. The first step is to approximate the function 
$f(x,y)$ over $\mathcal{I}_{i,j}$, the support of $\psi_{i,j}$. In 1-$D$, a function $f(x)$ is approximated as 
a piecewise linear function over 
the interval $\mathcal{I}_i$, the support of a 1-$D$ highest frequency wavelet, 
by linearly interpolating between the cell center values (see \S4 in \cite{RaReSh2018a}). 
The analogue of a line in the two-dimensional setting
is a plane, so a first attempt is to approximate $f(x,y)$ by a plane in each of the sub-cells 
$[x_r,x_{r+1}] \times [y_{s},y_{s+1}]$. However, this is not possible, since 3 points define a plane, whereas each of the sub-cells contains 4 cell center values.

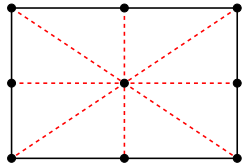
\begin{figure}[H]
\centering\scalebox{.5}{
\begin{tikzpicture}
\draw[very thick] (-3,4)--(3,4)--(3,0)--(-3,0)--cycle;
\draw[very thick,dashed,color=red] (-3,4)--(3,0);
\draw[very thick,dashed,color=red] (3,4)--(-3,0);
\draw[very thick,dashed,color=red] (0,4)--(0,0);
\draw[very thick,dashed,color=red] (-3,2)--(3,2);
\filldraw [black] (-3,4) circle (3pt);
\filldraw [black] (0,4) circle (3pt);
\filldraw [black] (3,4) circle (3pt);
\filldraw [black] (-3,2) circle (3pt);
\filldraw [black] (0,2) circle (3pt);
\filldraw [black] (3,2) circle (3pt);
\filldraw [black] (-3,0) circle (3pt);
\filldraw [black] (0,0) circle (3pt);
\filldraw [black] (3,0) circle (3pt);
\end{tikzpicture}}
\caption{Dividing $\mathcal{I}_{i,j}$ into 8 regions.}
\label{fig:2d-wavelet-region}
\end{figure}

Our solution is to divide $\mathcal{I}_{i,j}$ into 8 regions as shown in Fig.\ref{fig:2d-wavelet-region}.
Each of these regions contains precisely 3 cell-center values, and these 3 values define a plane. Thus, 
we can approximate $f(x,y)$ by a plane in each of these 8 regions, and define a piecewise
linear function $\tilde{f}(x,y)$ such that $\tilde{f}(x_i,y_j) = f(x_i,y_j)$. We may then 
approximate the 
$(i,j)$-th wavelet coefficient as
\begin{equation}\label{2d-wavelet-coefficient}
\begin{split}
\mathcal{C}_{i,j}(f) := \langle f \,, \psi_{i,j} \rangle_{L^2(\Omega)} &\approx \langle \tilde{f} \,, \psi_{i,j} \rangle_{L^2(\Omega)} \\[1em]
&= -\frac{\sqrt{6 \, \delta x \, \delta y}}{192} \,\,
\begin{Bmatrix}
f_{2i-1,2j+1} & + & 2f_{2i,2j+1} & + & f_{2i+1,2j+1} & + \\[0.5em]
2f_{2i-1,2j} & + & -12f_{2i,2j} & + &  2f_{2i+1,2j} & + \\[0.5em]
f_{2i-1,2j-1} & + & 2f_{2i,2j-1} & + & f_{2i+1,2j-1} & \phantom{+}
\end{Bmatrix},
\end{split}
\end{equation}
where $f_{m,n} = f(x_m,y_n)$. One can verify that the inner product of $\psi_{i,j}$ with an arbitrary
plane is identically zero i.e. if $f(x,y) = \alpha x + \beta y$ for some constants $\alpha$ and $\beta$, then 
$\mathcal{C}_{i,j}(u) = 0$ for every $i=1,\ldots,(M-1)/2$ and $j=1,\ldots,(N-1)/2$. 
This is analogous to the result in 1-$D$ that each of the highest frequency wavelets
is orthogonal to linear functions.

\subsection{Algorithm}\label{section-noise-indicator-2d-algorithm}
Given a noisy function $\tilde{u}(x,y)$, we now present an algorithm for first detecting and then
subsequently removing the noise. The algorithm is almost identical to that presented for the 1-$D$ case 
(see \S4 in \cite{RaReSh2018a}). 

\subsubsection{Noise detection in the presence of discontinuities}
We first calculate, using formula \eqref{2d-wavelet-coefficient}, the 
$\frac{M-1}{2} \times \frac{N-1}{2}$ wavelet coefficients, each associated with one of 
the highest frequency wavelets. The coefficients that are largest in magnitude should indicate
the location of the noise. However, as in the 1-$D$ case, it is possible that the wavelet coefficients 
that are largest in magnitude are actually indicating the location of the curve of discontinuity $\Gamma(t)$. 
Since the $C$-method is taking care of artificial diffusion in this region, one needs to manually 
``turn off'' the noise detection in a small region surrounding the shock curve $\Gamma(t)$. 
There are numerous ways to do this, but we employ one of the simplest methods, namely to turn off the noise 
detection if 
\begin{equation}\label{deltaoff}
\frac{C}{\max_{\Omega} C} > \delta_{\operatorname{off}}\,,
\end{equation}
where 
$\delta_{\operatorname{off}}$ is some value between 0 and 1. A typical range of 
values for $\delta_{\mathrm{off}}$ is $\delta_{\mathrm{off}} \in [0.05,0.25]$. 

\begin{figure}[H]
\centering
\subfigure[reference oscillation: a hat function]{\label{fig:hat-function-2d-1}\includegraphics[width=70mm]{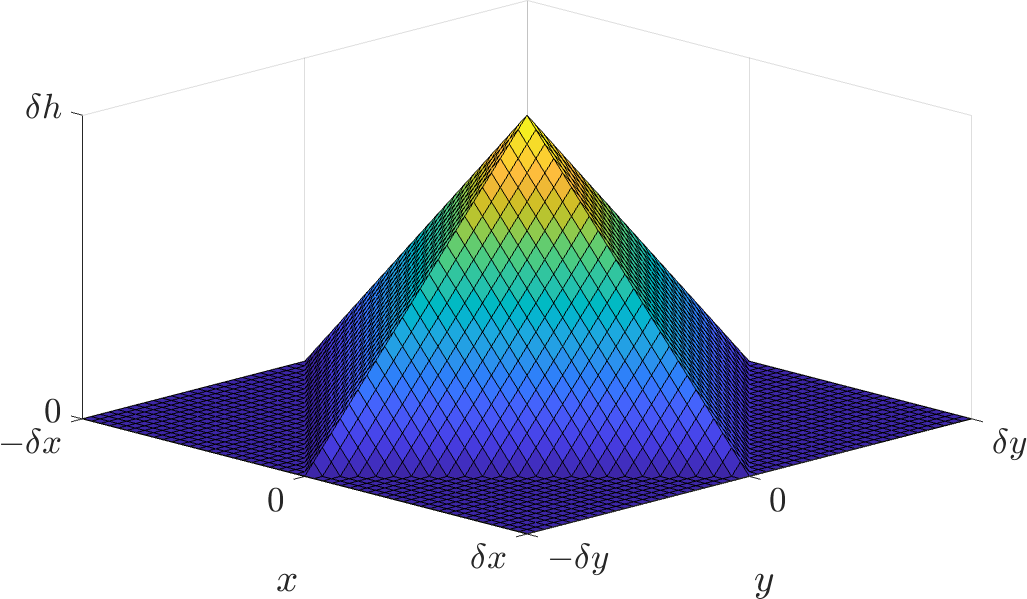}}
\hspace{1em}
\subfigure[cross sectional view along $y = 0$]{\label{fig:hat-function-2d-2}\includegraphics[width=70mm]{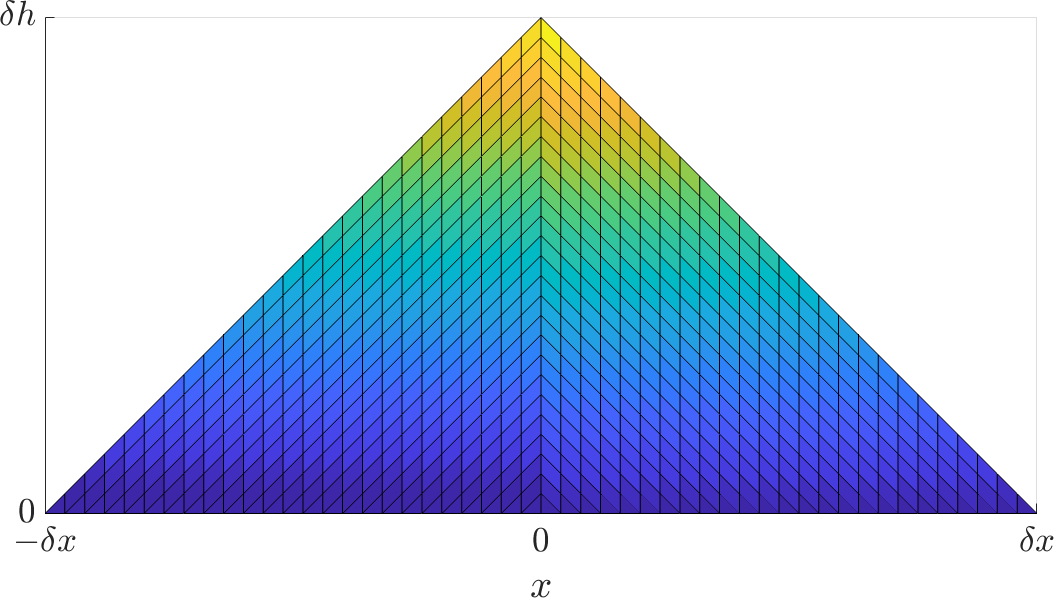}}
\caption{The 2-$D$ reference oscillation: a hat function with amplitude $\delta h$.}
\label{fig:2d-hat-function}
\end{figure}

With noise detection deactivated in the region surrounding $\Gamma(t)$, the wavelet coefficients that
are largest in magnitude
now indicate the location of the noise. We now ``turn on'' a noise detector function
$\mathbbm{1}_{\operatorname{noise}}(x,y)$ in the domain $\mathcal{I}_{i,j}$ if the associated wavelet 
coefficient $ \mathcal{C}_{i,j} $ satisfies $| \mathcal{C}_{i,j}| \geq \mathcal{C}_{\mathrm{ref}}$. 
The constant $\mathcal{C}_{\operatorname{ref}}$ is the wavelet coefficient obtained from a ``typical'' 
high-frequency oscillation, namely a hat function (see Fig.\ref{fig:2d-hat-function}) 
centered in the domain $[-\delta x, + \delta x] \times [-\delta y, + \delta y]$
with amplitude $\delta h$. The associated wavelet coefficient may then be calculated as
\begin{equation}\label{Cref}
\mathcal{C}_{\operatorname{ref}} = \delta h \frac{ \sqrt{6\, \delta x\,  \delta y}}{16}\,.
\end{equation}

\subsubsection{Noise removal with a localized heat equation}
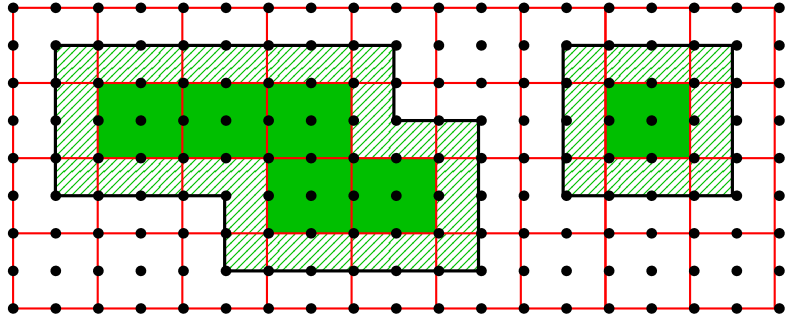
\begin{figure}[H]
\centering
\begin{tikzpicture}
\filldraw [pattern=north east lines, pattern color=black!25!green,very thick] (-4.4375,3.5)--(0.0625,3.5)--(0.0625,2.5)--(1.1875,2.5)--(1.1875,0.5)--(-2.1875,0.5)--(-2.1875,1.5)--(-4.4375,1.5)--cycle;
\filldraw [pattern=north east lines, pattern color=black!25!green,very thick] (2.3125,3.5)--(4.5625,3.5)--(4.5625,1.5)--(2.3125,1.5)--cycle;
\filldraw [black!25!green] (-3.875,3) --(-2.75,3) -- (-2.75,2) -- (-3.875,2);
\filldraw [black!25!green] (-2.75,3) --(-1.625,3) -- (-1.625,2) -- (-2.75,2);
\filldraw [black!25!green] (-1.625,3) --(-0.5,3) -- (-0.5,2) -- (-1.625,2);
\filldraw [black!25!green] (-1.625,2) --(-0.5,2) -- (-0.5,1) -- (-1.625,1);
\filldraw [black!25!green] (-0.5,2) --(0.625,2) -- (0.625,1) -- (-0.5,1);
\filldraw [black!25!green] (2.875,3) --(4,3) -- (4,2) -- (2.875,2);
\draw[thick,color=red] (-5,4)--(-5,0);
\draw[thick,color=red] (-3.875,4)--(-3.875,0);
\draw[thick,color=red] (-2.75,4)--(-2.75,0);
\draw[thick,color=red] (-1.625,4)--(-1.625,0);
\draw[thick,color=red] (-0.5,4)--(-0.5,0);
\draw[thick,color=red] (0.625,4)--(0.625,0);
\draw[thick,color=red] (1.75,4)--(1.75,0);
\draw[thick,color=red] (2.875,4)--(2.875,0);
\draw[thick,color=red] (2.875,4)--(2.875,0);
\draw[thick,color=red] (4,4)--(4,0);
\draw[thick,color=red] (5.125,4)--(5.125,0);
\draw[thick,color=red] (-5,4)--(5.125,4);
\draw[thick,color=red] (-5,3)--(5.125,3);
\draw[thick,color=red] (-5,2)--(5.125,2);
\draw[thick,color=red] (-5,1)--(5.125,1);
\draw[thick,color=red] (-5,0)--(5.125,0);
\draw [line width=4pt, line cap=round, dash pattern=on 0pt off 4\pgflinewidth] (-5,4)--(5.5,4);
\draw [line width=4pt, line cap=round, dash pattern=on 0pt off 4\pgflinewidth] (-5,3.5)--(5.5,3.5);
\draw [line width=4pt, line cap=round, dash pattern=on 0pt off 4\pgflinewidth] (-5,3)--(5.5,3);
\draw [line width=4pt, line cap=round, dash pattern=on 0pt off 4\pgflinewidth] (-5,2.5)--(5.5,2.5);
\draw [line width=4pt, line cap=round, dash pattern=on 0pt off 4\pgflinewidth] (-5,2)--(5.5,2);
\draw [line width=4pt, line cap=round, dash pattern=on 0pt off 4\pgflinewidth] (-5,1.5)--(5.5,1.5);
\draw [line width=4pt, line cap=round, dash pattern=on 0pt off 4\pgflinewidth] (-5,1)--(5.5,1);
\draw [line width=4pt, line cap=round, dash pattern=on 0pt off 4\pgflinewidth] (-5,0.5)--(5.5,0.5);
\draw [line width=4pt, line cap=round, dash pattern=on 0pt off 4\pgflinewidth] (-5,0)--(5.5,0);
\end{tikzpicture}
\caption{Construction of the domains $\mathcal{V}_k$ and $\tilde{\mathcal{V}}_k$. The squares
bounded by the red lines indicate the support of each of the highest frequency wavelets $\psi_{i,j}$. 
The shaded green regions indicate where the noise indicator algorithm detects noise, 
and represent the domains $\mathcal{V}_k$. The 
hatched regions indicate the extension of each $\mathcal{V}_k$ to $\tilde{\mathcal{V}}_k$. 
The domains $\tilde{\mathcal{V}}_k$ are the regions bounded by the solid black lines. }
\label{fig:2d-domain-construction}
\end{figure}

The noise removal process in the 2-$D$ case is identical to that in 1-$D$. We first construct the domain 
$\mathcal{V}$, given by the union of all domains $\mathcal{I}_{i,j}$ such that the noise detector function 
$\mathbbm{1}_{\operatorname{noise}}(x,y)$ is non-zero in $\mathcal{I}_{i,j}$. We write 
$\mathcal{V}$ as the union of its connected subsets $\mathcal{V} = \bigcup_k \mathcal{V}_k$, and then 
define the domains $\tilde{\mathcal{V}}_k$ as the domain $\mathcal{V}_k$ extended by one cell in each 
outward in each direction (see Fig.\ref{fig:2d-domain-construction}). 
For example, if $\mathcal{V}_k = [x_1,x_2] \times [y_1,y_2]$, then 
$\tilde{\mathcal{V}}_k = [x_1 - \delta x, x_2 +\delta x] \times [y_1 - \delta y, y_2 + \delta y]$. 

A localized heat equation with Dirichlet boundary conditions is then solved in each of the domains 
$\tilde{\mathcal{V}}_k$ for a ``de-noised'' solution $u(x,y,\tau)$, 
\begin{subequations}
\begin{alignat}{2}
\partial_{\tau} u(x,y,\tau) &= \eta \cdot \Delta u(x,y,\tau),\quad  &&\text{for } \bm{x} \in \tilde{\mathcal{V}}_k \text{ and } \tau > 0, \\
u(x,y,0) &= \tilde{u}(x,y), &&\text{for } \bm{x} \in \tilde{\mathcal{V}}_k, \label{noise-removal-init} \\
u(x,y,\tau) &= \tilde{u}(x,y), &&\text{for } \bm{x} \in \partial \tilde{\mathcal{V}}_k \text{ and } \tau > 0, \label{noise-removal-bc}
\end{alignat}
\end{subequations}
while $u(x,y,\tau) = \tilde{u}(x,y)$ for $\bm{x} \in \left( \bigcup_k \tilde{\mathcal{V}}_k \right)^{\mathcal{C}}$ 
and $\tau \geq 0$. The time $\tau$ is a fictitious time, introduced solely for the diffusion mechanism, while
$0 < \eta \ll 1$ is a small constant, which we refer to as the noise removal viscosity. Equation 
\eqref{noise-removal-init} is the initial condition, and \eqref{noise-removal-bc} is a Dirichlet boundary 
condition ensuring continuity of $u(x,y,\tau)$ over the domain $\Omega$. 

However, as in the 1-$D$ case, it is not necessary to explicitly construct the domains 
$\tilde{\mathcal{V}}_k$. Instead, 
one can use the noise detector function $\mathbbm{1}_{\operatorname{noise}}(x,y)$ and solve a
modified heat equation with Dirichlet boundary conditions, given by
\begin{subequations}\label{noise-removal-alt}
\begin{alignat}{2}
\partial_{\tau} u(x,y,\tau) &= \eta \cdot \mathbbm{1}_{\operatorname{noise}}(x,y) \cdot \Delta u(x,y,\tau),\quad &&\text{for } \bm{x} \in \Omega \text{ and } \tau > 0, \\
u(x,y,0) &= \tilde{u}(x,y), &&\text{for } \bm{x} \in \Omega, \\
u(x,y,\tau) &= \tilde{u}(x,y), &&\text{for } \bm{x} \in \partial \Omega \text{ and } \tau > 0.  
\end{alignat}
\end{subequations}  

In practice, the system \eqref{noise-removal-alt} is solved using a simple, explicit forward Euler time 
integration scheme, along with a second order central difference approximation for the spatial derivatives. 
Moreover, in the simulations shown below, a single time-step is sufficient to remove noise;
thus, our procedure is the equivalent of the inversion of a Helmholtz elliptic operator, and can hence be
 viewed as a \emph{filtering} process, in which high frequency noise is 
eliminated from the solution through a local averaging or localized frequency truncation. 
However, we remark that for certain simulations with very large amplitude high frequency noise, it 
may be necessary to solve the localized heat equation for several time steps to completely 
remove the oscillations.

\subsection{Implementation of the algorithm for the Euler equations}\label{sec:noise-implementation}

We now describe how we implement the noise detection and removal algorithm described above for the 
particular case of the Euler system \eqref{Euler-2d}. Suppose that we are given the solution 
$\bm{U} = \bm{U}(x,y,t_n)$ at time $t = t_n = n \cdot \delta t$, and we wish to 
calculate the solution $\bm{U}(x,y,t_{n+1})$ at time $t = t_{n+1} =(n+1) \cdot \delta t$. 
The implementation proceeds in two steps:
\begin{enumerate}
\item We first compute in the usual manner the 
(potentially noisy) solution at time $t=t_{n+1}$. We denote this solution by $\tilde{\bm{U}}(x,y)$. 
In the numerical studies below, a simplified WENO-based scheme for the spatial discretization 
and an explicit Runge-Kutta method for the time integration are used to calculate this solution.  
\item We then pass the potentially noisy velocity components $\tilde{u}(x,y)$ and $\tilde{v}(x,y)$ through the
noise detection and removal algorithm described in \S\ref{section-noise-indicator-2d-algorithm}, to 
produce de-noised velocity components $u(x,y,t_{n+1})$ and $v(x,y,t_{n+1})$. We then \emph{define} the 
solution $\bm{U}(x,y,t_{n+1})$ at time $t =t_{n+1}$ by 
$$
\bm{U}(x,y,t_{n+1}) \equiv 
\begin{bmatrix}
\rho(x,y,t_{n+1}) \\ \rho u(x,y,t_{n+1}) \\ \rho v(x,y,t_{n+1}) \\ E(x,y,t_{n+1})
\end{bmatrix} \coloneqq 
\begin{bmatrix}
\tilde{\rho}(x,y) \\ \tilde{\rho}(x,y) \cdot u(x,y,t_{n+1}) \\ \tilde{\rho}(x,y) \cdot v(x,y,t_{n+1}) \\ \tilde{E}(x,y)
\end{bmatrix} \,.
$$
\end{enumerate}

This algorithm mimics the 1-$D$ version in \cite{RaReSh2018a}. We apply this procedure 
to the Rayleigh-Taylor instability in \S\ref{sec:RT} and demonstrate its ability to suppress 
spurious high frequency noise that otherwise corrupts the solution. 

\begin{remark}
We note that the noise removal procedure for the horizontal velocity component
$\tilde{u}$ is completely independent from the noise removal procedure for the vertical velocity component $\tilde{v}$. 
It is also perhaps more useful to view our algorithm as a predictor-corrector method, in which we first compute 
auxiliary quantities using the WENO-based portion of the algorithm, and then ``correct'' these quantities by 
removing high frequency noise from the solutions in the corrector portion of the method. Extensive 
testing of the method in the 1-$D$ setting {\cite{RaReSh2018a}} shows that the noise reduction algorithm 
successfully eliminates high-frequency noise from the auxiliary solution, thereby producing 
``corrected'' solutions with smaller errors. In the 
2-$D$ setting, we provide error analysis and convergence tests for the 
Sedov  problem in Table {\ref{table:sedov}}. These tests indicate that the noise removal 
algorithm decreases the error of the computed solution;
furthermore, our numerical experiments for the Noh problem in \S{\ref{sec:Noh}} 
and Rayleigh-Taylor problem in \S{\ref{sec:RT}} demonstrate that, 
qualitatively, the noise removal algorithm greatly reduces the numerical error of solutions. Further 
quantitative evidence in the form of error analysis and convergence tests is presented in {\cite{RaReSh2019}}.
\end{remark}

\section{The numerical algorithm for the $C$-method}\label{sec:numerical-implementation}
The systems \eqref{Euler-2d}, \eqref{EulerC-2D}, \eqref{EulerCtau-2D}, and 
\eqref{EulerCW-2D} are discretized using a simplified finite-differencing 
WENO scheme for the nonlinear flux terms, and 
a standard central difference approximation for the diffusion terms. Time integration is done using a 
 $k^{\text{th}}$ order Runge-Kutta scheme. For each simulation, a fixed $\Delta t$ is used at every 
 time step, so that the CFL number can change at each time level; however, for each of the problems presented,  the time step $\Delta t$ is chosen so that the CFL condition is not violated at any time $t$ during 
 the simulation.
For convenience, we provide full details of the numerical discretization of these systems in Appendix \ref{appendix:WENO}. In particular, we refer the reader to Table \ref{table:schemes} listing the various
methods (and combinations of the methods) that we will use for the numerical tests. 

We stress that the WENO-type discretization we 
use is highly simplified, and is not meant to be representative of the class of full WENO solvers. 
However, we note that, for certain problems, our simplified WENO-type discretization produces 
solutions with similar errors and convergence rates to those produced using a standard WENO scheme 
(see \S5.2.5 in {\cite{RaReSh2018a}}).

As with any artificial viscosity scheme, parameters must be chosen for the particular problem 
under consideration. All of the relevant parameters for the schemes considered are listed in 
Table {\ref{table:parameters}}.In {\cite{RaReSh2018a}}, we suggest some practical guidelines 
on choosing these parameters; a brief summary of the discussion 
in {\cite{RaReSh2018a}} is the following. We choose the artificial viscosity parameters 
$\beta^{(\cdot)}$, $\beta_w^{(\cdot)}$, and $\eta$ large enough to damp post-shock oscillations and 
high frequency noise both pre and post shock-wall collision, while the parameters $\varepsilon$ and $\kappa$ 
in the $C$-equations control the support and smoothness of the $C$-functions. For a more thorough 
discussion, we refer the reader to \S5 of {\cite{RaReSh2018a}}.

{
\begin{table}[H]
\centering
{\small
\begin{tabular}{|M{4cm} | M{6cm}|} 
 \hline
 Parameter / Variable & Description \\ [0.0em] 
 \hline \hline 
$\beta^{u}$, $\beta^E$ & artificial viscosity coefficients for the momentum 
 and energy, respectively. \\[0.5em] 
\hline
$\beta^{u}_w$, 
 $\beta^E_w$ &  wall viscosity coefficients for the
 momentum and energy, respectively. \\[0.5em] 
 \hline 
 $\delta h$, $\delta_{\mathrm{off}}$, $\eta$ & amplitude of noise, noise detection deactivation parameter,
  and noise removal viscosity, respectively. \\[0.5em] 
 \hline 
 $\varepsilon$, $\varepsilon_w$ & parameters controlling support 
 of $C$ (and $C^{\tau_i}$) and $C_w$, respectively. \\[0.5em]
 \hline
 $\kappa$, $\kappa_w$ & parameters controlling smoothness 
 of $C$ (and $C^{\tau_i}$) and $C_w$, respectively. \\[0.5em]
 \hline
\end{tabular}}
\caption{Relevant parameters and variables used in the numerical tests.}
\label{table:parameters}
\end{table}}

\subsection{Accuracy study: linear advection}

For the purposes of demonstrating the high order convergence of the base WENO-type scheme, 
we consider the following linear advection equation \mbox{\cite{LiWe2003,Jiang1996202}}
 \begin{subequations}
\label{advection-2d}
\begin{alignat}{2}
\partial_t \varphi(\bm{x},t)+ \divv ( \bm{a}  \varphi(\bm{x},t))) = {0},& && \ \ \  \bm{x} \in [-1,1]^2 \,,   t > 0,  \label{advection-2d-motion} \\
\varphi (\bm{x},0)  = 1+0.2 \sin(\pi (x+y)),& && \ \ \ \bm{x} \in [-1,1]^2 \,,   t = 0,
\end{alignat}
\end{subequations}
with $\bm{a} =(1,-0.5)$. Periodic boundary conditions are employed, and 
the exact solution at time $t$ is given by $\varphi(\bm{x},t) = 1+0.2 \sin(\pi (x+y - 0.5 t ))$.
The problem is run on grids with $25 \times 25$, $50 \times 50$, $100 \times 100$, and 
$200 \times 200$ cells 
until the final time $t=4$, at which time the sinusoidal wave has been 
advected one full wavelength. We choose the time-step so that CFL=0.8. 
Following {\cite{LiWe2003}}, the relative $L^1$ and relative $L^{\infty}$ errors are listed in percentage form in 
Table {\ref{table:advection}}. Our simplified WENO-type scheme achieves the advertised fifth-order 
convergence rate, and the errors are similar to those produced with an ``industry-standard'' WENO method 
{\cite{LiWe2003}}. 

\begin{table}[H]
\centering
\renewcommand{\arraystretch}{1.0}
\scalebox{0.8}{
\begin{tabular}{|lcc|cccc|}
\toprule
\midrule
\multirow{2}{*}{\textbf{Scheme}} & \multirow{2}{*}{\textbf{Error}} & & \multicolumn{4}{c|}{\textbf{Cells}}\\

{}  & & & $25 \times 25$   & $50 \times 50$    & $100 \times 100$   & $200 \times 200$ \\
\midrule
\multirow{2}{*}{WENO} & \multirow{2}{*}{$L^1$} & Error & 
$2.894 \times 10^{-2}$  & $9.014 \times 10^{-4}$  & $2.820 \times 10^{-5}$ & $8.821 \times 10^{-7}$\\
				  &  & Order & -- & 5.005   & 4.998  & 4.999 \\
\midrule
\multirow{2}{*}{WENO} & \multirow{2}{*}{$L^{\infty}$} & Error & 
$5.254 \times 10^{-2}$  & $1.929 \times 10^{-3}$  & $6.253 \times 10^{-5}$ & $1.970 \times 10^{-6}$\\
				  &  & Order & -- & 4.767   & 4.947  & 4.988 \\
\midrule
\bottomrule
\end{tabular}}
\caption{Relative $L^1$ and $L^{\infty}$ errors at $t=4$ 
of the computed solution minus the exact solution at and convergence for the 
linear advection problem. The errors are given in percentage form.}
\label{table:advection}
\end{table}

\section{The Noh infinite strength shock problem}\label{sec:Noh}
We begin our numerical experiments by considering a 
radially symmetric version of a classic test of Noh \cite{Noh1987,LiWe2003}. 
This problem simulates an infinite strength shock wave formed by uniformly compressing a cold gas with 
 constant velocity 1 directed towards the origin. The initial pressure is identically zero, but following 
 \cite{LiWe2003} we use the initial value $p_0 = 10^{-6}$. 
The 
domain is $\Omega = [-1,1]^2 \subset \mathbb{R}^2$, the adiabatic constant is $\gamma = 5/3$, and the 
initial data is 
\begin{equation}\label{noh_initialdata1}
\begin{bmatrix}
\rho_0 \\ (\rho u)_0 \\ (\rho v)_0 \\ E_0 
\end{bmatrix}
=
\begin{bmatrix}
1 \\ -\cos(\theta) \\ -\sin(\theta) \\ 0.5+10^{-6}/(\gamma-1)
\end{bmatrix}\,,
\end{equation}
where $\theta \in [0,2\pi)$ is the polar angle. The final time is $t=2.0$. 

The exact solution consists of a shock front moving radially outwards into the cold gas with speed $1/3$. 
The shock is of ``infinite strength'', since the sound speed satisfies $c=0$ in the cold gas. 
The numerical solution is computed in the 
positive quadrant $[0,1]^2$ and then reflected appropriately to yield the 
solution on all of $[-1,1]^2$. Consequently, reflecting boundary conditions are employed on the 
$x$ and $y$ axes. 
The exact solution is used to enforce the boundary conditions at the boundaries $x=1$ and $y=1$ (see
\cite{LiWe2003} for the details). 

The Noh test is a difficult problem; among the numerical methods 
considered in \cite{LiWe2003}, only the PPM and CFLFh schemes produce somewhat satisfactory 
solutions with sharp shock fronts, though the solutions are still noisy and have large errors at the origin 
due to the phenomenon of anomalous \emph{wall-heating}
\cite{Rider2000,RaReSh2018a,LiWe2003,Noh1987}. 
In particular, the WENO scheme considered in 
\cite{LiWe2003} fails for this problem. Consequently, it is not so surprising that our grossly simplified 
WENO scheme also fails for this problem. That is to say, our stand-alone WENO scheme is unable to 
run until the final time $t=2.0$; the solution develops noise in the region with the cold gas, which eventually
causes a violation of the positivity of the density and, subsequently, blow-up of the solution. This is the case 
even for a very small time-step $\delta t$; using CFL=$5.0 \times 10^{-3}$ still results in blow-up. 
We propose the 
use of the $C$-method and the noise detection and removal algorithm to deal with these issues. 

\subsection{Application of WENO-$C$-$N$ to the Noh problem}\label{Noh-weno-c-n}

We will apply the WENO-$C$-$N$ scheme on a grid with 200$\times$200 cells in the 
domain $[0,1]^2$ with a time-step $\delta t = 5 \times 10^{-4}$, giving a CFL number of approximately
0.25. A modified noise detection algorithm is employed, which 
we now describe. The ``input functions'' in the noise detection procedure 
described in \S\ref{sec:noise-implementation} are the (potentially noisy) horizontal and vertical velocities 
$\tilde{u}(x,y)$ and $\tilde{v}(x,y)$. The noise detection procedure for each of these functions 
produces two different ``noise indicator functions'', denoted by 
$\mathbbm{1}^{u}_{\operatorname{noise}}(x,y)$ and $\mathbbm{1}^{v}_{\operatorname{noise}}(x,y)$. 
The function $\mathbbm{1}^{u}_{\operatorname{noise}}(x,y)$ is then used to de-noise the 
velocity field $\tilde{u}$, while $\mathbbm{1}^{v}_{\operatorname{noise}}(x,y)$ is used to de-noise $\tilde{v}$. 
For the Noh problem, we exploit the radial symmetry available by using the radial velocity
$\tilde{u}_r(x,y) \coloneqq \cos(\theta) \tilde{u}(x,y) + \sin(\theta) \tilde{v}(x,y)$
to produce a noise indicator function $\mathbbm{1}^{r}_{\operatorname{noise}}(x,y)$, where $\theta$ is the 
polar angle. The de-noising procedure for $\tilde{u}$ and $\tilde{v}$ then uses the function 
$\mathbbm{1}^{r}_{\operatorname{noise}}(x,y)$, but is otherwise identical to the algorithm described in 
\S\ref{section-noise-indicator-2d-algorithm} and \S\ref{sec:noise-implementation}. 

The parameters for the WENO-$C$-$N$ scheme are chosen as
\begin{gather*}
\beta^u=50.0, \qquad  \beta^E=350.0, \qquad \varepsilon=200.0, \qquad  \kappa=0.5,  \\
\eta \cdot \delta \tau /|\delta \bm{x}|^2=5 \times 10^{-2}, \qquad  \delta h=10^{-5}, \qquad \delta_{\mathrm{off}}=0.2\,.
\end{gather*}
The artificial viscosity term 
on the right-hand side of the energy equation \eqref{EulerC-energy}
serves to correct for the wall-heating error. The local heat equation solver for noise removal is iterated for 
a single time step only. 

\begin{figure}[H]
\centering
\subfigure[]{\label{fig:noh-rho1}\includegraphics[width=40mm]{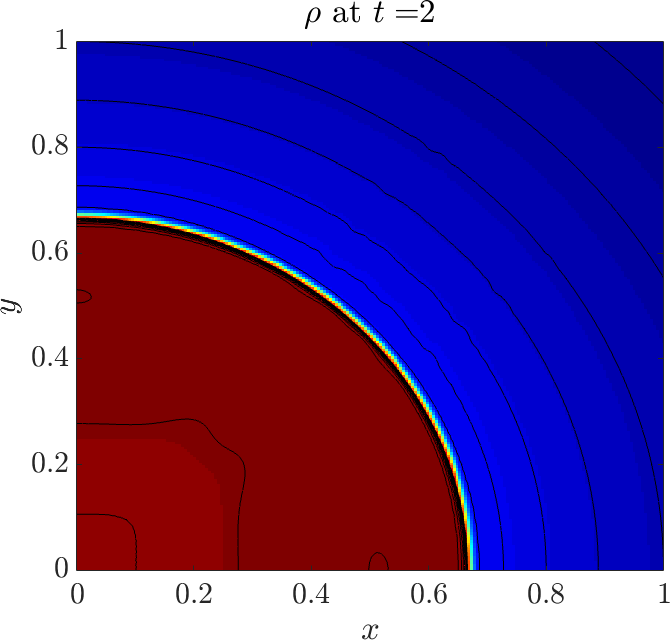}}
\hspace{2em}
\subfigure[]{\label{fig:noh-rho2}\includegraphics[width=70mm]{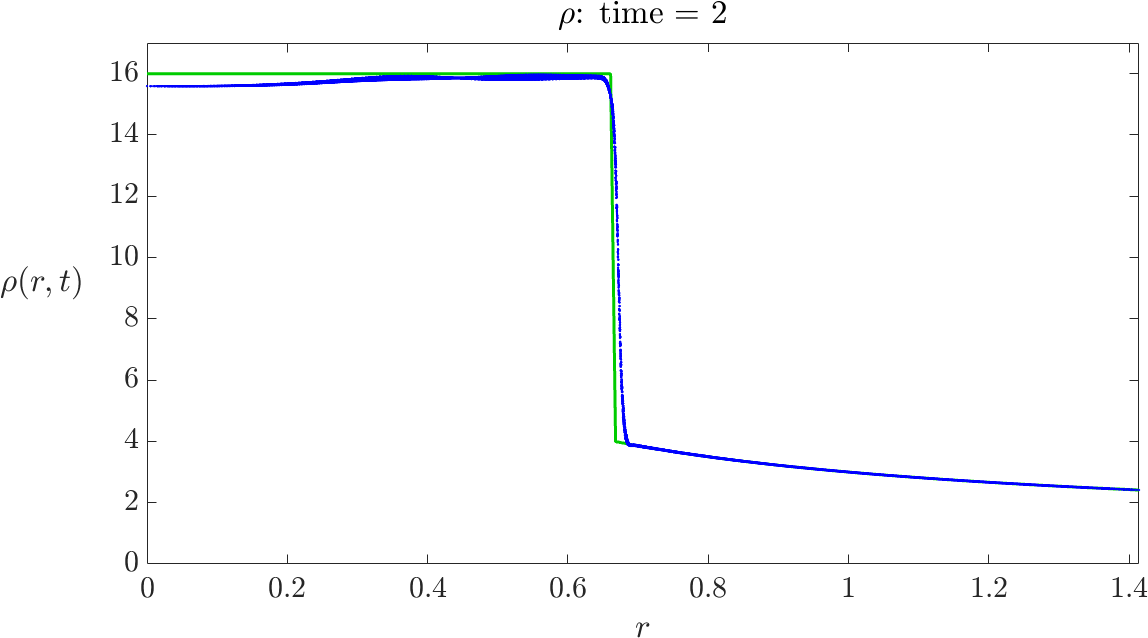}}
\caption{Application of WENO-$C$-$N$ to the Noh problem on the grid $[0,1]^2$ 
with 200$\times$200 cells. Shown on
the left is a heatmap of the density $\rho$ at time $t=2.0$. This is overlaid by 23 density contours, 
from 2.5 to 4 with step 0.25, and 14 to 17 with step 0.2. On the right is a scatter plot of the density 
versus radius. The green curve is the exact solution.}
\label{fig:noh-rho}
\end{figure}

We provide in Fig.\ref{fig:noh-rho} a heatmap plot of the density computed using WENO-$C$-$N$, 
as well as a scatter plot of the density 
versus radius. We see that the space-time smooth artificial diffusion provided by the $C$-method stabilizes
the strong shock wave and prevents spurious oscillations from developing behind the solution, while 
the artificial heat conduction term on the right-hand side of the energy equation \eqref{EulerC-energy} 
significantly reduces the wall-heating error. The noise
detection and removal procedure prevents high frequency noise from corrupting the solution and does not 
affect the sharpness of the shock front. Moreover, the solution maintains, for the most part, 
angular symmetry, though there are minor variations in the azimuthal direction; this should be contrasted
with the results presented in {\cite{LiWe2003}}, which show a much more obvious lack of symmetry by other schemes.

\subsection{Comparison with Noh's artificial viscosity scheme}

For the purposes of comparison, we also implement our WENO scheme together with a modification of Noh's 
artificial viscosity scheme \cite{Noh1987}, which is designed specifically for the Noh problem. The resulting scheme is referred to as WENO-Noh. 
 Noh's scheme couples the classical artificial viscosity method 
of Von Neumann and Richtmeyer with a heat conduction term for the energy equation. In particular, the 
diffusion terms on the right-hand sides of \eqref{EulerC-momentum-u}, \eqref{EulerC-momentum-v}, 
and \eqref{EulerC-energy} are replaced with the terms
\begin{subequations}\label{weno-noh-scheme}
\begin{gather}
\divv \left(  \tilde{\beta}^u_N \, \rho \, | \nabla u_r | \nabla u \right) + \tilde{\alpha}^u_N \Delta u\,,\\
\divv \left(  \tilde{\beta}^u_N \, \rho \, | \nabla u_r | \nabla v  \right) + \tilde{\alpha}^u_N \Delta v\,, \\
 \divv \left(  \tilde{\beta}^e_N \, \rho \, | \nabla u_r | \nabla e \right) + \tilde{\alpha}^e_N \Delta e\,,
\end{gather}
\end{subequations}
respectively, where $u_r$ is the radial velocity, $e = p/\rho(\gamma-1)$ is the internal energy of the system, 
and the artificial viscosity parameters $\tilde{\beta}_N$ and $\tilde{\alpha}_N$ are defined by 
$$
\tilde{\beta}_N = \frac{|\delta \bm{x}|^2}{\max_{\Omega} | \nabla u_r |} \beta_N \quad \text{ and } 
\quad \tilde{\alpha}_N = |\delta \bm{x}| \alpha_N\,.
$$ 
The parameter $\beta_N$ controls the amount of \emph{classical artificial viscosity} added to the system,
 while the parameter $\alpha_N$ controls the amount of \emph{linear viscosity} added to the system. 
 We refer the reader to \cite{RaReSh2018a} for a discussion on the differences between the 
 diffusion terms used in WENO-Noh and the diffusion terms used in WENO-$C$-$N$. 

\begin{figure}[H]
\centering
\subfigure[]{\label{fig:noh-classical1}\includegraphics[width=40mm]{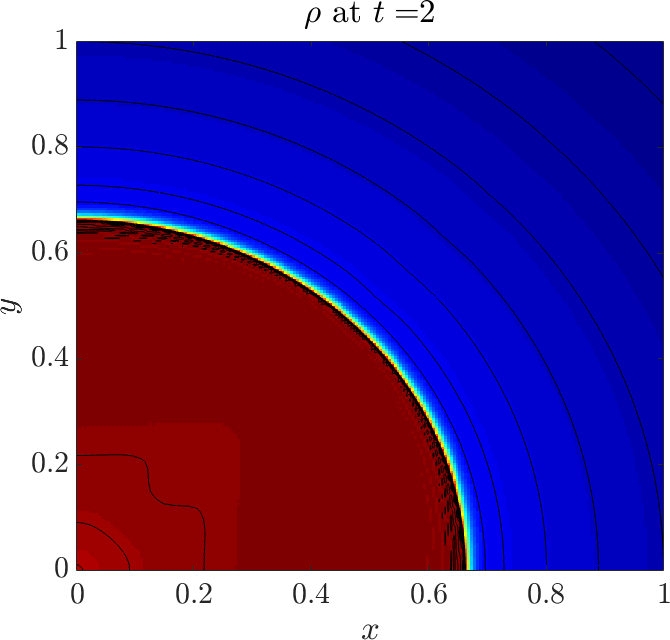}}
\hspace{2em}
\subfigure[]{\label{fig:noh-classical2}\includegraphics[width=70mm]{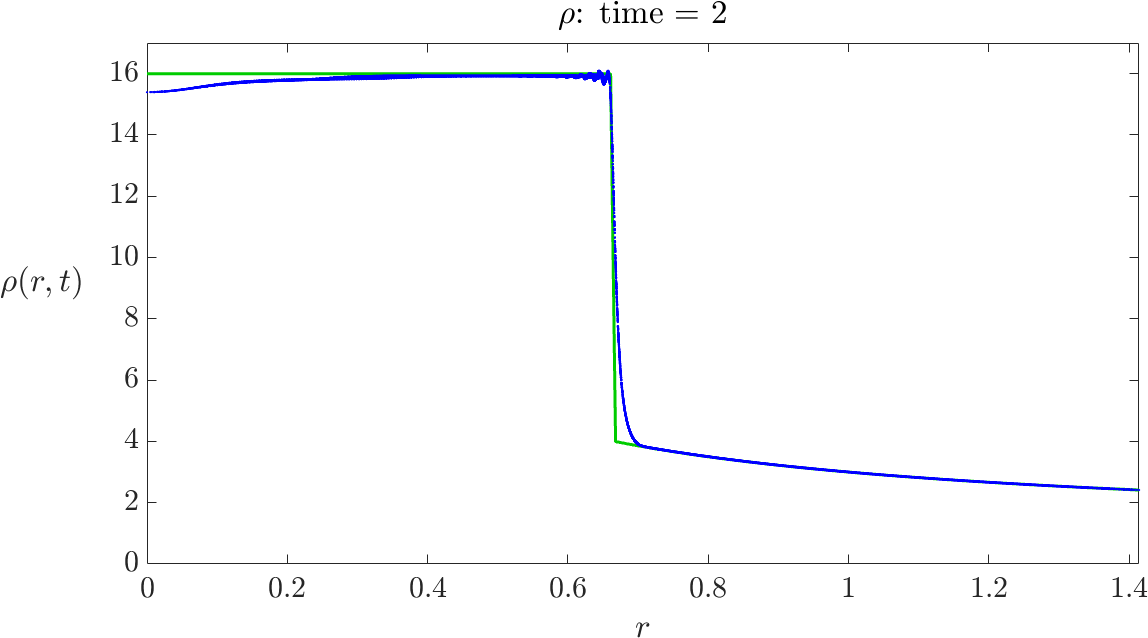}}
\caption{Application of WENO-Noh to the Noh problem on the grid $[0,1]^2$ 
with 200$\times$200 cells. Shown on
the left is a heatmap of the density $\rho$ at time $t=2.0$. This is overlaid by 23 density contours, 
from 2.5 to 4 with step 0.25, and 14 to 17 with step 0.2. On the right is a scatter plot of the density 
versus radius. The green curve is the exact solution.}
\label{fig:noh-classical}
\end{figure} 

We implement the above WENO-Noh scheme for the Noh problem, with the following choices of viscosity 
parameters: $\beta^u_{N} = 50.0$, $\beta^e_N=350.0$, $\alpha^u_N=0.5$, and 
$\alpha^e_N=1.5$. The results are shown in Fig.\ref{fig:noh-classical}. 
The WENO-Noh scheme produces a solution that is, for the most part, oscillation-free;
however, there are still some oscillations in the density profile behind the shock, and the shock curve is overly smeared. We remark 
here that the use of the linear viscosity term in \eqref{weno-noh-scheme} was needed to allow the WENO-Noh scheme to run; indeed, 
when that linear viscosity term was removed, the numerical simulation could not run as the solution blew-up.
 This is due primarily to the extremely
oscillatory nature of the localizing function $|\nabla u_r|$; the additional linear viscosity stabilized the
solution, at the cost of a very smeared shock curve and a loss of accuracy.  

The solution produced with 
WENO-$C$-$N$ has a much 
sharper shock front. In fact, a simple computation shows that $\max_{\Omega} | \nabla \rho(x,y,2)|  \approx 856$ and 
$\max_{\Omega} | \nabla \tilde{\rho}(x,y,2)| \approx 774$, where $\rho$ is the WENO-$C$-$N$ solution for the 
density and 
$\tilde{\rho}$ is the WENO-Noh solution for the density. Consequently, we see that 
WENO-$C$-$N$ produces a less oscillatory, more accurate solution with a sharper shock front. 

\section{The Sedov blast wave}
The Sedov problem \mbox{\cite{Sedov1946,Taylor1950b,Bethe1947}} 
models the self-similar evolution of a cylindrical blast 
wave,  arising from a point-source explosion in a cold, uniform density fluid. The computational 
domain is $\Omega = [0,1.2] \times [0,1.2] \subset \mathbb{R}^2$. The initial density and velocity are, respectively,
 $\rho_0 = 1$, $u_0 = v_0 = 0$, and the initial energy is set to $E_0 = 10^{-12}$ everywhere except in 
 the lower left corner cell, where it takes the value $\frac{0.244816}{\delta x \cdot \delta y}$.  
 The adiabatic constant is $\gamma=1.4$, reflecting boundary conditions are employed at the 
 bottom and left boundaries, while inhomogeneous Dirichlet boundary conditions are enforced on the
 top  boundary and the right boundary; in particular, the velocity, density, and energy are set equal to the corresponding values of the initial data on those boundaries. 
 
 The solution consists of an infinite strength radially symmetric shock front propagating outwards 
 from the origin, leaving behind a near vacuum state. 
 The simulation is run until time $t=1.0$, at which time the shock front is a circle with radius equal to 1. 
 A semi-exact solution exists for this problem 
 \mbox{\cite{Sedov1959,Kamm2000,Kamm2007}}, and we shall use the code 
 made available by {\cite{cococubed}} to compute this solution. Our WENO and 
 WENO-$C$-$N$ schemes are employed on a grid with 96 cells in both
directions, and with a time-step $\delta t = 10^{-4}$. This fixed time-step was 
chosen as the largest possible value for which the WENO simulation runs on the desired time interval. 
For the noise detection portion of our WENO-$C$-$N$ scheme, we  utilize the modified 
algorithm for radially symmetric flow, described in \S{\ref{Noh-weno-c-n}} for the Noh problem. The 
parameters for the $C$-$N$-method are as follows:
\begin{gather*}
\beta^u=1.0, \qquad  \beta^E=10.0, \qquad \varepsilon=1.0, \qquad  \kappa=0.5,  \\
\eta \cdot \delta \tau /|\delta \bm{x}|^2=10^{-2}, \qquad  \delta h=10^{-4}, \qquad \delta_{\mathrm{off}}=0.02\,.
\end{gather*}

\begin{figure}[H]
\centering
\subfigure[]{\label{fig:sedov-weno-u1}\includegraphics[width=40mm]{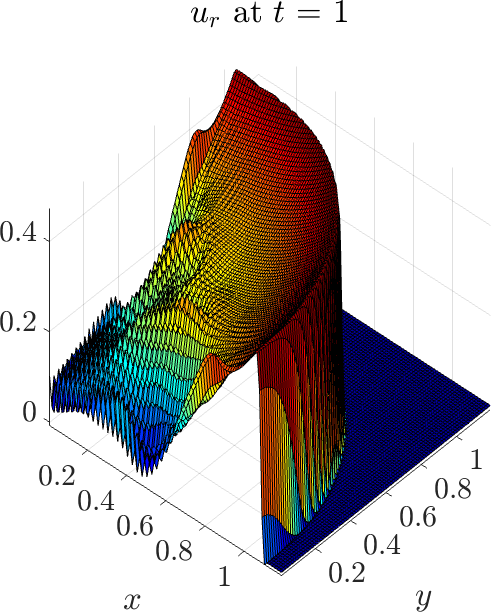}}
\hspace{2em}
\subfigure[]{\label{fig:sedov-weno-u2}\includegraphics[width=70mm]{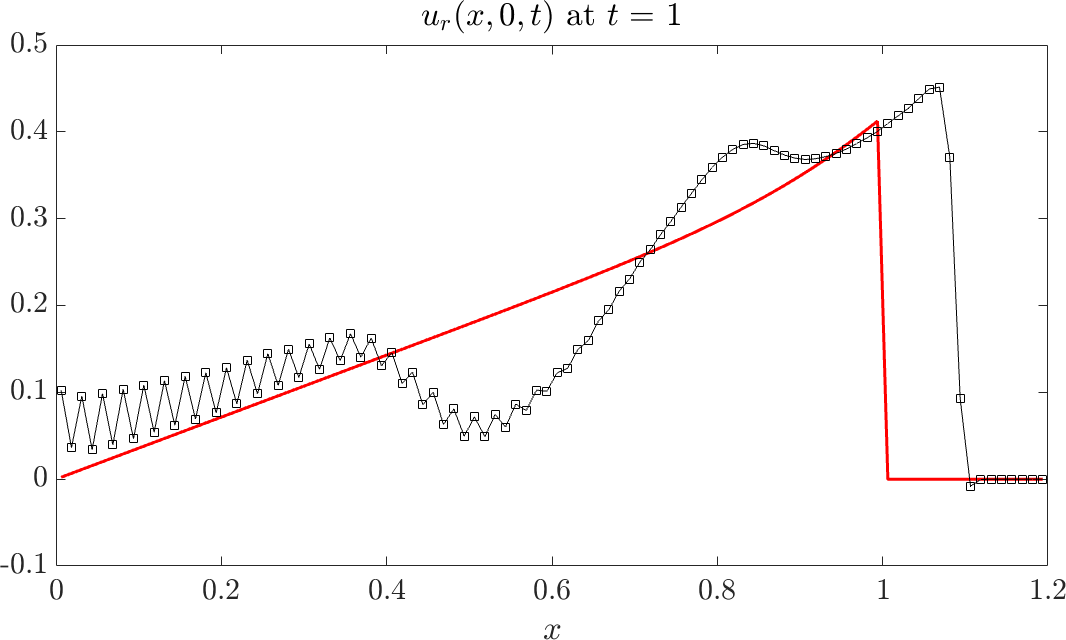}}
\caption{Application of our standalone WENO scheme to the Sedov problem on the grid $[0,1.2]^2$ 
with 96$\times$96 cells. Shown are
(a) a surface plot of the radial velocity $u_r$, (b) $u_r$ 
along the cut $y=0$ The red curves are the exact solutions.}
\label{fig:sedov-weno}
\end{figure}

\begin{figure}[H]
\centering
\subfigure[]{\label{fig:sedov-weno-c-n-u1}\includegraphics[width=40mm]{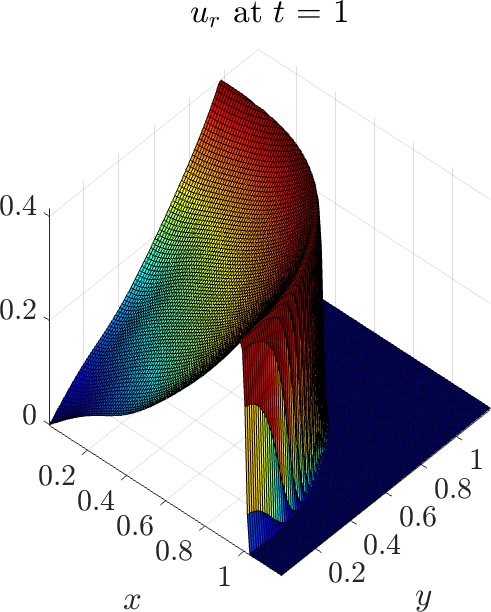}}
\hspace{2em}
\subfigure[]{\label{fig:sedov-weno-c-n-u2}\includegraphics[width=70mm]{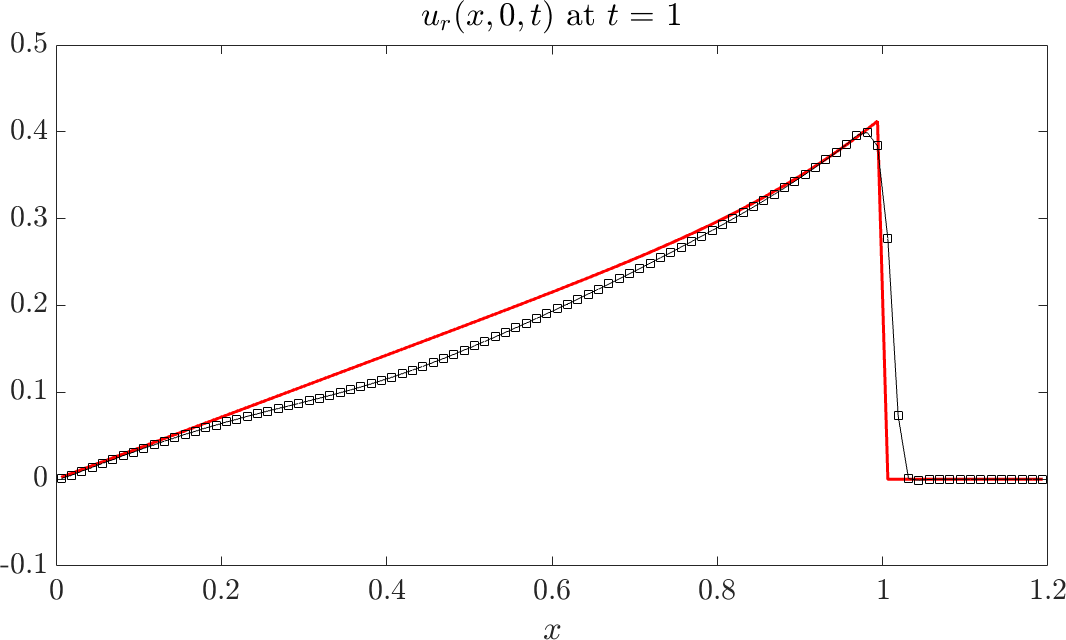}}
\caption{Application of WENO-$C$-$N$ to the Sedov problem on the grid $[0,1.2]^2$ 
with 96$\times$96 cells. Shown are
(a) a surface plot of the radial velocity $u_r$, (b) $u_r$ 
along the cut $y=0$ The red curves are the exact solutions.}
\label{fig:sedov-weno-c-n}
\end{figure}

The results are shown in Figs.{\ref{fig:sedov-weno}} and {\ref{fig:sedov-weno-c-n}}. 
Our stand-alone  WENO scheme produces  a solution with inaccurate shock speed
and location, and is corrupted by a large amount of high-frequency oscillations (or noise) behind the shock front. 
On the other hand,
our WENO-$C$-$N$ algorithm provides a solution with accurate wave speed and location, stabilizes the dynamics during early
time-steps,  and removes high-frequency oscillations 
behind the shock front.

We next conduct $L^1$ error analysis and convergence tests for the Sedov problem. Due to the fact that 
the exact solution is known only in radial coordinates, and our solution is defined on a rectangular mesh, we 
shall compute the errors of the density $\rho$ and radial velocity $u_r$ along the cut $y=0$. We thus define
the quantities $\tilde{\rho} = \rho(x,0,t) - \rho^*(x,t)$ and $\tilde{u}_r = u_r(x,0,t) - u_r^*(x,t)$, where 
$\rho^*$ and $u_r^*$ are the exact solutions. The $L^1$ norm for a function $f(x)$ defined on a 
one-dimensional computational grid of $M$ cells with cell centers $x_i$  is defined as
\begin{equation}
\lVert f \rVert_{L^1} = \frac{1}{M} \sum_{i=1}^M | f(x_i) |\,.
\end{equation}

In Table {\ref{table:sedov}}, we list the $L^1$ norms and order of convergence for the errors $\tilde{\rho}$ and 
$\tilde{u}_r$. For the solution produced with the stand-alone WENO algorithm, the presence 
of the computational noise and 
the incorrect shock speed result in large errors and poor convergence rates\footnote{We note the ``super-convergence'' {\cite{Jiang1996202}} of the WENO solutions on the coarse grids; this  is 
due to large errors on coarser meshes, rather than smaller errors on finer meshes, and is therefore 
superficial.}. 
The shock stabilization and noise removal provided by the WENO-$C$-$N$ algorithm
produces solutions with much smaller errors than those computed with stand-alone WENO, as well as 
better rates of convergence.

\begin{table}[H]
\centering
\renewcommand{\arraystretch}{1.0}
\scalebox{0.8}{
\begin{tabular}{|llc|ccc|}
\toprule
\midrule
\multirow{2}{*}{\textbf{Norm}} & \multirow{2}{*}{\textbf{Scheme}} &  & \multicolumn{3}{c|}{\textbf{Cells}}\\

& {}  &  &24 & $48$ & $96$ \\
\midrule
 \multirow{4}{*}{\vspace{-1.25em}$ \lVert \tilde{\rho} \rVert_{L^1}$} & \multirow{2}{*}{WENO} & Error & 
 $6.347 \times 10^{-1}$ & $4.722 \times 10^{-1}$ & $4.648 \times 10^{-1}$  \\
				  &  & Order & -- & 0.427 & 0.023  \\[0.5em]

& \multirow{2}{*}{WENO-$C$-$N$} & Error & 
 $3.939 \times 10^{-1}$  & $1.081 \times 10^{-1}$ & $5.765 \times 10^{-2}$  \\
				  &  & Order & --  & 1.866 & 0.907   \\
\midrule
 \multirow{4}{*}{\vspace{-1.25em}$ \lVert \tilde{u}_r \rVert_{L^1}$} & \multirow{2}{*}{WENO} & Error & 
 $2.113 \times 10^{-1}$  & $8.993 \times 10^{-2}$ & $7.266 \times 10^{-2}$ \\
				  &  & Order & -- & 1.232 & 0.308  \\[0.5em]

& \multirow{2}{*}{WENO-$C$-$N$} & Error & 
 $4.695 \times 10^{-2}$ & $1.979 \times 10^{-2}$ & $1.482 \times 10^{-2}$  \\
				  &  & Order & --  & 1.247 & 0.417  \\
\midrule
\bottomrule
\end{tabular}}
\caption{$L^1$ error analysis and convergence tests for the Sedov problem at $t=1.0$.} 
\label{table:sedov}
\end{table} 

\begin{figure}[H]
\centering
\subfigure[WENO-$C$-$N$]{\label{fig:sedov-convergence-rho}\includegraphics[width=35mm]{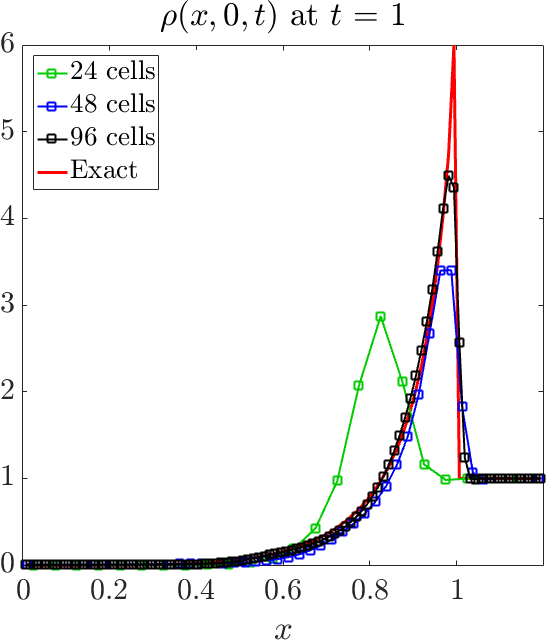}}
\hspace{.01em}
\subfigure[WENO]{\label{fig:sedov-convergence-rho2}\includegraphics[width=35mm]{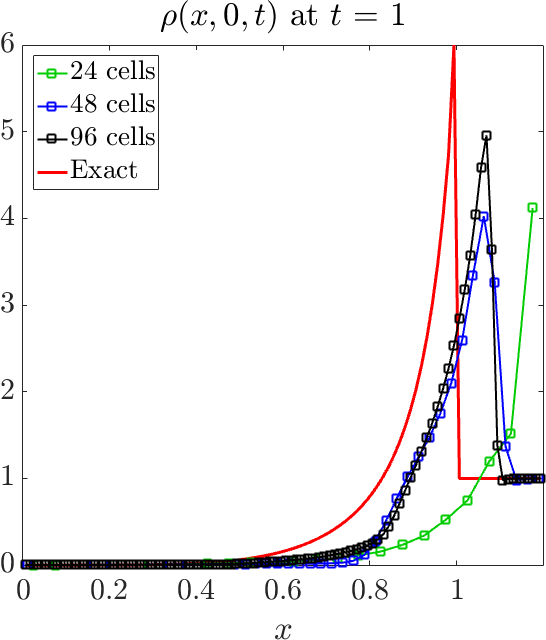}}
\hspace{0.25em}
\subfigure[WENO-$C$-$N$]{\label{fig:sedov-convergence-u}\includegraphics[width=36.5mm]{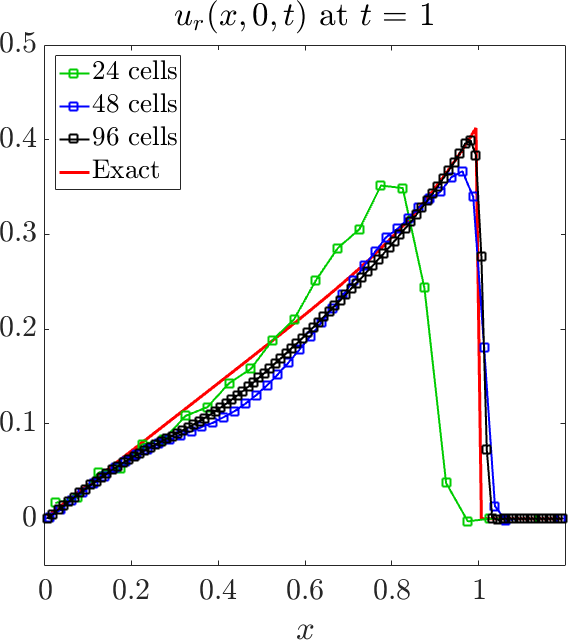}}
\hspace{.01em}
\subfigure[WENO]{\label{fig:sedov-convergence-u2}\includegraphics[width=37.2mm]{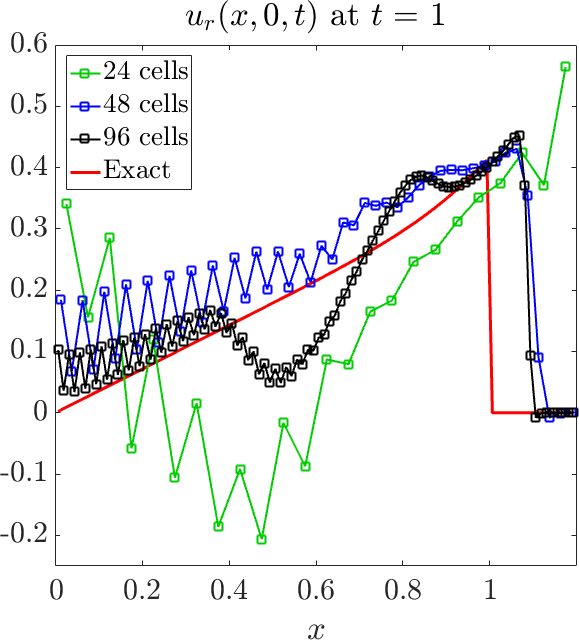}}
\caption{Plots of (a,b) the density and (c,d) the radial velocity along the cut
$y=0$. Subfigures (a) and (c) show the solutions computed with WENO-$C$-$N$, while Subfigures 
(b) and (d) show the solutions computed with WENO.}
\label{fig:sedov-convergence}
\end{figure}

In Fig.{\ref{fig:sedov-convergence}}, we show plots of the computed density 
$\rho$ and radial velocity $u_r$ along the cut $y=0$ at various grid resolutions. It is clear
from these plots, as well as the data in Table {\ref{table:sedov}}, that WENO-$C$-$N$ produces solutions that 
are both quantitatively and qualitatively better than those produced with WENO.

\section{The Sod circular explosion problem}\label{sec:Sod}
The explosion problem proposed in \cite{Toro2009,LiWe2003} 
is a radially symmetric version of the classic Sod shock tube problem \cite{Sod19781}. 
\begin{equation}\label{sod_initialdata1}
\begin{bmatrix}
\rho_0 \\ (\rho u)_0 \\ (\rho v)_0 \\ E_0 
\end{bmatrix}
=
\begin{bmatrix}
1 \\ 0 \\ 0 \\ 2.5
\end{bmatrix}
\mathbbm{1}_{[0,0.4)}(r)
+
\begin{bmatrix}
0.125 \\ 0\\ 0 \\ 0.25
\end{bmatrix}
\mathbbm{1}_{[0.4,\infty)}(r)  \,,
\end{equation}
where $r = \sqrt{x^2+y^2}$, and $\mathbbm{1}_{\Sigma}$ denotes the indicator function on the 
set $\Sigma$. We remark that the initial conditions are implemented in \cite{LiWe2003,Toro2009} by 
assigning area-weighted initial values for the cells which are crossed by the initial curve of discontinuity. We 
omit this modification, noting that an instability develops at the contact surface regardless of whether the 
area-weighted initial conditions are implemented. Since we wish to capture the evolution of the 
unstable contact, we do not smear the initial data, and instead employ our anisotropic artificial viscosity 
method.

Reflecting boundary conditions are employed on the $x$ and $y$ axes, and free-flow boundary conditions are
employed at the boundaries $x=1.5$ and $y=1.5$. The free-flow boundary conditions are implemented using 
the characteristic form of the Euler equations so as to minimize the reflection from the outgoing waves (see
\cite{Thompson1987} for the details). Nonetheless, there are numerical boundary effects occurring at the 
top and right boundaries that we were unable to eliminate. 

The solution consists of a circular shock front and contact curve traveling outwards from the origin, and a 
circular expansion wave traveling inwards to the origin. The shock front and contact surface become weaker 
as time evolves, with the contact eventually coming to rest before traveling inwards, while the
shock front passes through the free-flow boundary. The rarefaction 
wave traveling inwards collides with itself at the origin 
and reflects as an outwards traveling expansion wave. This 
results in an inward traveling shock forming and subsequently imploding into the origin. This shock then reflects
outwards from the origin and eventually passes through the contact curve.

We shall demonstrate the ability of the $C$-method to allow the artificial viscosity operator associated with  the shock wave to
``communicate''  with the artificial viscosity operator associated to the contact discontinuity.  The objective of the scheme is to allow the
shock wave to pass through the contact discontinuity, while leaving the small-scale KH structure of the contact undisturbed by over-diffusion.
Our results compare favorably to those
produced by PPM, CLAW, and WAFT {\cite{LiWe2003}}.

\subsection{The WENO-$C$-$\hat{C}$ scheme applied to the Sod explosion problem}\label{subsec:Sod-exp}
We employ a combination of our WENO-$C$ and WENO-$C^{\tau}$ schemes for this problem. More precisely, 
we use isotropic artificial viscosity (as provided by the WENO-$C$ scheme) to stabilize shock fronts, and 
anisotropic tangential artificial viscosity (as provided by the WENO-$C^{\tau}$ scheme) to add diffusion
to the unstable contact curve. We utilize a combination of compression and expansion switches to track 
the contact curve and shock fronts. In particular, we remark that, since shock implosion is a highly singular
phenomenon, the shock that reflects from the origin requires stabilization. However, this shock 
passes through the contact curve, and consequently we ensure that the stabilization of the shock does not result in 
an overly diffused contact discontinuity.    We do so by turning-off the artificial viscosity on the shock front as it
passes through the contact curve by using compression 
and expansion switches (which are detailed below).

More precisely, we consider the following Euler-$C$-$\hat{C}$ system:
\begin{subequations}\label{Euler-C-Ctau-2D}
\begin{align}
\partial_t \rho + \divv (\rho \bm{u}) &=0 \,, \\[0.5em] 
\partial_t (\rho u) + \partial_x (\rho u^2 + p) + \p_y (\rho uv) &= \divv \left( \tilde{\beta}^u \, \rho \, C \, \nabla u \right) + \p_i \left( \tilde{\alpha} \, \rho \, \hat{C} \, C^{\tau_i} \, C^{\tau_j} \, \p_j u \right)\, , \\[0.5em]
\partial_t (\rho v) + \p_x(\rho uv) + \partial_y (\rho v^2 + p) &= \divv \left( \tilde{\beta}^u \, \rho \, C \, \nabla v \right) +  \p_i \left( \tilde{\alpha} \, \rho \, \hat{C} \, C^{\tau_i} \, C^{\tau_j} \, \p_j v \right) \,, \\[0.5em]
\partial_t E + \divv (\bm{u}(E + p)) &=  \divv \left( \tilde{\beta}^E \, \rho \, C \, \nabla (E/\rho) \right) \,, 
\end{align}
\begin{gather}
\p_t {C} - \mathscr{L}[{C}\,; \varepsilon,\kappa] 
 = \frac{S(\bm{u})}{\varepsilon | \delta \bm{x} |} {G_\rho} \\
\p_t \hat{C} - \mathscr{L}[\hat{C}\,; \varepsilon,\kappa] 
 = \frac{S(\bm{u})}{\varepsilon | \delta \bm{x} |} \hat{G_\rho}  \,, \quad  
 \p_t C^{\tau_i} - \mathscr{L}[C^{\tau_i} \,; \varepsilon,\kappa] 
 = \frac{S(\bm{u})}{\varepsilon | \delta \bm{x} |} \hat{\tau_i}  \text{ for } i=1,2\, .
\end{gather}
\end{subequations}
The  forcing functions to the $C$-equations are given by 
\begin{subequations}\label{forcing-C-Ctau}
\begin{align}
G_{\rho} &= \left[1 - \mathbbm{1}_{(-\infty,0)}(\p_r e \, \p_r \rho) \right] \cdot \mathbbm{1}_{(-\infty,0)}(\divv \bm{u}) \cdot |\nabla \rho| \,, \label{forcing-C-Ctau-1} \\
\hat{G_{\rho}} &= \mathbbm{1}_{(-\infty,0)}(\p_r e \, \p_r \rho) \cdot |\nabla \rho| \,, \\
\hat{\tau_1} &= - \mathbbm{1}_{(-\infty,0)}(\p_r e \, \p_r \rho) \cdot \p_y \rho \,, \\ 
\tilde{\tau_2} &= \mathbbm{1}_{(-\infty,0)}(\p_r e \, \p_r \rho) \cdot \p_x \rho\,,
\end{align}
\end{subequations}
where $\p_r = \cos \theta \, \p_x + \sin \theta \, \p_y$ denotes the radial derivative. The function 
$\mathbbm{1}_{(-\infty,0)}(\divv \bm{u})$ is a compression switch that localizes $C$ to shocks, while the 
function $\mathbbm{1}_{(-\infty,0)}(\p_r e \, \p_r \rho)$ localizes $\hat{C}$ and $\hat{\tau_i}$ to the contact 
curve. Consequently, the use of the function $\left[1 - \mathbbm{1}_{(-\infty,0)}(\p_r e \, \p_r \rho) \right]$ 
in \eqref{forcing-C-Ctau-1}
ensures that $C$ deactivates during the short time interval that the shock front passes through the contact, so that isotropic diffusion is not added during this time interval, which prevents the smearing of the contact curve.  

The artificial viscosity parameters $\tilde{\beta}$ and $\tilde{\alpha}$ are defined by 
$$
\tilde{\beta^{\cdot}} = \frac{|\delta \bm{x}|^2}{\max_{\Omega} C} \beta^{\cdot} \quad \text{and} \quad 
\tilde{\alpha} = \frac{|\delta \bm{x}|^2}{\mu^2 \max_{\Omega} \hat{C}} \alpha\,,
$$
with $\mu = \max_{\Omega} \left\{ \max \left\{ |C^{\tau_1}| \,, |C^{\tau_2}|  \right\} \right\}$. 

The Euler-$C$-$\hat{C}$ system is numerically discretized in an identical fashion to the other schemes 
presented; we will refer to the discretized method as the WENO-$C$-$\hat{C}$ scheme. We employ the 
method to the problem on a grid with 400$\times$400 cells with a time-step of $\delta t = 6.4 \times 10^{-4}$, 
giving an initial CFL number of approximately 0.75. The parameters for the WENO-$C$-$\hat{C}$ method are 
chosen as
\begin{gather*}
\beta^u=15.0, \qquad  \beta^E=200.0, \qquad \alpha=2.0, \qquad \varepsilon = 1.0, \qquad \kappa = 10.0\,. 
\end{gather*}

\begin{figure}[H]
\centering
\subfigure[WENO-$C$-$\hat{C}$]{\label{fig:Sodexp2-rho1}\includegraphics[width=50mm]{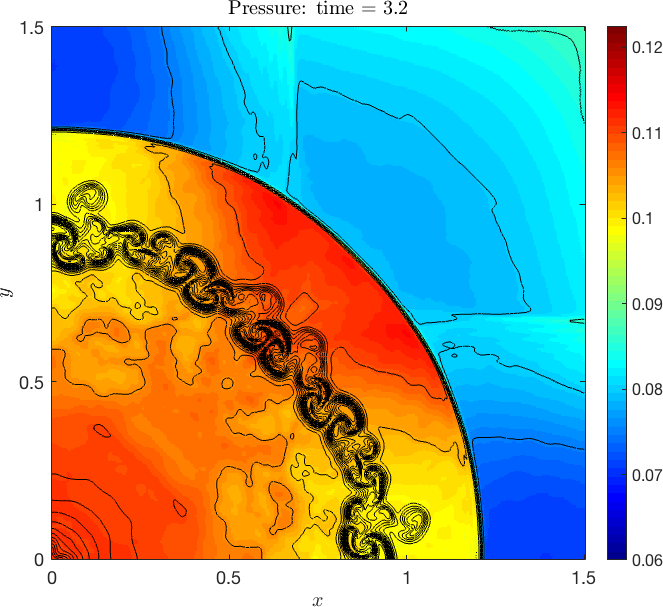}}
\hspace{2em}
\subfigure[WENO]{\label{fig:Sodexp2-rho2}\includegraphics[width=50mm]{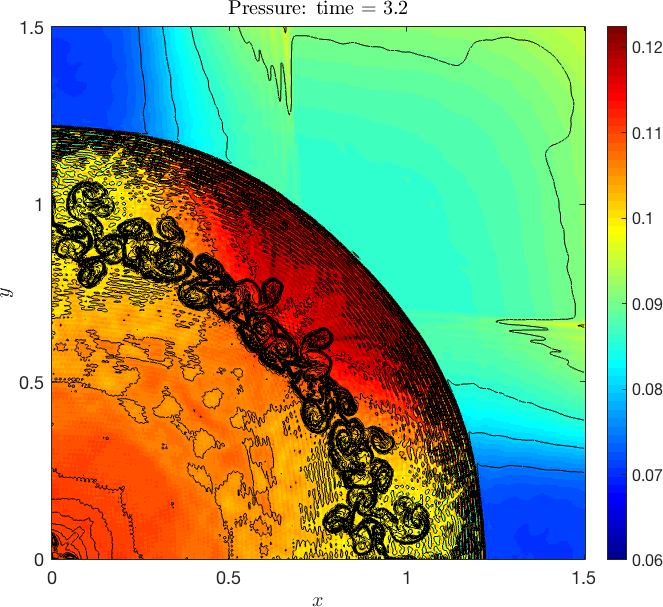}}
\hspace{2em}
\subfigure[WENO-$C$-$\hat{C}$ with $C$ active during shock-contact collision]{\label{fig:Sodexp2-rho3}\includegraphics[width=50mm]{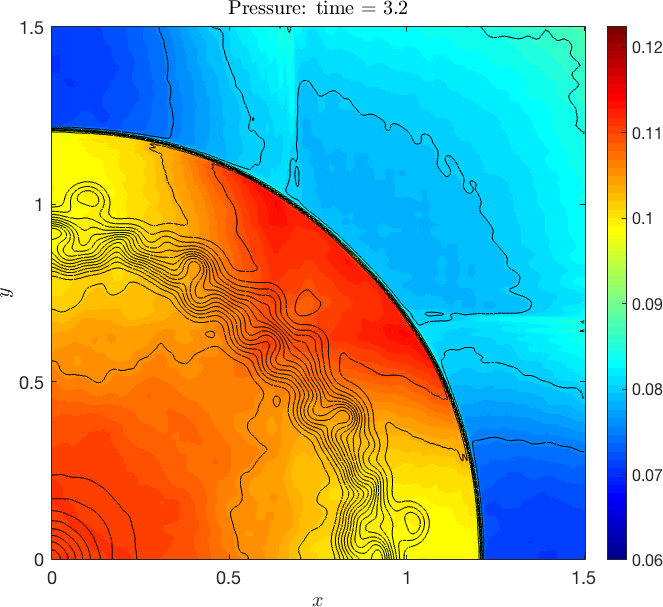}}
\hspace{2em}
\subfigure[isotropic diffusion for the contact and shocks]{\label{fig:Sodexp2-rho4}\includegraphics[width=50mm]{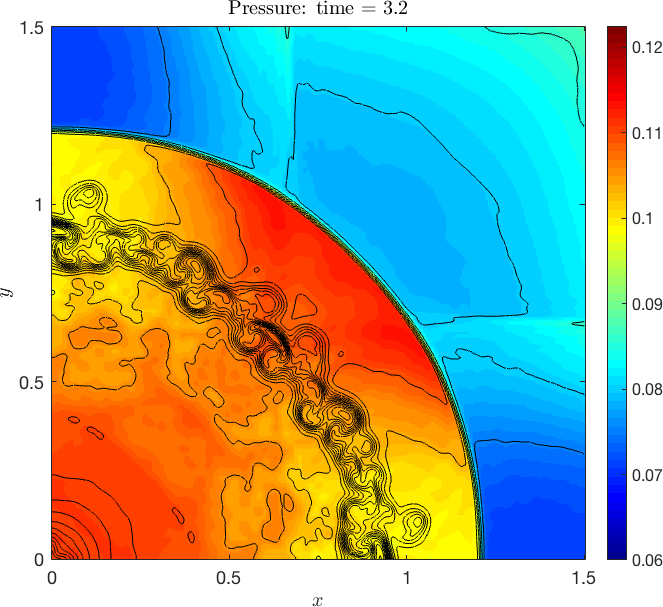}}
\caption{Comparison of WENO and WENO-$C$-$\hat{C}$ for the Sod explosion problem. The 
figures shown are heatmap plots of the pressure $p$ overlaid with 27 density contours from 
0.08 to 0.21 with step 0.005. Results are presented at time $t=3.2$.}
\label{fig:Sodexp2-rho}
\end{figure} 

The results are presented in Fig.\ref{fig:Sodexp2-rho}, which show heatmap plots of the pressure function $p$
overlaid with density contours. These figures should be contrasted with those presented in 
\cite{LiWe2003}. The stand-alone WENO scheme produces a highly oscillatory solution behind the shock
front (see Fig.\ref{fig:Sodexp2-rho2}), whereas the $C$-method allows for the stabilization of the shock front. 
Additionally, we may see the role that the function $\left[1 - \mathbbm{1}_{(-\infty,0)}(\p_r e \, \p_r \rho) \right]$
plays by comparing with the solution computed without the deactivation of $C$ during shock-contact collision; 
this solution, shown in Fig.\ref{fig:Sodexp2-rho3}, is noticeably more smeared at the contact discontinuity.
 Finally, in
Fig.\ref{fig:Sodexp2-rho4} we show the solution computed using isotropic diffusion for both the contact 
discontinuity as well as shock fronts, with all the relevant parameters identical to those used in the
WENO-$C$-$\hat{C}$ simulation. It is clear that the contact discontinuity is not as sharp as the contact curve
for the solution computed
using WENO-$C$-$\hat{C}$; this is due to the addition of diffusion in the direction normal to the contact curve.
The WENO-$C$-$\hat{C}$ scheme produces a non-oscillatory solution with minimal noise, a sharp shock front, 
and a sharp contact curve.  

\subsection{Shock-wall collision for a Sod-type explosion problem}\label{subsec:sod-collision}
To test the shock-collision scheme presented in \S\ref{sec:C-wall-collision}, we consider a modified version of
 the Sod circular 
explosion problem. The domain is 
$\Omega = [-1,1] \times [-0.7,1.3] \subset \mathbb{R}^2$, the 
adiabatic constant is $\gamma = 1.4$, and the initial data is 
\begin{equation}\label{sod_initialdata2}
\begin{bmatrix}
\rho_0 \\ (\rho u)_0 \\ (\rho v)_0 \\ E_0 
\end{bmatrix}
=
\begin{bmatrix}
1 \\ 0 \\ 0 \\ 2.5
\end{bmatrix}
\mathbbm{1}_{[0,0.1)}(r)
+
\begin{bmatrix}
0.125 \\ 0\\ 0 \\ 0.25
\end{bmatrix}
\mathbbm{1}_{[0.1,\infty)}(r)  \,,
\end{equation}
where $r = \sqrt{x^2+y^2}$, and $\mathbbm{1}_{\Sigma}$ denotes the indicator function on the 
set $\Sigma$. 

We are interested in the problem of shock-wall collision, and so treat the bottom boundary as a fixed
wall, so that solid-wall (or reflecting) boundary conditions are enforced at $y=-0.7$. Free-flow (or symmetric)
boundary conditions are implemented at the other three boundaries. The final time is $t=0.6$.
The outwards traveling shock front collides
with the bottom boundary $y=-0.7$ at time $t \approx 0.4$. As the gas is compressed, an increase in the
density leads to a reversal in the direction of travel of the shock front. 
The collision with the bottom boundary breaks the radial symmetry of the problem; since the first point of 
contact of the shock front with the bottom boundary occurs at $x=0$, the gas is forced outwards along the 
bottom boundary $y=-0.7$. 

We employ our WENO-$C$-$W$ scheme (see \S\ref{sec:C-wall-collision}), which couples the $C$-method with a shock collision scheme. Due to the symmetry of the problem, we compute the solution 
in the half domain $[0,1] \times [-0.7,1.3]$, and then reflect appropriately to obtain the solution 
on all of $\Omega$. The WENO-$C$-$W$ scheme is applied on a grid with 200$\times$400 cells in the 
half domain $[0,1] \times [-0.7,1.3]$, with a time-step $\delta t = 5 \times 10^{-4}$, giving a CFL number of 
approximately 0.4. The parameters in the WENO-$C$-$W$ method are chosen as
\begin{gather*}
\beta^u=20.0, \qquad  \beta^E=0.0, \qquad \varepsilon = 1.0, \qquad \kappa = 1.0, \\
\beta^u_w=50.0, \qquad  \beta^E_w=100.0, \qquad \varepsilon_w = 10.0, \qquad \kappa_w = 1.0\,. 
\end{gather*}

For the purposes of comparison, we also implement our stand-alone WENO scheme as well as the WENO-Noh
scheme with the following modification: due to the loss of radial symmetry when the shock front collides with 
the bottom boundary $y=0$, we do not use the (normalized) gradient of the radial velocity 
$| \nabla u_r| / \max_{\Omega} | \nabla u_r|$ to track the shock front; instead, we use the function 
$ \mathbbm{1}_{\divv \bm{u} < 0}|\nabla \rho | / \max_{\Omega} | \nabla \rho|$, where  
$\mathbbm{1}_{\divv \bm{u} < 0}$ is a compression switch. The artificial viscosity parameters in \eqref{weno-noh-scheme} are chosen as 
$$
\alpha^u_N= \alpha^e_N = 0 \quad \text{ and } \quad \beta^u_N = 70.0 \,,
\beta^e_N = 200.0 \,.
$$

\begin{figure}[H]
\centering
\subfigure[WENO-$C$-$W$]{\label{fig:sod-rho_surf1}\includegraphics[width=70mm]{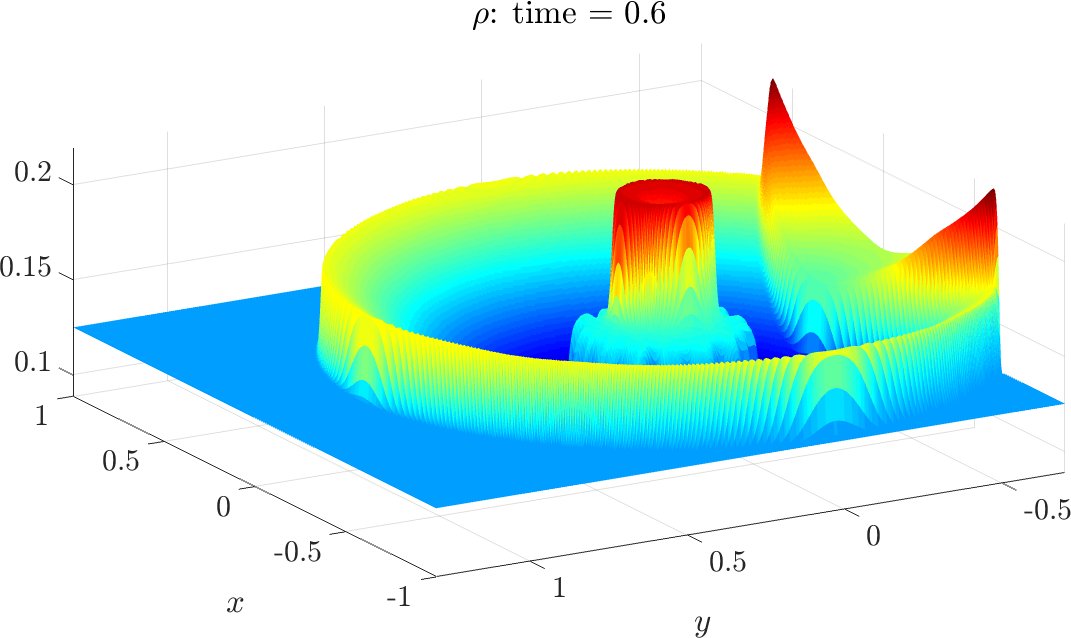}}
\hspace{2em}
\subfigure[WENO]{\label{fig:sod-rho_surf2}\includegraphics[width=70mm]{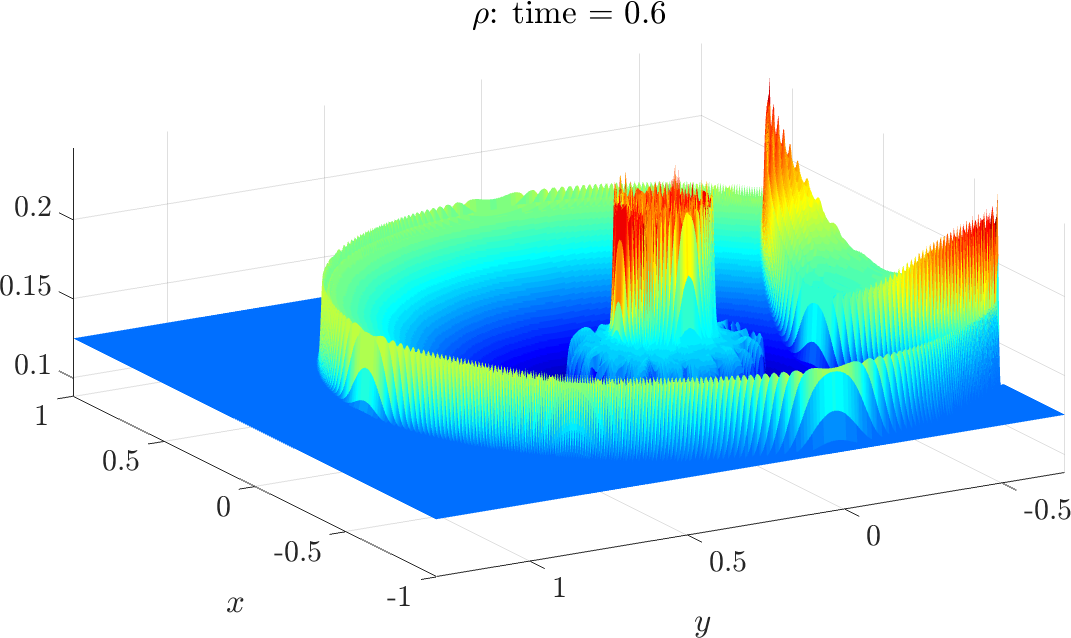}}
\caption{Comparison of WENO and WENO-$C$-$W$ for the Sod explosion and bounce-back problem. The 
figures shown are surface plots of the density $\rho$ after shock-wall collision.}
\label{fig:Sod}
\end{figure}

\begin{figure}[H]
\centering
\includegraphics[width=100mm]{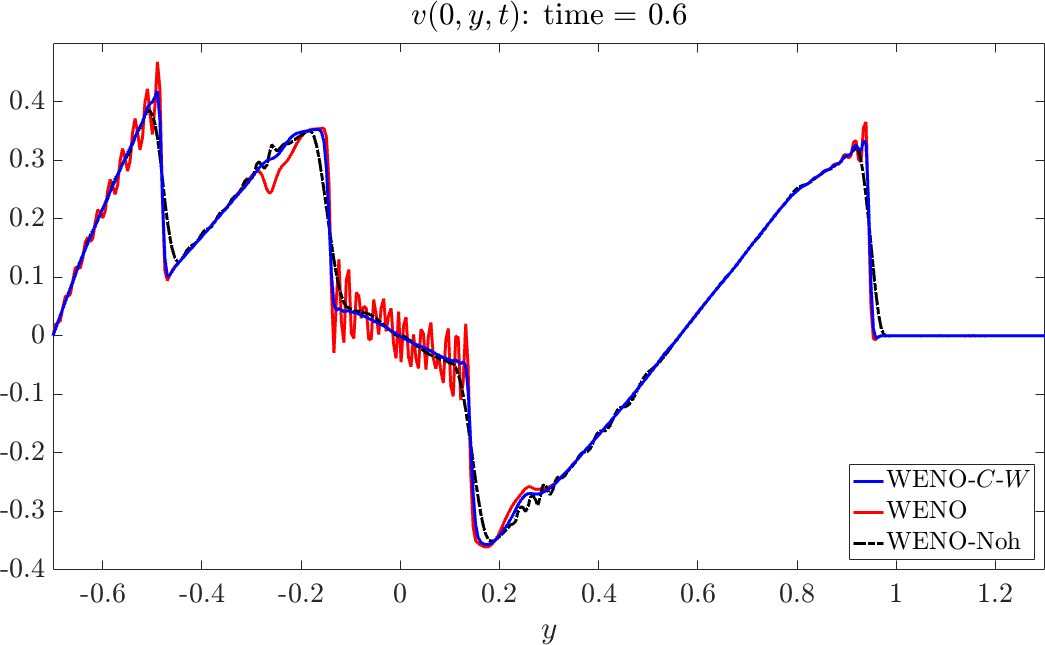}
\caption{Comparison of WENO, WENO-Noh, 
and WENO-$C$-$W$ for the Sod explosion and bounce-back problem. Shown
are cross-sections of the vertical velocity $v(0,y,t)$ along $x=0$, at time $t=0.6$ after the shock-wall collision.}
\label{fig:Sod2}
\end{figure}

The results are provided in Fig.\ref{fig:Sod}, which shows surface plots of the density, and Fig.\ref{fig:Sod2}, 
which is a plot of the vertical velocity $v(0,y,t)$ along $x=0$. 

The stand-alone WENO scheme produces an oscillatory solution both pre and post shock-wall collision. 
In particular, the shock implosion at the origin results in a large amount of noise (see Fig.\ref{fig:Sod2}). 
The solution computed using WENO-Noh is also oscillatory, albeit to a lesser extent, and the shock fronts
are overly smeared. 
On the other hand, the WENO-$C$-$W$ scheme produces a solution that is completely noise-free while
retaining sharp shock fronts and correct wave speeds. 

\section{The Mach 10 shock reflection problem}
The double Mach shock reflection problem introduced by Woodward \& Colella {\cite{CoWo1984}} features a 
Mach 10 shock in a $\gamma =1.4$ gas reflecting from a wedge inclined at an angle of $60^{\circ}$. 
The computational domain is $\Omega = [0,3.25] \times [0,1]$, and the initial data is given by
\begin{equation}\label{mach10-initialdata}
\begin{bmatrix}
\rho_0 \\ (\rho u)_0 \\ (\rho v)_0 \\ E_0 
\end{bmatrix}
=
\begin{bmatrix}
8 \\ 66 \cos(\pi/6) \\ -66 \sin(\pi/6) \\ 563.5 
\end{bmatrix}
\mathbbm{1}_{\Sigma}(x,y)
+
\begin{bmatrix}
1.4 \\ 0\\ 0 \\ 1.0
\end{bmatrix}
\mathbbm{1}_{\Omega \setminus \Sigma}(x,y)  \,, 
\end{equation}
where $\Sigma = \left\{ (x,y) \in \Omega \mid  y > \sqrt{3} (x - 1/6)  \right\}$ is the 
region behind the initial 
position of the shock. On the left  boundary, the density, velocity, and energy are prescribed the values of the corresponding
initial conditions, while free-flow boundary 
conditions are imposed at the right boundary. The conditions at the top boundary are set to describe the 
exact motion of the initial Mach 10 shock. At time $t$, the shock intersects with the top boundary $y=1$ at 
the point $(s(t),1)$, where $s(t) = 1/6 + (1+20t)/\sqrt{3}$. The pre- and post-shock conditions are then 
imposed for $x \geq s(t)$ and $x < s(t)$, respectively. At the bottom boundary, reflecting boundary conditions
 are prescribed, except for the short region $0 \leq x < 1/6$, along which the exact initial conditions are 
 imposed. This condition forces the reflected shock to remain ``attached'' to the wedge.
 
\begin{figure}[H]
\centering
\includegraphics[width=120mm]{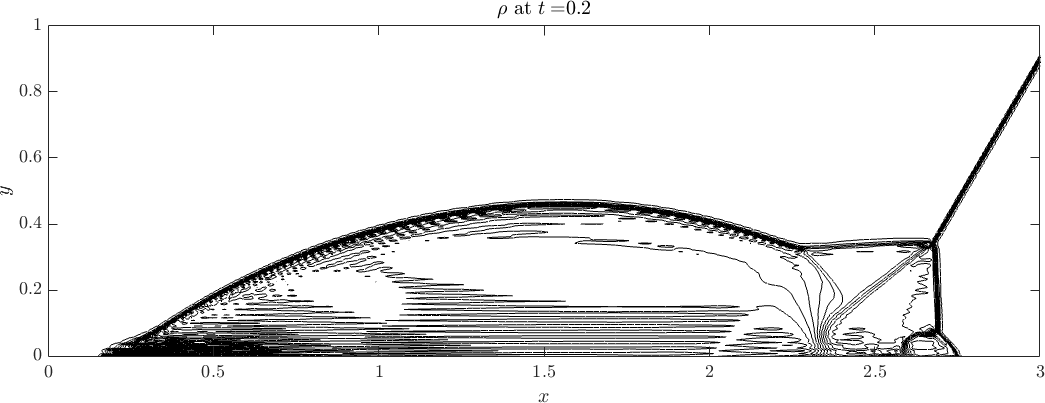}
\caption{A density contour plot of the stand-alone WENO solution computed on a grid with $\delta x = \delta y = 1/120$. Shown are 30 equally spaced density contours from $\rho=1.5$ to $\rho=22.9705$. 
The carbuncle instability leads to a kink at the leading shock front, causing a spurious triple point to form. 
There are high-frequency oscillations behind the reflected shock front, especially in the regions where 
shock curves intersect with the reflecting wall.}
\label{fig:mach-weno}
\end{figure}

 The solution develops a complex self-similar flow structure featuring two triple points and a jet attached to 
 bottom wall. Numerical methods applied to this problem often produce solutions suffering from the 
 ``carbuncle phenomenon'' \mbox{\cite{Quirk1994555}}. In such solutions, the wave speed of the 
 leading Mach stem is incorrect, leading to a ``kink'' in the stem, and often the formation of a spurious 
 additional triple point. In fact, our simplified WENO-type scheme produces a solution, 
 shown in Fig.{\ref{fig:mach-weno}}, exhibiting this error; 
 moreover, the solution is corrupted by a large amount of high frequency noise, especially in the regions 
 where the shock fronts meet with the reflecting wall.
 
 To eliminate the carbuncle error and the high-frequency noise in the computed solution near the reflecting
 wall, we shall employ our WENO-$C$-$W$ algorithm for this problem, with the following simple modification: since 
 there is noise in the solution near the wall-jet, we do not use the compression switch 
 $\mathbbm{1}_{\mathrm{div} \bm{u} < 0}$ in the $C$-equation {\eqref{CW-Sod}}. 
 This allows the $C$-function to remain active at the wall-jet, so that the artificial viscosity damps the 
 high-frequency oscillations in this region. 
 The discontinuities are stabilized through the use of  directionally isotropic 
 viscosity, and the wall $C$-method is used to implement additional viscosity in the regions where the 
 shock fronts meet with the bottom reflecting boundary. 
 We note that the compression switch 
 $\mathbbm{1}_{\mathrm{div} \bm{u} < 0}$ remains active for the wall $C$ equation  {\eqref{CW-Sod2}}; this 
 forces the wall function $\overline{C}(x,t)$ to vanish at the wall-jet, which subsequently prevents the 
 over-smearing of the contact discontinuity. 
 
We implement the WENO-$C$-$W$ scheme, with the small modification mentioned above,
  on a grid with $781 \times 241$ cells, giving 
 a mesh resolution of $\delta x = \delta y = 1/240$. The time-step is set as $\delta t = 4 \times 10^{-5}$, giving 
 CFL$\approx 0.4$. This time-step was chosen as the largest possible value for which the stand-alone 
 WENO simulation runs until the final time $t=0.2$. 
 The relevant parameters for the WENO-$C$-$W$ method are chosen as
 \begin{gather*}
\beta^u=3 \times 10^2, \qquad  \beta^E=0.0, \qquad \varepsilon = 10.0, \qquad \kappa = 1.0, \\
\beta^u_w=1 \times 10^3, \qquad  \beta^E_w=0, \qquad \varepsilon_w = 10.0, \qquad \kappa_w = 1.0\,. 
\end{gather*}

\begin{figure}[H]
\centering
\includegraphics[width=120mm]{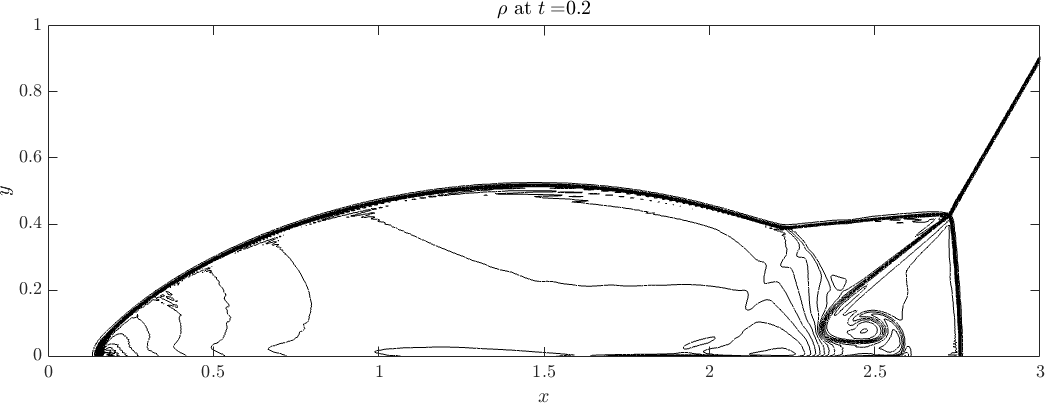}
\caption{A density contour plot of the WENO-$C$-$W$ solution computed on a grid with 
$\delta x = \delta y = 1/240$. Shown are 30 equally spaced density contours from 
$\rho=1.5$ to $\rho=22.9705$.}
\label{fig:mach-weno-c}
\end{figure}

A contour plot of the density computed using WENO-$C$-$W$ is 
shown in Fig.{\ref{fig:mach-weno-c}}. The artificial viscosity stabilizes the shock fronts and corrects the 
wave speeds, resulting in the elimination of the spurious carbuncle instability at the leading shock. Moreover, 
the high-frequency oscillations at the intersections of shock fronts with the bottom boundary, and 
near the wall-jet, are suppressed 
by the additional viscosity provided by the wall $C$-method active in those regions. 
Our result is comparable to the simulations presented, for example, by  Shi, Zhang, \& Shu {\cite{ShiZhangShu2003}}, for their developed WENO schemes.

\section{The Rayleigh-Taylor instability}\label{sec:RT}

The Rayleigh-Taylor (RT) instability \cite{Rayleigh1882,Taylor1950} is an instability of a heavy fluid layer supported by a light one.
 Specifically, RT  occurs when a perturbed interface, between two fluids of different density, is subjected to a normal 
pressure gradient; the light fluid {\it bubbles} into the heavy fluid, while the heavy fluid {\it spikes} into the lighter fluid, which in turn
initiates the Kelvin-Helmholtz (KH) instability.  The RT
 instability occurs  in a wide range of physical phenomena; see
\cite{Sharp1984,Kull1991} and the references therein for an overview.

We use the following setup for the RT problem considered by Liska \& Wendroff
 \cite{LiWe2003}: the domain
is $\Omega = [-1/6,+1/6] \times [0,1]$, the adiabatic constant is $\gamma=1.4$, and the initial data is 
\begin{equation}\label{RT-initialdata}
\begin{bmatrix}
\rho_0 \\ (\rho u)_0 \\ (\rho v)_0 \\ E_0 
\end{bmatrix}
=
\begin{bmatrix}
1 \\ 0 \\ 0 \\ p_0/(\gamma-1) 
\end{bmatrix}
\mathbbm{1}_{[0,h_0)}(y)
+
\begin{bmatrix}
2 \\ 0\\ 0 \\ p_0/(\gamma-1)
\end{bmatrix}
\mathbbm{1}_{[h_0,1]}(y)  \,. 
\end{equation}
Here, the initial interface $\Gamma_0$ is parameterized by $(x,h_0(x))$ with $h_0(x)=0.5 + 0.01\cos(6 \pi x)$, and $p_0$ is 
the initial pressure, defined as
\begin{equation}\label{initial-pressure-hydrostatic}
 p_0 = \begin{cases}
 	   P_0 + g \, (h_0(x) - y) + 2g\,(1-h_0(x)) &, \text{ if } y < h_0(x) \\
 	   P_0 +2g \, (1 - y) &, \text{ if } y \geq h_0(x) \\	
 	   \end{cases} \,,
\end{equation}
with $g = 0.1$ the gravitational acceleration, 
and $P_0$ the reference pressure.   While Liska \& Wendroff 
\cite{LiWe2003} do not provide $P_0$, we follow the ATHENA 
code \cite{Athena} and use the value
$P_0 = 2.4$ so as to produce a sound speed  $c = \sqrt{3.5}$ in the lower density gas at the interface.
The equations of motion governing the flow are given by the Euler system \eqref{Euler-2d} with the 
addition of the gravity forcing on the right-hand side of \eqref{Euler-2d-motion}: $\bm{f} = [ 0 \,, 0 \,, -g \rho \,, -g \rho v ]^{\mathcal{T}}$. 
Due to the symmetry of the problem, it is sufficient to 
calculate the solution in the half-domain $[0,1/6] \times [0,1]$, and then use reflection.

For a very short time, when the amplitude of the perturbation remains very small  compared with the wavelength of the perturbation, 
the motion of the interface can be analyzed 
using linear stability analysis 
\cite{Rayleigh1882,Taylor1950,Chandra1961,Livescu2004}.  Quickly, however, the nonlinearity is activated, and 
the interface evolves to the classical ``mushroom-shaped'' profile \cite{WaNiJa2001}.
The flow at later times is characterized by the development of ``roll-up'' regions, driven by 
the KH shear instability \cite{Kull1991}. 

The particular problem set up described above results in a low Mach number flow. For explicit 
time-integration schemes, the CFL stability condition requires the time-step to be very small for 
such flows, due to the fact that, in this regime, sound waves are much faster than the 
advection of the flow. We do not switch to implicit time-integration, but note 
that one of the authors has extensively compared implicit and explicit temporal differencing for low 
Mach number flows and found little differences between the two approaches, provided 
the time step size is below the
sound speed in the explicit approach; the time step size can be at least a factor of 10 bigger in the implicit approach.   Likewise, for 
time-split methods, it has been shown that time-split errors grow as the time step begins to exceed the fastest
 time scale of the problem, i.e. the speed of sound waves. 
 Hence, it is presumed for the current algorithm, in which the time 
 steps utilized are below this fast time scale, that the temporal error growth is small. 
 Moroever, we note that explicit time integration 
is used by Liska \& Wendroff {\cite{LiWe2003}}, and so for the purposes of comparison we 
utilize the same approach. We also note that the Navier-Stokes equations 
(rather than the Euler equations) are often used for low Mach flow calculations; however, 
for the numerical tests presented here, the mesh 
resolution is so low that the kinematic viscosity coefficient appearing in the Navier-Stokes equations is 
negligible, and consequently algorithms for either sets of equations will produce similar results.

To highlight the applicability of our noise reduction algorithm to RT problems, we shall artificially generate 
noise in the solution by using explicit forward Euler time integration, rather than the usual 
third order 
Runge-Kutta method used for the other test problems. As we shall show in \S{\ref{subsubsec:noise-RT}}, 
the use of the explicit first order in time method results in the generation of spurious 
high frequency oscillations 
in solutions computed without any noise reduction algorithm. This test 
allows us to model a typical scenario in computational physics, in which high resolution and long time 
simulations require the use of an extremely small $\Delta t$ and, consequently, a
prohibitively large number of time steps. Our aim is to show that the noise generated by 
using a larger time step (or lower order time integration method) can be removed by means of our 
noise algorithm, resulting in a noise-free and accurate solution.

Numerical simulations of the RT instability often suffer from the the development of spurious 
small-scale structure due to the discretization of the problem (to be described in more detail below); see
\cite{LiWe2003} for example. On the one hand, as described in  \cite{LiWe2003}, the numerical methods with the least amount of implicit 
diffusion, such as the Piecewise Parabolic Method (PPM), 
Virginia Hydrodynamics 1 (VH1) and WENO schemes, 
produce interface break-up at early times which  corrupts the solution (even if the initial density and pressure are smoothed over a few
cells).
 The solutions produced using these schemes generally do not possess the classical 
mushroom-shaped profile; the spurious small-scale structures give rise to a complex interface and a 
significant amount of mixing.   On the other hand,
the more dissipative methods in \cite{LiWe2003}, such as the scheme of Liu and Lax (LL) and the 
CFLF hybrid (CFLFh) scheme, suppress this small-scale spurious structure from developing but, in doing so,
 produce solutions with overly smeared interfaces and overly diffused mixing zones with  very little KH roll-up. 

Our aim, therefore, is to produce a noise-free solution with a sharp interface and the classical mushroom-shaped profile, 
while ensuring that the KH-driven roll-up regions are not significantly affected. We do so by implementing 
the \emph{anisotropic} version of the $C$-method (that we detail below).

\subsection{Tangential spikes and the need for anisotropic diffusion}\label{subsec:oscillations}
\begin{figure}[H]
\centering
\subfigure[tangential spikes in the  vertical velocity $v$.]{\label{fig:RT-weno1}\includegraphics[width=70mm]{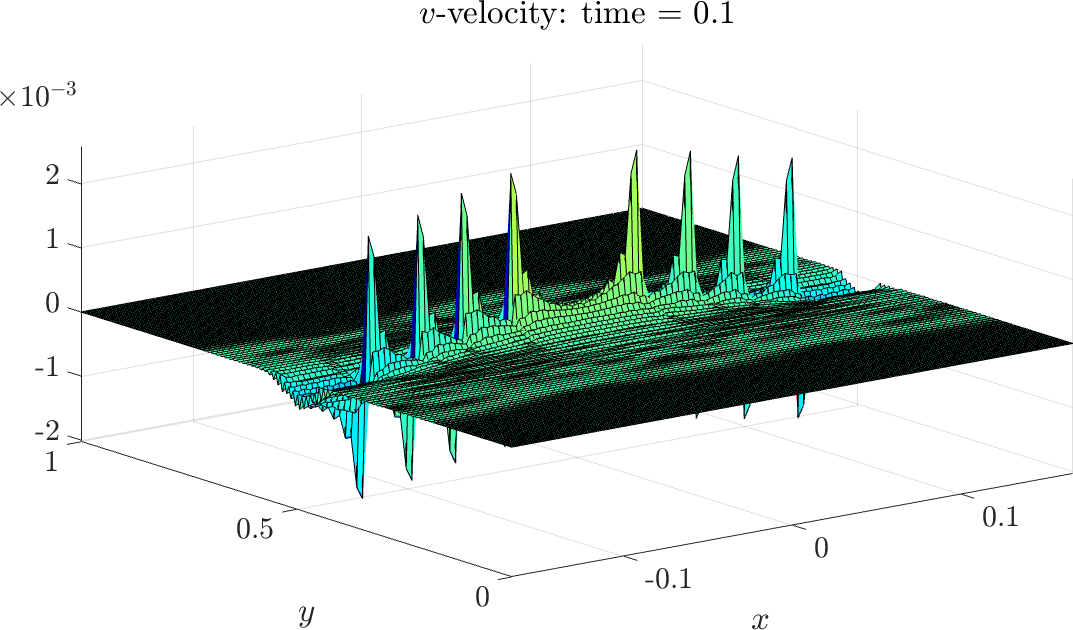}}
\hspace{2em}
\subfigure[``jumps'' in the initial data for the density $\rho$. Figure is zoom-in on the initial interface $\Gamma_0$.]{\label{fig:RT-weno2}\includegraphics[width=70mm]{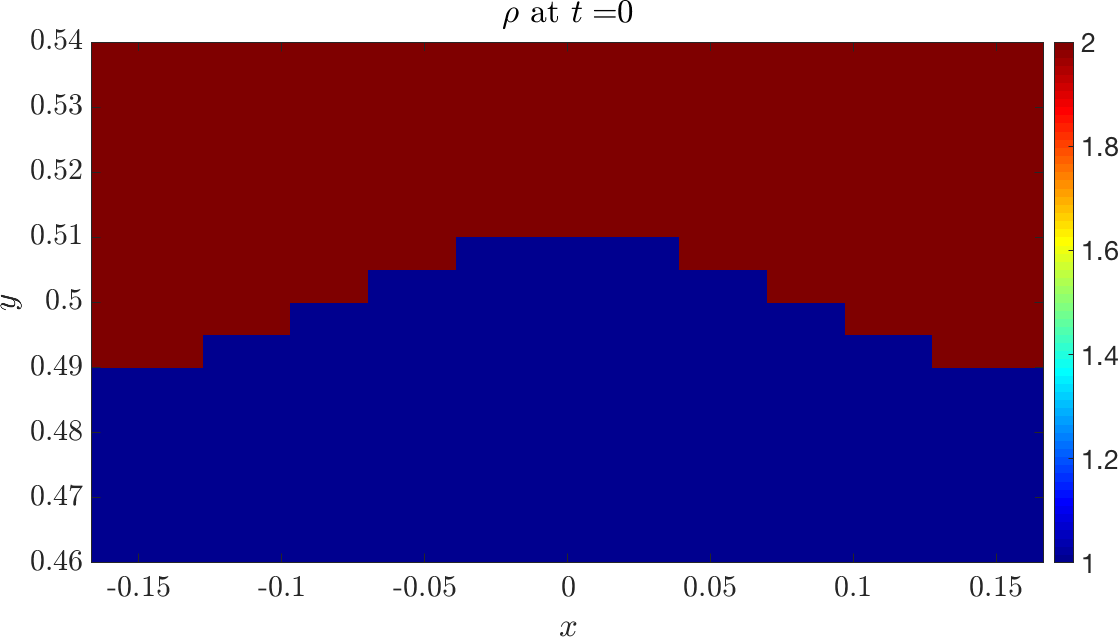}}
\par\bigskip
\subfigure[nooneline][development of fingers, caused by tangential spikes. Shown is a heatmap of the density $\rho$.]{\label{fig:RT-weno3}\includegraphics[width=70mm]{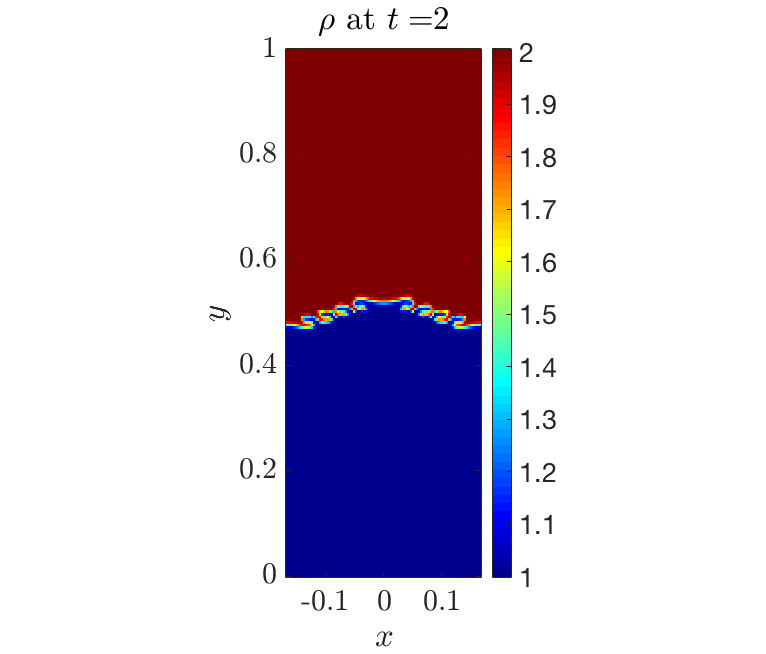}}
\hspace{2em}
\subfigure[mixing caused by interaction between fingers. Shown is a heatmap of the density $\rho$.]{\label{fig:RT-weno4}\includegraphics[width=70mm]{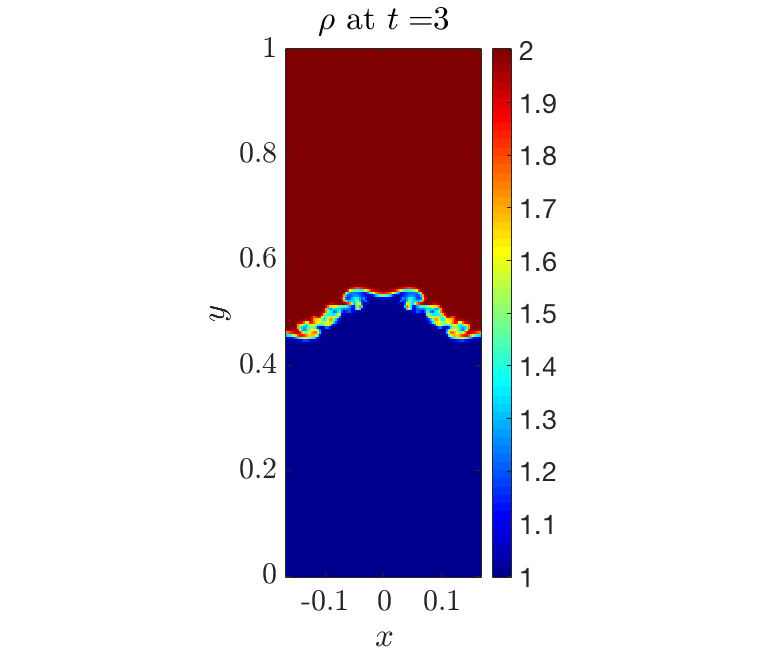}}
\caption{Figures demonstrating the occurrence of tangential spikes due
to the discretization of the initial conditions, which leads to the development of fingers and subsequent mixing.
The solution is computed on a 50$\times$200 grid on $[0,1/6] \times [0,1]$
using stand-alone WENO i.e. without any artificial diffusion.}
\label{fig:RT-weno}
\end{figure}

We now consider numerical simulations of the RT problem with data given by \eqref{RT-initialdata}.
Let us motivate the need for  strictly tangential artificial viscosity operators for the long-time motion of contact discontinuities. 
In Fig.\ref{fig:RT-weno}, we show results of RT simulations, run with our simplified WENO scheme (without  the use of any artificial viscosity);
due to interpolation errors of the cosine function onto the uniform mesh, together with a lack of artificial viscosity, extremely large 
{\it tangential spikes} are
produced in the vertical velocity, shown in Fig.\ref{fig:RT-weno1}. These oscillations are, of course, non-physical.  The interpolation error is demonstrated in Fig.\ref{fig:RT-weno2}; the initial density
profile has a tangential ``staircase'' whose jumps produce 
 the tangential spikes in the vertical velocity, which in turn cause the 
interface to develop thin fingers as shown in Fig.\ref{fig:RT-weno3}. The fingers continue to grow and,
in the absence of any artificial diffusion, eventually interact with each other and corrupt the solution as shown in
Fig.\ref{fig:RT-weno4}. 

Liska and Wendroff \cite{LiWe2003} smooth the initial density and pressure functions
over a region of width $\mathcal{O}(\delta y)$ in the  vertical direction; using smoothed initial data mitigates the development of the { tangential
spikes} (although not entirely), but has the undesirable effect of overly diffusing the contact discontinuity, and modifying 
wave speeds in a non-transient manner. Moreover, this smoothing is not enough to suppress the 
development of spurious small-scale structure for the highest order schemes with the least amount of
 numerical diffusion; in particular, as shown in \cite{LiWe2003}, both WENO and PPM break-up the interface.

Our strategy, therefore, is to implement the explicit anisotropic diffusion term in \eqref{EulerCtau-2D}
for use with our simplified WENO scheme. The tangential artificial viscosity operators smooth the 
solution only in  tangential directions with no diffusion added in directions normal to the contact curve.   This is extremely important for 
RT  problems which take a very long time to fully develop (for example, in the runs presented here,
425,000 time-steps are used for the numerical simulation).
 The equations of motion are thus given by the Euler-$C^{\tau}$ system 
\eqref{EulerCtau-2D} with the addition of  the gravity term
$\bm{f} = [ 0 \,, 0 \,, -g \rho \,, -g \rho v ]^{\mathcal{T}}$ to the right-hand side  of the equations.

\begin{figure}[H]
\centering
\subfigure[vertical velocity $v$ computed with WENO-$N$ at time $t=1.0$; 
there are oscillations in the tangential direction.]{\label{fig:RT-Ctau-early3}\includegraphics[width=70mm]{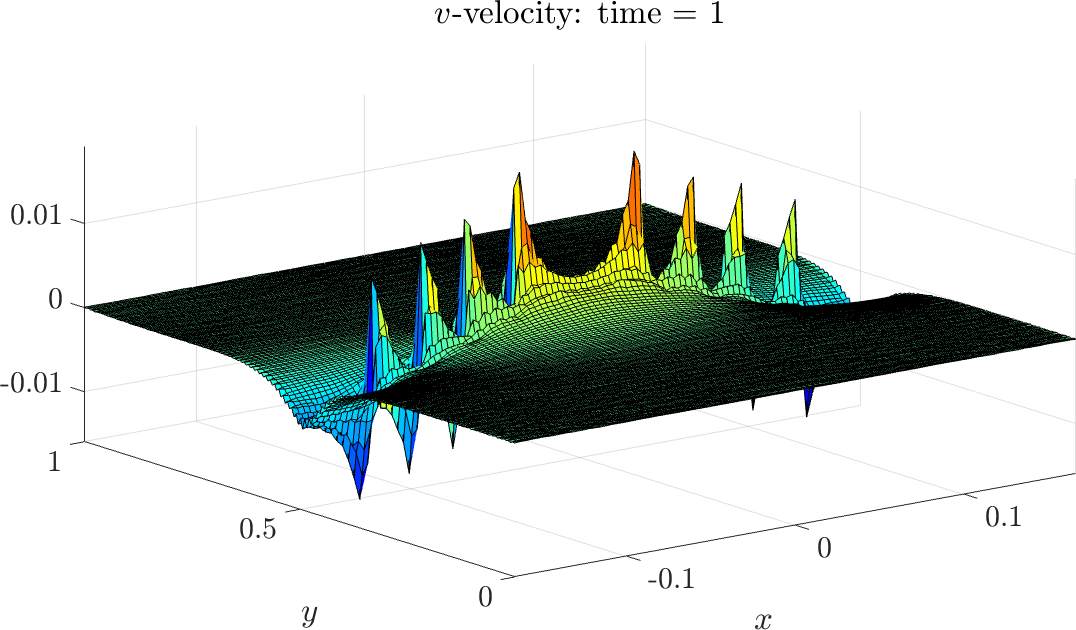}}
\hspace{1.5em}
\subfigure[vertical velocity $v$ computed with WENO-$C^{\tau}$-$N$ at time $t=1.0$; the tangential spikes are removed.]{\label{fig:RT-Ctau-early4}\includegraphics[width=70mm]{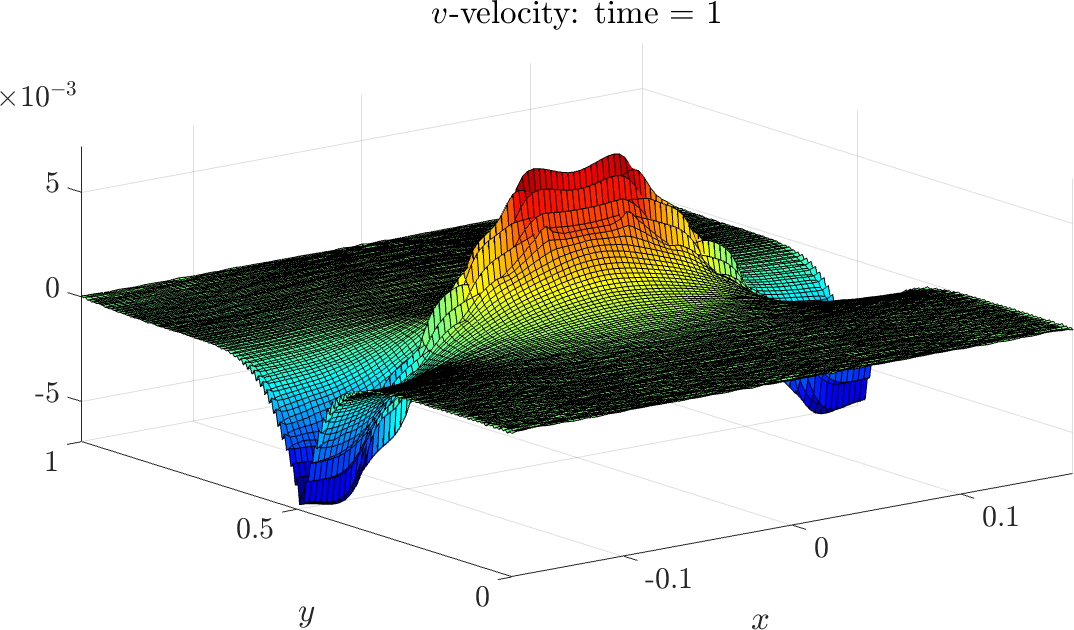}}
\caption{Demonstration of the removal of tangential spikes using anisotropic diffusion. Solutions are 
computed using WENO-$N$ and WENO-$C^{\tau}$-$N$ in the domain $[0,1/6] \times [0,1]$ 
with 50$\times$200 cells.}
\label{fig:RT-Ctau-early}
\end{figure}

Fig.\ref{fig:RT-Ctau-early3} and Fig.\ref{fig:RT-Ctau-early4} show the effect of tangential artificial viscosity:
the contact discontinuity remains sharp, while the spurious tangential spikes are suppressed.

\subsubsection{Boundary conditions for the RT problem}

We implement reflecting boundary conditions for the RT problem  on all four boundaries. Let $\vec{\nu}$
denote the normal vector to the boundary $\partial \Omega$. The reflecting boundary condition mean 
that $\vec{\nu} \cdot \bm{u} |_{\partial \Omega} = 0$ for all time $t \geq 0$. In the numerical discretization 
of the problem, this condition is enforced through the choice of \emph{ghost node} values via an even
or odd extension of each of the conservative variables. 
We provide further details in Appendix \ref{appendix:BCs}, 
but mention here that the presence of the gravity terms in 
\eqref{EulerCtau-2D} requires that the pressure be extended in a 
linear fashion at the top and bottom boundaries so as to ensure that the vertical velocity $v$ satisfies
$v = 0$ there. 

\subsection{Application of WENO-$C^{\tau}$-$N$ to the RT instability}
We now apply the WENO-$C^{\tau}$-$N$ scheme (detailed in Appendix 
\ref{appendix:WENO-Ctau}) to the RT instability. The domain $[0,1/6] \times [0,1]$ is discretized using a
50$\times$200 cell grid with a time-step of $\delta t = 2 \times 10^{-5}$, and the problem is run up to 
a final time of $t = 8.5$. The artificial viscosity parameter in \eqref{artificial-viscosity-Ctau} is 
set as $\beta = 20.0$, the $C$-equation parameters are chosen as $\varepsilon = 0.4$ and $\kappa = 10.0$, 
and we employ our noise detection and removal algorithm, with 
$\delta h = 2 \times 10^{-5}$ in \eqref{Cref}, $\delta_{\mathrm{off}} = 0.05$ in \eqref{deltaoff},
the noise removal viscosity $\eta$ in 
\eqref{noise-removal-alt} chosen such that 
$\eta \cdot \delta\tau / |\delta \bm{x}|^2 = 5 \times 10^{-4}$, and the local 
heat equation solver iterated for only a single time-step. 

\begin{figure}[H]
\centering
\subfigure[$t=4.0$]{\label{fig:RT-anisotropic-rho1}\includegraphics[width=30mm]{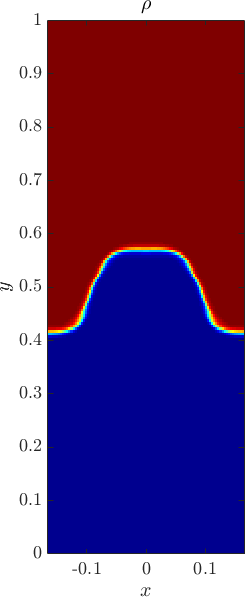}}
\hspace{1.5em}
\subfigure[$t=5.5$]{\label{fig:RT-anisotropic-rho2}\includegraphics[width=30mm]{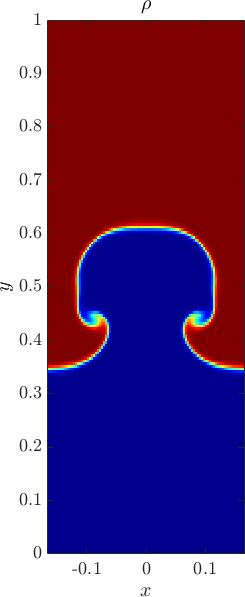}}
\hspace{1.5em}
\subfigure[$t=7.0$]{\label{fig:RT-anisotropic-rho3}\includegraphics[width=30mm]{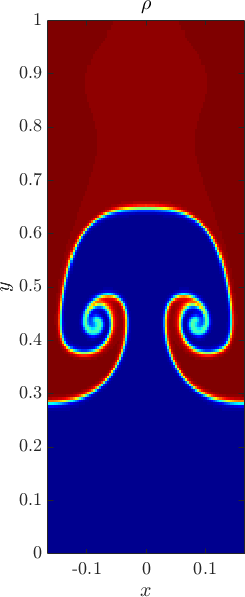}}
\hspace{1.5em}
\subfigure[$t=8.5$]{\label{fig:RT-anisotropic-rho4}\includegraphics[width=30mm]{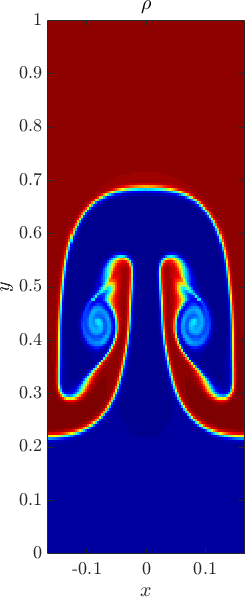}}
\caption{Application of WENO-$C^{\tau}$-$N$ to the RT instability in the domain $[0,1/6] \times [0,1]$ with 
50$\times$200 cells; figures are heatmap plots of the density $\rho$ at various instances of time.}
\label{fig:RT-anisotropic}
\end{figure}  

The density heatmap at various instances of time are presented in Fig.\ref{fig:RT-anisotropic}. At time 
$t=4.0$, the anisotropic diffusion acting in the tangential direction to the interface $\Gamma$ 
prevents grid noise
from developing into fingers and corrupting the solution, while ensuring that the interface $\Gamma$ itself
remains sharp, see Fig.\ref{fig:RT-anisotropic-rho1}. By time $t=5.5$, the solution begins to assume the 
classical ``mushroom''-shaped profile (Fig.\ref{fig:RT-anisotropic-rho2}), and at time $t=7.0$, the roll-up begins
to occur, see Fig.\ref{fig:RT-anisotropic-rho3}.
The anisotropic diffusion ensures that the solution in the roll-up region does not become overly diffused, 
and this is demonstrated in Fig.\ref{fig:RT-anisotropic-rho4}. This final figure should be compared with those
figures in \cite{LiWe2003}, which were produced with a finer mesh of 100$\times$400 cells. 

The WENO-$C^{\tau}$-$N$ solution shown in Fig.\ref{fig:RT-anisotropic}
compares very favorably with those in \cite{LiWe2003}, which are either 
overly diffused (such as LL, CFLFh), or are corrupted by the small-scale structure at the interface
(PPM, VH1). 
We note here that low resolution simulations generally exhibit a significant amount of mixing due to numerical 
diffusion \cite{Calder2002}. This is in contrast with the results presented here, where the low resolution 
50$\times$200 WENO-$C^{\tau}$-$N$ simulation produces a solution with a 
sharper interface and more development
in the roll-up region than the solutions obtained from the high resolution 100$\times$400 simulations 
in \cite{LiWe2003}. 

%

\subsubsection{Comparison with isotropic artificial viscosity}
\begin{figure}[H]
\centering
\subfigure[$\beta=1.0$]{\label{fig:RT-isotropic1}\includegraphics[width=30mm]{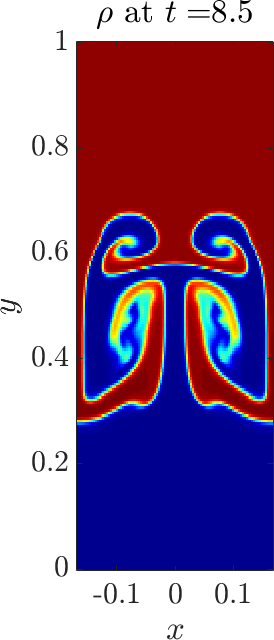}}
\hspace{1.5em}
\subfigure[$\beta=5.0$]{\label{fig:RT-isotropic2}\includegraphics[width=30mm]{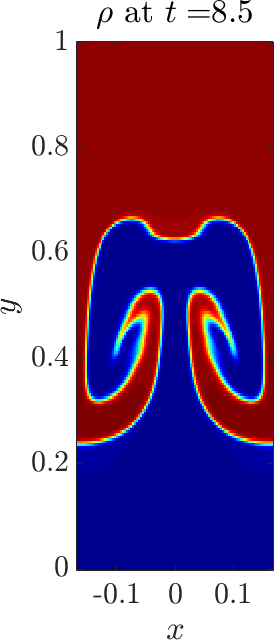}}
\hspace{1.5em}
\subfigure[$\beta=10.0$]{\label{fig:RT-isotropic3}\includegraphics[width=30mm]{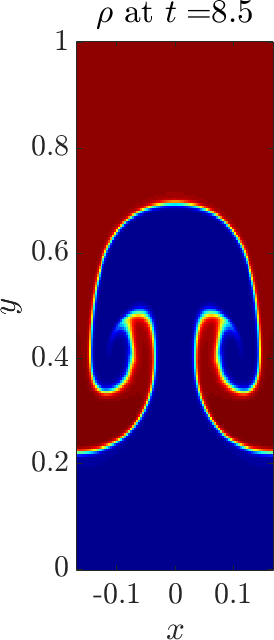}}
\hspace{1.5em}
\subfigure[$\beta=20.0$]{\label{fig:RT-isotropic4}\includegraphics[width=30mm]{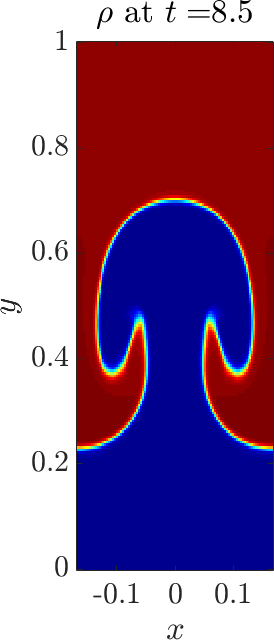}}
\caption{Application of WENO-$C$-$N$ to the RT instability in $[0,1/6] \times [0,1]$ with 
50$\times$200 cells; figures are of the density $\rho$ at the final time $t=8.5$.}
\label{fig:RT-isotropic}
\end{figure}

If we instead use isotropic artificial viscosity, we cannot produce an RT simulation which is as good as that produce by the anisotropic scheme.
Fig.\ref{fig:RT-isotropic} shows the density profile for the RT simulation 
using the \emph{isotropic} $C$-method as described by the
Euler-$C$ system \eqref{EulerC-2D} with
$\beta^E = 0$ and $\beta^u = \beta^v = \beta$ with four different values of $\beta$.

The choice of $\beta=1.0$
is insufficient to smooth the density oscillations in the tangential direction, so that the fingers develop, interact with 
each other and eventually corrupt the solution. Setting 
$\beta=5.0$ removes the density oscillations and produces a solution that is closer to the classic mushroom-shaped
profile, but still results in a depression at the top of the mushroom.  Moreover, there is very little roll-up
occurring, as can be seen from Fig.\ref{fig:RT-isotropic2}. Increasing to $\beta=10.0$ removes the 
depression at the top of the mushroom, but results in almost no roll-up, see Fig.\ref{fig:RT-isotropic3}. 
In Fig.\ref{fig:RT-isotropic4}, we show the result computed with $\beta=20.0$, which is  the value of $\beta$ used for the anisotropic
$C$-method simulation shown Fig.\ref{fig:RT-anisotropic}. The isotropic solution with $\beta=20$ shows no roll-up at all.

\begin{figure}[H]
\centering
\subfigure[$\Vert \omega(\cdot,\cdot,t) \Vert_{L^1(\Omega)}$]{\label{fig:RT-vorticity1}\includegraphics[width=50mm]{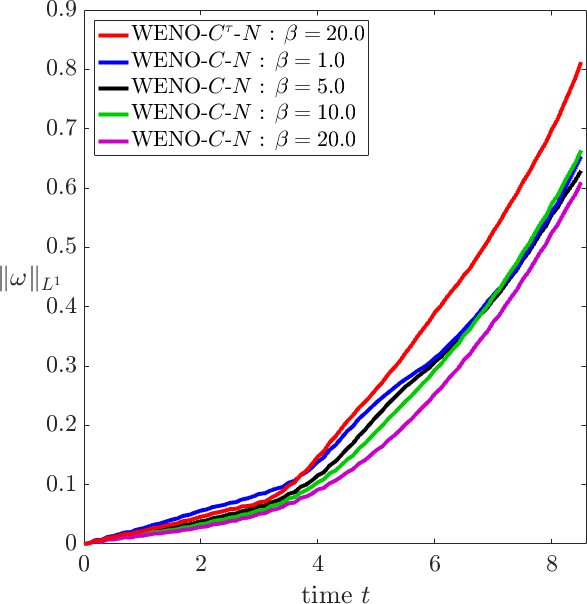}}
\hspace{3em}
\subfigure[$\Vert \omega(\cdot,\cdot,t) \Vert_{L^2(\Omega)}$]{\label{fig:RT-vorticity2}\includegraphics[width=50mm]{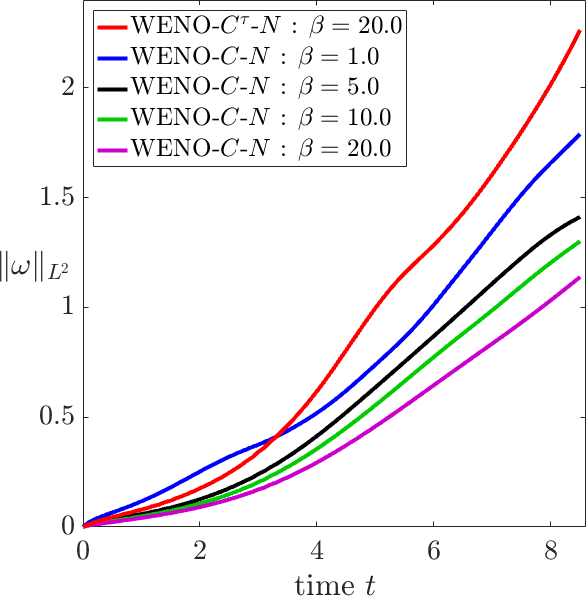}}
\caption{Comparison of the $L^1$ and $L^2$ norms of 
vorticity $\omega$ for solutions computed with anisotropic or isotropic diffusion.}
\label{fig:RT-vorticity}
\end{figure}

In Fig.\ref{fig:RT-vorticity}, we compute the $L^q$ norm (with $q=1,2$) of the \emph{vorticity} 
$\omega \coloneqq \p_x v - \p_y u$, which is a measure of the local rotation of the fluid, and 
hence provides a measure of the amount of interface roll-up in the flow. It is evident that the solution computed with the 
anisotropic $C$-method displays the largest growth (in time) of the norm $\Vert \omega(\cdot,\cdot,t) \Vert_{L^q(\Omega)}$. 
The simulation with isotropic diffusion 
and $\beta=1.0$ has a distinct increase in $\Vert \omega(\cdot,\cdot,t) \Vert_{L^1(\Omega)}$ 
at around time $t=4.5$. This is due to the fingers that 
develop and roll-up; the growth then slows down due to the interaction between the fingers which causes
mixing. The isotropic $\beta=5.0$, $\beta=10.0$ and 
$\beta=20.0$ simulations share similar profiles for $\Vert \omega(\cdot,\cdot,t) \Vert_{L^q(\Omega)}$, with 
all three displaying significantly less growth of vorticity than the simulation run using WENO-$C^{\tau}$-$N$. 

\subsubsection{Suppression of noise with the noise detection and removal algorithm}\label{subsubsec:noise-RT}
\begin{figure}[H]
\centering
\subfigure[WENO-$C^{\tau}$]{\label{fig:RT-noise1}\includegraphics[width=70mm]{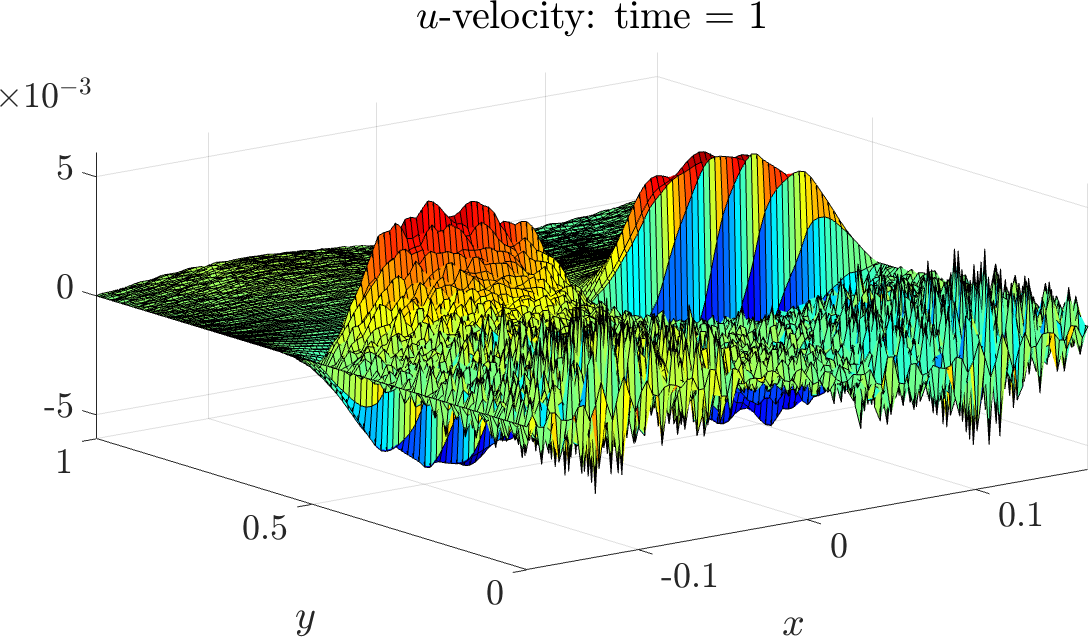}}
\hspace{1.5em}
\subfigure[WENO-$C^{\tau}$-$N$]{\label{fig:RT-noise3}\includegraphics[width=70mm]{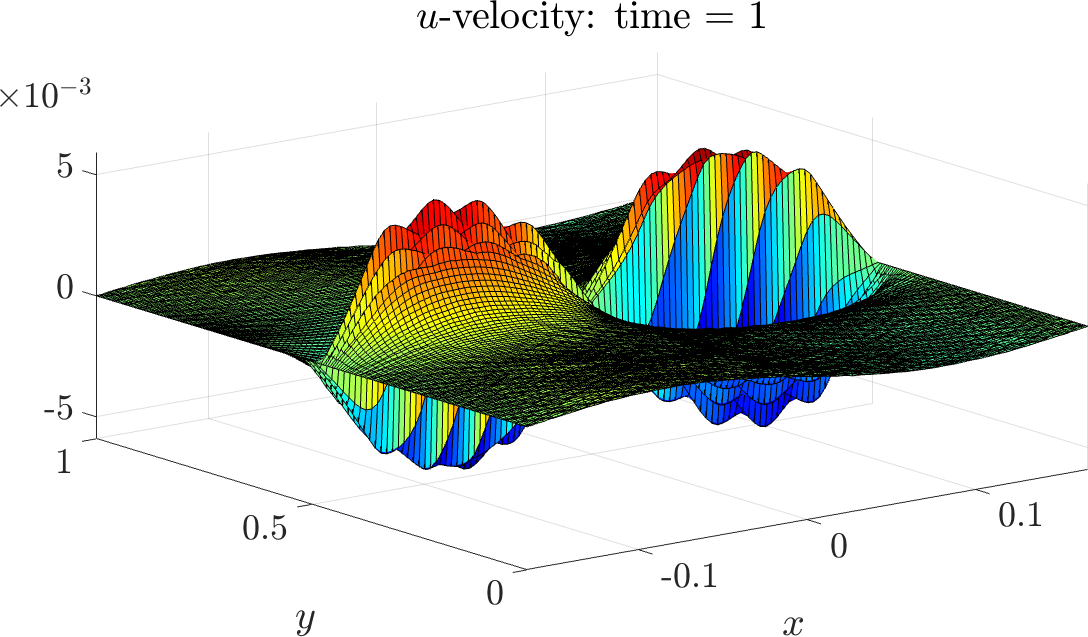}}
\caption{Figures demonstrating the ability of the wavelet-based noise detection and removal 
algorithm to suppress high-frequency spurious noise that can occur in the solution. Figures are of the horizontal
velocity $u$. Fig.\ref{fig:RT-noise1} shows the solution
 computed with WENO-$C^{\tau}$, 
 while Fig.\ref{fig:RT-noise3} shows the solution computed
with WENO-$C^{\tau}$-$N$. Both solutions were 
solved in $[0,1/6] \times [0,1]$ with 50$\times$200 cells and explicit forward Euler time integration, 
and are shown at time $t=1.0$. }
\label{fig:RT-noise}
\end{figure}

In this section, we briefly discuss the role of the noise detection and removal algorithm 
described in \S\ref{sec:noise-ind} in suppressing 
noise that would otherwise occur and corrupt the solution. In Fig.\ref{fig:RT-noise}, we compare the results
from two simulations: one run using WENO-$C^{\tau}$-$N$, which includes the noise detection and removal algorithm, and one run using WENO-$C^{\tau}$, which does not include the noise detection and removal algorithm. The figures shown are of the horizontal velocity $u$ at time $t=1.0$. 
We recall that explicit forward Euler time integration is used for both simulations.

It is clear that the WENO-$C^{\tau}$-$N$ solution is much better than the WENO-$C^{\tau}$ solution. The 
noise that develops\footnote{The use of explicit forward Euler time-stepping to 
simulate low Mach flow results in extremely severe stability constraints and, although 
the CFL condition is not violated, 
solutions are quickly corrupted by high-frequency noise.} 
corrupts the WENO-$C^{\tau}$ solution at time $t=1.0$, as
shown in Fig.\ref{fig:RT-noise1}.
This is in contrast to the WENO-$C^{\tau}$-$N$ solution, 
which remains noise-free. We note here that the WENO-$C^{\tau}$-$N$ solution is unaffected at the interface 
$\Gamma$, due to the
deactivation of the noise detection near the interface. 

\begin{figure}[H]
\centering
\subfigure[$\beta_l=1.0$: noise develops at early times. Shown is the horizontal velocity $u$]{\label{fig:RT-noise-linear1}\includegraphics[width=60mm]{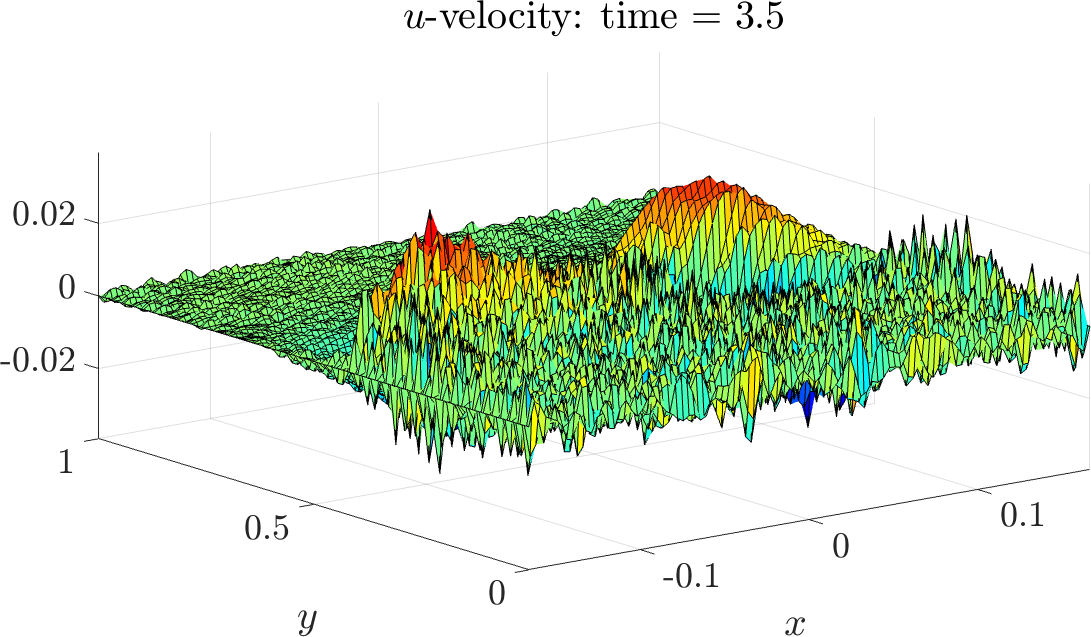}}
\hspace{2em}
\subfigure[$\beta_l=2.0$: noise develops at late times, particularly near the bottom boundary $y=0$; shown is a cross-section of the horizontal velocity $u$ along $x=0$]{\label{fig:RT-noise-linear2}\includegraphics[width=60mm]{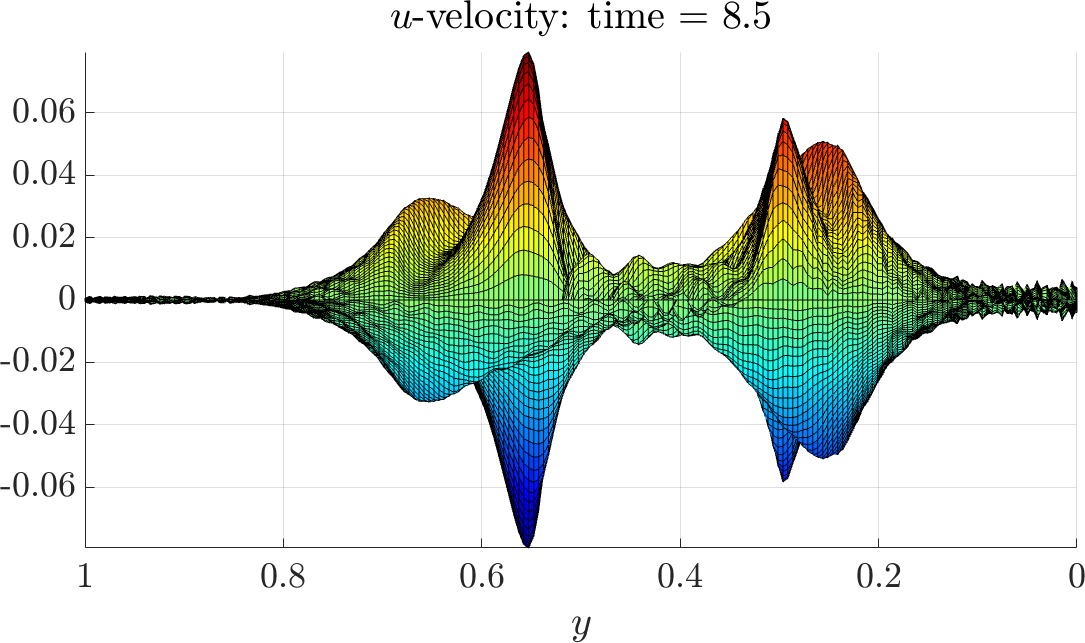}}
\par\bigskip
\subfigure[nooneline][$\beta_l=3.0$: heatmap plot of the density $\rho$]{\label{fig:RT-noise-linear3}\includegraphics[width=20mm]{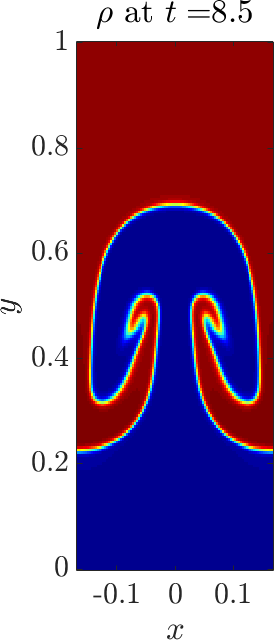}}
\hspace{8em}
\subfigure[$\beta_l=25.0$: heatmap plot of the density $\rho$]{\label{fig:RT-noise-linear4}\includegraphics[width=20mm]{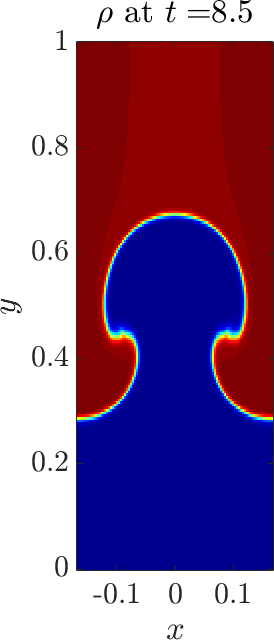}}
\caption{Figures showing solutions computed with linear viscosity 
replacing the noise detection and removal algorithm. 
Solutions computed in $[0,1/6] \times [0,1]$ with 
50$\times$200 cells using WENO-$C^{\tau}$ and a linear viscosity term to suppress the noise.}
\label{fig:RT-noise-linear}
\end{figure} 

For the purpose of benchmarking our noise detection and removal procedure, we compare the 
algorithm with a
simple alternative, namely the addition of a linear viscosity term to the 
right-hand side of the momentum equations \eqref{EulerCtau-momentum-u} and 
\eqref{EulerCtau-momentum-v} of the form 
\begin{equation}\label{noise-linear}
\beta_{l} |\delta \bm{x}|^2 \Delta \bm{u}\,,
\end{equation}
with $\beta_l$ the linear viscosity parameter. The resulting scheme is referred to as the 
WENO-$C^{\tau}$-$\Delta \bm{u}$ scheme. 

Setting $\beta_l=25.0$ corresponds to the choice of 
$\eta \cdot \delta \tau/|\delta \bm{x}|^2 = 5 \times 10^{-4}$, which is the value used for the 
simulation shown in Fig.\ref{fig:RT-anisotropic}. We present in Fig.\ref{fig:RT-noise-linear}
the solutions obtained when the noise detection and removal algorithm is replaced by the linear
viscosity \eqref{noise-linear}, with various choices of $\beta_l$.   

The issue with the uniform application of linear viscosity \eqref{noise-linear} 
is that the parameter $\beta_l$ controlling the 
amount of diffusion needs to be large enough so as to suppress noise, while simultaneously small 
enough to allow some physical structure to develop in the KH mixing zones. For the RT problem 
considered here, this 
is not possible; for instance, 
the simulation run with $\beta_l=25.0$, shown in Fig.\ref{fig:RT-noise-linear4}, is overly diffused with no structure
at all in the roll-up region. The same is true, albeit to a lesser extent, with the choice of $\beta_l=3.0$, shown 
in Fig.\ref{fig:RT-noise-linear3}. On the other hand, setting $\beta_l$ too small by choosing, for example, 
$\beta_l=1.0$ means that there is not enough viscosity to suppress the high frequency oscillations. This is 
demonstrated in Fig.\ref{fig:RT-noise-linear1}. 
The same is true, again to a lesser extent, when $\beta_l$ is set as 
$\beta_l=2.0$, see Fig.\ref{fig:RT-noise-linear2}. In this case, the noise is suppressed at the early stages, but
begins to develop near the boundaries at later times, and again corrupts the solution. 
On the other hand, 
the use of the noise detection and removal algorithm allows for the use of a large amount of viscosity, due to 
the localized nature of the heat equation solver. The solution produced is noise-free, does not require the 
use of a smaller time-step, and retains the physical structure in the KH roll-up regions. 

To quantitatively demonstrate the improvement in accuracy, we compare the 
solutions shown above with a 
``reference'' solution $\bm{U^{*}}$. 
This reference solution is computed using WENO-$C^{\tau}$ with $50 \times 200$ 
cells and explicit third order Runge-Kutta time integration; 
the use of a higher order in time method results in a noise-free solution. 
We define the $L^1$ error norm of a computed quantity $f$ as
\begin{equation}
E_{f}(t) = \Delta x \cdot \Delta y \sum_{i=1}^M \sum_{j=1}^N  {|f(x_i,y_j,t) - f^{*}(x_i,y_j,t) |}  \,,
\end{equation}
where $f^{*}$ is the reference solution, and $M$ and $N$ are the number of cells in the $x$ 
and $y$ directions, respectively. For this test, $M=50$ and $N=200$.

\begin{figure}[H]
\centering
\subfigure[\scriptsize{WENO-$C^{\tau}$-$N$}]{\label{fig:RT-best}\includegraphics[width=26mm]{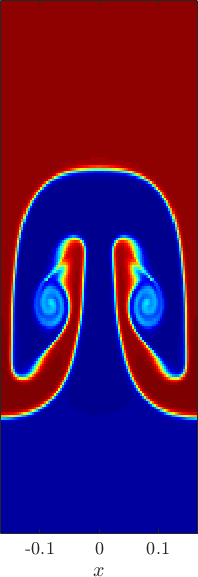}}
\hspace{1.5em}
\subfigure[\scriptsize{WENO-$C^{\tau}$}]{\label{fig:RT-RK3}\includegraphics[width=26mm]{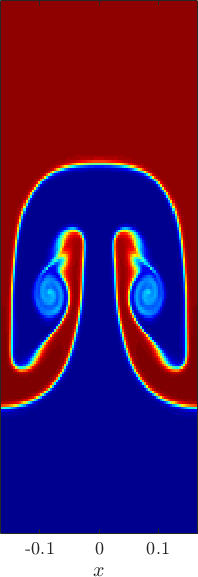}}
\hspace{1.5em}
\subfigure[\scriptsize{WENO-$C^{\tau}$-$\Delta \bm{u}$}]{\label{fig:RT-linear2-rho}\includegraphics[width=26mm]{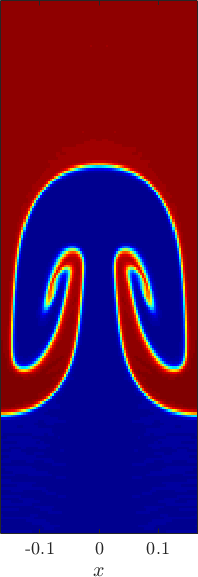}}
\caption{Comparison of (a) WENO-$C^{\tau}$-$N$, (b) WENO-$C^{\tau}$ with 3\textsuperscript{rd} order Runge-Kutta, and (c) WENO-$C^{\tau}$-$\Delta \bm{u}$ with $\beta_l=2.0$. Figures are of the density $\rho$ at the final time $t=8.5$.}
\label{fig:RT-final-comparison}
\end{figure}

In Table {\ref{table:noise-RT-comparison}}, we provide $E_{\rho}(t)$, $E_{u}(t)$, and 
$E_{v}(t)$ at various times $t$ for 
solutions computed using either WENO-$C^{\tau}$, WENO-$C^{\tau}$-$N$, or 
WENO-$C^{\tau}$-$\Delta \bm{u}$. We see that the WENO-$C^{\tau}$-$N$ method produces solutions 
with smaller errors in all of the quantities and at each of the times $t$, which establishes the
high accuracy of the method. Finally, we compare in
Fig.{\ref{fig:RT-final-comparison}} WENO-$C^{\tau}$-$N$ with WENO-$C^{\tau}$-$\Delta \bm{u}$ for 
$\beta_l=2$, which is the optimal artificial viscosity parameter value for the linear viscosity scheme. 
It is clear from the figure that the WENO-$C^{\tau}$-$N$ solution is much closer in appearance to the 
reference solution than the WENO-$C^{\tau}$-$\Delta \bm{u}$ solution is; the latter solution is overly smeared 
in the mixing regions due to the uniform application of viscosity, whereas the former solution retains the 
KH roll-up and matches almost exactly with the reference solution. These qualitative and quantitative 
observations establish the superiority of the noise detection and removal algorithm over the simple 
linear viscosity method. 

\begin{table}[H]
\centering
\renewcommand{\arraystretch}{1.0}
\scalebox{0.8}{
\begin{tabular}{|ll|ccc|}
\toprule
\midrule
\multirow{2}{*}{\textbf{Error}} & \multirow{2}{*}{\textbf{Scheme}}  & \multicolumn{3}{c|}{\textbf{Time}}\\

& {}  &3.0 & $7.0$ & $8.5$ \\
\midrule
 \multirow{6}{*}{\vspace{-0.0em}$E_{\rho}(t)$} 
 
 & {WENO-$C^{\tau}$-$N$} & 
 $1.066 \times 10^{-4}$ & $6.286 \times 10^{-3}$ & $9.180 \times 10^{-3}$  \\[0.25em]

& {WENO-$C^{\tau}$} & 
 $5.812 \times 10^{-3}$  & fail & fail  \\[0.25em]
 
 & {WENO-$C^{\tau}$-$\Delta \bm{u}$: $\beta_l =1.0$} & 
 $3.914 \times 10^{-4}$  & $8.033 \times 10^{-3}$ & fail  \\[0.25em]
 
 & {WENO-$C^{\tau}$-$\Delta \bm{u}$: $\beta_l =2.0$}  & 
 $4.475 \times 10^{-4}$  & $6.349 \times 10^{-3}$ & $9.222 \times 10^{-3}$  \\[0.25em]
 
 & {WENO-$C^{\tau}$-$\Delta \bm{u}$: $\beta_l =3.0$} & 
 $5.694 \times 10^{-4}$  & $7.197 \times 10^{-3}$ & $9.479 \times 10^{-3}$  \\[0.25em]
 
 & {WENO-$C^{\tau}$-$\Delta \bm{u}$: $\beta_l =25.0$}  & 
 $1.852 \times 10^{-3}$  & $2.238 \times 10^{-2}$ & $2.952 \times 10^{-2}$  \\[0.25em]

\midrule

 \multirow{6}{*}{\vspace{-0.0em}$E_{u}(t)$} 
 
 & {WENO-$C^{\tau}$-$N$} & 
 $1.938 \times 10^{-5}$ & $5.753 \times 10^{-4}$ & $8.094 \times 10^{-4}$  \\[0.25em]

& {WENO-$C^{\tau}$} & 
 $5.569 \times 10^{-3}$  & fail & fail  \\[0.25em]
 
 & {WENO-$C^{\tau}$-$\Delta \bm{u}$: $\beta_l =1.0$} & 
 $1.302 \times 10^{-4}$  & $2.691 \times 10^{-3}$ & fail  \\[0.25em]
 
 & {WENO-$C^{\tau}$-$\Delta \bm{u}$: $\beta_l =2.0$}  & 
 $6.694 \times 10^{-5}$  & $6.013 \times 10^{-4}$ & $8.819 \times 10^{-4}$  \\[0.25em]
 
 & {WENO-$C^{\tau}$-$\Delta \bm{u}$: $\beta_l =3.0$} & 
 $7.921 \times 10^{-5}$  & $6.205 \times 10^{-4}$ & $9.393 \times 10^{-4}$  \\[0.25em]
 
 & {WENO-$C^{\tau}$-$\Delta \bm{u}$: $\beta_l =25.0$}  & 
 $2.040 \times 10^{-4}$  & $1.085 \times 10^{-3}$ & $1.160 \times 10^{-3}$  \\[0.25em]

\midrule

 \multirow{6}{*}{\vspace{-0.0em}$E_{v}(t)$} 
 
 & {WENO-$C^{\tau}$-$N$} & 
 $8.603 \times 10^{-5}$ & $1.046 \times 10^{-3}$ & $1.355 \times 10^{-3}$  \\[0.25em]

& {WENO-$C^{\tau}$} & 
 $4.794 \times 10^{-3}$  & fail & fail  \\[0.25em]
 
 & {WENO-$C^{\tau}$-$\Delta \bm{u}$: $\beta_l =1.0$} & 
 $2.469 \times 10^{-4}$  & $2.776 \times 10^{-3}$ & fail  \\[0.25em]
 
 & {WENO-$C^{\tau}$-$\Delta \bm{u}$: $\beta_l =2.0$}  & 
 $1.015 \times 10^{-4}$  & $1.109 \times 10^{-3}$ & $1.480 \times 10^{-3}$  \\[0.25em]
 
 & {WENO-$C^{\tau}$-$\Delta \bm{u}$: $\beta_l =3.0$} & 
 $1.137 \times 10^{-4}$  & $1.172 \times 10^{-3}$ & $1.587 \times 10^{-3}$  \\[0.25em]
 
 & {WENO-$C^{\tau}$-$\Delta \bm{u}$: $\beta_l =25.0$}  & 
 $2.651 \times 10^{-4}$  & $2.253 \times 10^{-3}$ & $2.820 \times 10^{-3}$  \\[0.25em]

\midrule
\bottomrule
\end{tabular}}
\caption{$L^1$ error analysis for the RT problem at various times $t$.} 
\label{table:noise-RT-comparison}
\end{table} 

\section{Concluding remarks}

This paper presents the two-dimensional generalization of the 1-$D$  $C$-method described in 
\cite{RaReSh2018a}.   The 2-$D$ $C$-method is a  space-time smooth artificial viscosity 
scheme for adding {\it isotropic} diffusion to shock curves and {\it anisotropic} diffusion to contact discontinuities.   
We have applied the isotropic $C$-method to
four difficult problems, namely
the Noh problem, the Sedov point blast test, the Sod circular explosion problem, and the Mach 10 reflection 
test, and 
demonstrated that our method compares favorably 
to stand-alone WENO as well as the WENO-Noh scheme, the latter using the artificial viscosity method of Noh \cite{Noh1987}.

We have also demonstrated the efficacy of the anisotropic $C$-method.   For problems
requiring long-time evolution of contact discontinuities with hundreds of thousands of time-steps, the \emph{anisotropic} $C$-method
works extremely well. This method was applied to the 
RT instability problem of Liska-Wendroff \cite{LiWe2003}, and was shown to be able to suppress tangential spikes, while keeping the interface
sharp and not overly diffusing the KH roll-up regions. 

For problems with shock-contact interaction, our $C$-method uses a $C$-function to track the shock and another $\hat C$-function to
track the contact discontinuity. These two functions interact with one another: when the shock interacts with the contact, the $C$ tracking the
shock goes to zero, allowing the $\hat C$ of the contact to stabilize the contact curve without dissipation from the shock front.  

We have also described a simple shock-wall collision scheme
for stabilizing shock fronts during wall collision and bounce-back, and demonstrated its ability to suppress 
post-collision noise for a Sod-type explosion problem and the double Mach 10 reflection test.

For high-frequency noise detection and removal, we have devised and implemented a 
 simple 2-$D$ wavelet-based scheme for detecting these oscillations followed by a highly localized (in both space and time)
 removal algorithm based on a local explicit heat equation solver.   This noise detection and removal scheme has 
been applied  to the RT and Noh problems, wherein it
suppressed  high-frequency noise while retaining sharp shock and contact 
fronts. 

The qualitative and quantitative observations presented in this paper, 
together with the extensive accuracy studies presented in 
\cite{RaReSh2018a} in the 1-$D$ case, indicate that the schemes are high-order and produce accurate
solutions, even with the use of our highly simplified finite-difference WENO method.

\appendix

\section{Numerical discretization and schemes}\label{appendix:WENO}

In this section, we provide details for the WENO-based schemes used to produce the results in this paper. 
We first describe the simplified WENO procedure used for the spatial discretization of the nonlinear flux terms
in the Euler equations. We then describe the implementation for the Euler-$C$ and Euler-$C^{\tau}$ systems.

For readability, we will use Table \ref{table:schemes} to refer to these 
various schemes. 
{
\begin{table}[H]
\centering
\scalebox{0.8}{
\begin{tabular}{| M{4cm} | M{6cm}|} 
 \hline
 Scheme & Description \\ [0.0em] 
 \hline \hline 
 WENO & standard fifth-order WENO procedure for the usual Euler equations \eqref{Euler-2d}. \\[0.5em] 
\hline
 WENO-Noh &  WENO scheme with Noh's artificial viscosity method \eqref{weno-noh-scheme}. \\[0.5em]
 \hline
 WENO-$C$ & WENO scheme with the \emph{isotropic} $C$-method \eqref{EulerC-2D}. \\[0.5em] 
 \hline 
 WENO-$C^{\tau}$ & WENO scheme with the \emph{anisotropic} $C$-method \eqref{EulerCtau-2D}. \\[0.5em] 
 \hline 
 WENO-$C$-$N$ and WENO-$C^{\tau}$-$N$ & WENO-$C$ and WENO-$C^{\tau}$ with the noise detection and removal algorithm described in \S\ref{sec:noise-ind}. \\[0.5em]
 \hline
 WENO-$C$-$\hat{C}$ & WENO scheme with the \emph{isotropic} $C$-method for shock fronts and the \emph{anisotropic} $C$-method for contact discontinuities (see \S\ref{subsec:Sod-exp}). \\[0.5em]
 \hline
 WENO-$C$-$W$ &  WENO scheme with the shock-wall collision scheme \eqref{EulerCW-2D}. \\[0.5em]
 \hline
\end{tabular}}
\caption{Various numerical schemes used in the simulations.}
\label{table:schemes}
\end{table}} 

\subsection{WENO reconstruction procedure}
Our WENO reconstruction procedure is formally fifth-order with upwinding performed based only on the 
sign of the velocity at the cell edges. In particular, no approximate Riemann solvers are used. 

The spatial domain $\Omega = [x_1,x_M] \times [y_1,y_N]$ is subdivided into $M \cdot N$ cells, each with area
$\delta x \cdot \delta y$. We employ the notation
$$
x_{i+\frac{1}{2}} = \frac{x_i + x_{i+1}}{2} \ \text{ and } \ y_{j+\frac{1}{2}} = \frac{y_j + y_{j+1}}{2}\,.
$$
Any quantity $w$ evaluated at a cell center $(x_i,y_j)$ shall be denoted by $w_{i,j}$. Similarly, any quantity 
$w$ evaluated at the center of the cell edge $(x_{i\pm\frac{1}{2}}\,,y_j)$ shall be denoted by 
$w_{i \pm \frac{1}{2},j}$, and at the cell edge $(x_i\,,y_{j \pm \frac{1}{2}})$ by $w_{i,j \pm \frac{1}{2}}$. The 
quantity $w$ evaluated at the cell corners $(x_{i \pm \frac{1}{2}}\,,y_{j \pm \frac{1}{2}})$ is denoted by 
$w_{i \pm \frac{1}{2}, j \pm \frac{1}{2}}$.

Given an array $(a_{r,s})$
corresponding to cell-center values of the quantity $a$, and arrays $f_{r \pm \frac{1}{2},s}$
and $g_{r,s \pm \frac{1}{2}}$ corresponding to cell-edge values of the quantities $f$ and $g$, respectively,
 we define the $(i,j)^{\mathrm{th}}$ component 
of the array $\mathrm{WENO}\left(a_{r,s},f_{r \pm \frac{1}{2},s},g_{r,s\pm \frac{1}{2}}\right)$ by 
$$
\left[ \mathrm{WENO}\left(a_{r,s},f_{r \pm \frac{1}{2},s},g_{r,s\pm \frac{1}{2}}\right) \right]_{i,j} = \left[ \mathrm{WENO}_x\left(a_{r,s},f_{r \pm \frac{1}{2},s}\right) \right]_{i,j} + \left[ \mathrm{WENO}_y\left(a_{r,s},g_{r,s\pm \frac{1}{2}}\right) \right]_{i,j}\,,
$$
with
\begin{align*}
\left[ \mathrm{WENO}_x\left(a_{r,s},f_{r \pm \frac{1}{2},s}\right) \right]_{i,j} &= 
\frac{1}{\delta x} \left( \tilde{a}_{i + \frac{1}{2},j}f_{i + \frac{1}{2},j} - \tilde{a}_{i - \frac{1}{2},j}f_{i - \frac{1}{2},j} \right) \\ 
\left[ \mathrm{WENO}_y\left(a_{r,s},g_{r ,s\pm \frac{1}{2}}\right) \right]_{i,j} &=
 \frac{1}{\delta y} \left( \tilde{a}_{i,j + \frac{1}{2}}g_{i,j + \frac{1}{2}} - \tilde{a}_{i,j - \frac{1}{2}}g_{i,j - \frac{1}{2}} 
\right)\,,
\end{align*}
where the cell-edge values $\tilde{a}_{i \pm \frac{1}{2},j}$ and $\tilde{a}_{i,j \pm \frac{1}{2}}$ are calculated
using a standard fifth-order WENO reconstruction procedure (see \cite{Jiang1996202,Shu2003}) with 
upwinding performed based on the sign of $f_{i \pm \frac{1}{2},j}$ and $g_{i,j \pm \frac{1}{2}}$, 
respectively. 

Then, defining the vector $\bm{U} = [\rho,\rho u, \rho v, E]^{\mathcal{T}}$, we construct the operator 
$\mathcal{F}_{\mathrm{WENO}}$ as
$$
\left[ \mathcal{F}_{\operatorname{WENO}}(\bm{U}_{r,s}) \right]_{i,j}  =  -
	\begin{bmatrix}
         \left[ \operatorname{WENO}\left( \rho_{r,s},\hat{u}_{r \pm \frac{1}{2},s},\hat{v}_{r,s \pm \frac{1}{2}} \right) \right]_{i,j}  \\[1.5em]
         \left[ \operatorname{WENO}\left( (\rho u )_{r,s}, \hat{u}_{r \pm \frac{1}{2},s},\hat{v}_{r,s \pm \frac{1}{2}} \right) \right]_{i,j}+ \tilde{\partial}_{x,4} p_{i,j}  \\[1.5em]
         \left[ \operatorname{WENO}\left( (\rho v )_{r,s}, \hat{u}_{r \pm \frac{1}{2},s},\hat{v}_{r,s \pm \frac{1}{2}} \right) \right]_{i,j} + \tilde{\partial}_{y,4} p_{i,j}  \\[1.5em]
         \left[ \operatorname{WENO}\left( (E + p )_{r,s}, \hat{u}_{r \pm \frac{1}{2},s},\hat{v}_{r,s \pm \frac{1}{2}} \right) \right]_{i,j}
        \end{bmatrix} \,.
$$
Here, we use the notation $\tilde{\p}_{x,4} p _{i,j}$ and $\tilde{\p}_{y,4} p _{i,j}$ to denote the fourth-order
central difference approximation for $\p_x p $ and $\p_y p$, respectively, at the cell-center $(x_i,y_j)$. The 
cell-edge velocities $\hat{u}_{i \pm \frac{1}{2},j}$ and $\hat{v}_{i, j \pm \frac{1}{2}}$ are calculated using 
a fourth-order averaging:
\begin{align*}
\hat{u}_{i - \frac{1}{2},j} &= \frac{-u_{i-2,j} + 7u_{i-1,j} + 7u_{i,j} - u_{i+1,j}}{12} \,, \\
\hat{v}_{i,j - \frac{1}{2}} &= \frac{-v_{i,j-2} + 7v_{i,j-1} + 7v_{i,j} - v_{i,j+1}}{12} \,.
\end{align*}

The operator $\mathcal{F}_{\operatorname{WENO}}$ provides the discretization of the nonlinear 
flux terms in the Euler systems \eqref{Euler-2d}, \eqref{EulerC-2D}, \eqref{EulerCtau-2D} and \eqref{EulerCW-2D}. 
Below, we describe the discretization and implementation of the various forms of artificial viscosity used in 
this paper. If no artificial viscosity is implemented, then we have the semi-discrete equations
$$
\frac{\mathrm{d}}{\mathrm{d}t} \bm{U}_{i,j} = \left[ \mathcal{F}_{\mathrm{WENO}} (\bm{U}_{r,s}) \right]_{i,j}
$$
corresponding to the Euler system \eqref{Euler-2d}. Given the solution at a time-step $t_n = n \cdot \delta t$, denoted by $\bm{U}^{n}_{i,j}$, time integration is done using an explicit $k^{\text{th}}$ 
order Runge-Kutta method:
\begin{equation}\label{WENO-eqn}
\bm{U}^{n+1}_{i,j} = \mathrm{RK}(\bm{U}^{n}_{i,j},\left[ \mathcal{F}_{\mathrm{WENO}} (\bm{U}^n_{r,s}) \right]_{i,j};k)\,.
\end{equation}
We will refer to the scheme \eqref{WENO-eqn} as ``WENO'', or ``stand-alone WENO''.  

\subsection{The WENO-$C$ and WENO-$C$-$N$ schemes}
We first describe the discretization of the Euler-$C$ system \eqref{EulerC-2D}. Given arrays $(w_{r,s})$, 
$(C_{r,s})$ and $(\rho_{r,s})$ 
corresponding to cell-center values of the quantities $w$, $C$ and $\rho$,  
and a time-dependent function $\tilde{\beta}$, 
the $(i,j)^{\text{th}}$ component of the array 
${\operatorname{DIFF}}(w_{r,s},C_{r,s},\rho_{r,s};\tilde{\beta})$ is defined as
\begin{align*}
\left[ {\operatorname{DIFF}}(w_{r,s},C_{r,s},\rho_{r,s};\tilde{\beta}) \right]_{i,j} \coloneqq 
&\frac{\tilde{\beta}}{\delta x} \left(  \rho_{i + \frac{1}{2},j} C_{i + \frac{1}{2},j} \tilde{\p} w_{i + \frac{1}{2},j} - \rho_{i - \frac{1}{2},j} C_{i - \frac{1}{2},j} \tilde{\p} w_{i - \frac{1}{2},j} \right) \\
+ &\frac{\tilde{\beta}}{\delta y} \left(  \rho_{i,j + \frac{1}{2}} C_{i,j + \frac{1}{2}} \tilde{\p} w_{i,j + \frac{1}{2}} - \rho_{i,j - \frac{1}{2}} C_{i,j - \frac{1}{2}} \tilde{\p} w_{i,j - \frac{1}{2}} \right)\,,
\end{align*}
where
$$
\tilde{\p} w_{i + \frac{1}{2},j} = \frac{w_{i+1,j} - w_{i,j}}{\delta x} \quad \text{ and } \quad 
\tilde{\p} w_{i,j + \frac{1}{2}} = \frac{w_{i,j+1} - w_{i,j}}{\delta y}\,,
$$
and the notation $z_{i \pm \frac{1}{2},j}$ and $z_{i, j \pm \frac{1}{2}}$ denote the quantity evaluated at the 
cell edges $(x_{i \pm \frac{1}{2}},y_j)$ and $(x_i , y_{j \pm \frac{1}{2}})$, respectively, using a standard 
averaging e.g.
$$
z_{i + \frac{1}{2},j} = \frac{z_{i+1,j} + z_{i,j} }{2}\,.
$$

The operator $\mathcal{D}_{\operatorname{DIFF}}$ is then defined by 
$$
\left[ \mathcal{D}_{\operatorname{DIFF}}(\bm{U}_{r,s},C_{r,s}) \right]_{i,j}  =
	\begin{bmatrix}
         0  \\[0.5em]
         \left[ \operatorname{DIFF}\left( u_{r,s}, C_{r,s}, \rho_{r,s} ; \tilde{\beta}^u \right) \right]_{i,j}  \\[1.5em]
         \left[ \operatorname{DIFF}\left( v_{r,s}, C_{r,s}, \rho_{r,s} ; \tilde{\beta}^u \right) \right]_{i,j}  \\[1.5em]
         \left[ \operatorname{DIFF}\left( (E/\rho)_{r,s}, C_{r,s}, \rho_{r,s} ; \tilde{\beta}^E \right) \right]_{i,j}
        \end{bmatrix} \,,
$$
where the artificial viscosity coefficients are given by \eqref{C-artificial-viscosity}. 

We also define the operator $\mathcal{L}(C_{r,s};\varepsilon,\kappa)$ by 
$$
\left[ \mathcal{L}(C_{r,s};\varepsilon,\kappa) \right]_{i,j} =  \frac{S(\bm{u}_{r,s})}{\varepsilon |\delta \bm{x}|} 
([G_{\rho}]_{i,j}- C_{i,j}) +  
\kappa S(\bm{u}_{r,s}) | \delta \bm{x} | \left( \tilde{\p} C_{i+\frac{1}{2},j} - \tilde{\p} C_{i-\frac{1}{2},j} +  
\tilde{\p} C_{i,j+\frac{1}{2}} - \tilde{\p} C_{i,j-\frac{1}{2}} \right)\,,
$$
where $S(\bm{u}_{r,s})$ is given by \eqref{S-wave-speed}. 

The isotropic Euler-$C$ system \eqref{EulerC-2D} is then approximated by the semi-discrete equations 
$$
\begin{cases}
\frac{\mathrm{d}}{\mathrm{d}t} \bm{U}_{i,j} = \left[ \mathcal{F}_{\operatorname{WENO}}(\bm{U}_{r,s}) + \mathcal{D}_{\operatorname{DIFF}}(\bm{U}_{r,s},C_{r,s}) \right]_{i,j} \,, \\
\frac{\mathrm{d}}{\mathrm{d}t} {C}_{i,j} = \left[ \mathcal{L}(C_{r,s};\varepsilon,\kappa) \right]_{i,j} \,, 
\end{cases}
$$
and time integration is again done using a $k^{\text{th}}$ order Runge-Kutta method:
\begin{subequations}\label{WENO-C-eqn}
\begin{align}
\bm{U}^{n+1}_{i,j} &= \mathrm{RK} \left(\bm{U}^{n}_{i,j},\left[ \mathcal{F}_{\mathrm{WENO}} (\bm{U}^n_{r,s}) + \mathcal{D}_{\operatorname{DIFF}}(\bm{U}^n_{r,s},{C}^n_{r,s})\right]_{i,j};k \right)\,, \\ 
{C}^{n+1}_{i,j} &= \mathrm{RK} \left({C}^{n}_{i,j},\left[ \mathcal{L} ({C}^n_{r,s} ; \varepsilon,\kappa) \right]_{i,j};k \right)\,.
\end{align}
\end{subequations}
We will refer to the scheme \eqref{WENO-C-eqn} as the WENO-$C$ scheme. When coupled with the noise
detection and removal algorithm as detailed in \S \ref{sec:noise-implementation}, the scheme will be referred to as
the WENO-$C$-$N$ scheme. 

\subsection{The WENO-$C^{\tau}$ and WENO-$C^{\tau}$-$N$ schemes}\label{appendix:WENO-Ctau}
Next, we describe the discretization procedure for the Euler-$C^{\tau}$ system \eqref{EulerCtau-2D}. 
The anisotropic diffusion operator $\operatorname{DIFF}_{\tau}$ is defined by 
\begin{align*}
& \left[ {\operatorname{DIFF}_{\tau}}(w_{r,s},C_{r,s},C^{\tau_1}_{r,s}, C^{\tau_2}_{r,s},\rho_{r,s};\tilde{\beta}) \right]_{i,j} \coloneqq  \\
&\frac{\tilde{\beta}}{\delta x} \left(  \rho_{i + \frac{1}{2},j} C_{i + \frac{1}{2},j} C^{\tau_1}_{i + \frac{1}{2},j} C^{\tau_1}_{i + \frac{1}{2},j} \tilde{\p} w_{i + \frac{1}{2},j} - \rho_{i - \frac{1}{2},j} C_{i - \frac{1}{2},j} C^{\tau_1}_{i - \frac{1}{2},j} C^{\tau_1}_{i - \frac{1}{2},j} \tilde{\p} w_{i - \frac{1}{2},j} \right) \\
+ &\frac{\tilde{\beta}}{\delta y} \left(  \rho_{i,j + \frac{1}{2}} C_{i,j + \frac{1}{2}} C^{\tau_2}_{i,j + \frac{1}{2}} C^{\tau_2}_{i,j + \frac{1}{2}} \tilde{\p} w_{i,j + \frac{1}{2}} - \rho_{i,j - \frac{1}{2}} C_{i,j - \frac{1}{2}} C^{\tau_2}_{i,j - \frac{1}{2}} C^{\tau_2}_{i,j - \frac{1}{2}}  \tilde{\p} w_{i,j - \frac{1}{2}} \right) \\
+ & \frac{\tilde{\beta}}{\delta x} \left( \rho_{i + \frac{1}{2},j} C_{i + \frac{1}{2},j} C^{\tau_1}_{i + \frac{1}{2},j} C^{\tau_2}_{i + \frac{1}{2},j} \hat{\p} w_{i + \frac{1}{2},j} - \rho_{i - \frac{1}{2},j} C_{i - \frac{1}{2},j} C^{\tau_1}_{i - \frac{1}{2},j} C^{\tau_2}_{i - \frac{1}{2},j} \hat{\p} w_{i - \frac{1}{2},j} \right) \\ 
+ &\frac{\tilde{\beta}}{\delta y} \left(  \rho_{i,j + \frac{1}{2}} C_{i,j + \frac{1}{2}} C^{\tau_1}_{i,j + \frac{1}{2}} C^{\tau_2}_{i,j + \frac{1}{2}} \hat{\p} w_{i,j + \frac{1}{2}} - \rho_{i,j - \frac{1}{2}} C_{i,j - \frac{1}{2}} C^{\tau_1}_{i,j - \frac{1}{2}} C^{\tau_2}_{i,j - \frac{1}{2}}  \hat{\p} w_{i,j - \frac{1}{2}} \right) \,,
\end{align*}
where $\hat{\p} w_{i \pm \frac{1}{2},j}$ and $\hat{\p} w_{i,j \pm \frac{1}{2}}$ are approximations to the 
derivatives $\p_y w$ at $(x_{i \pm \frac{1}{2}},y_j)$ and $\p_x w$ at $(x_i, y_{j \pm \frac{1}{2}})$, respectively:
\begin{align*}
\hat{\p} w_{i \pm \frac{1}{2},j} &= \frac{ w_{i \pm \frac{1}{2}, j +\frac{1}{2}} - w_{i \pm \frac{1}{2}, j - \frac{1}{2}} }{\delta y}\,, \\
\hat{\p} w_{i,j \pm \frac{1}{2}} &= \frac{ w_{i + \frac{1}{2},j \pm \frac{1}{2}} - w_{i - \frac{1}{2},j \pm \frac{1}{2}} }{\delta x} \,.
\end{align*}
We use the notation $w_{i \pm \frac{1}{2},j \pm \frac{1}{2}}$ to mean the quantity $w$ evaluated at the cell
corner $(x_{i \pm \frac{1}{2}},y_{j \pm \frac{1}{2}})$ using a simple averaging. For example, 
$$
w_{i + \frac{1}{2},j + \frac{1}{2}} = \frac{w_{i,j} + w_{i+1,j} + w_{i,j+1} + w_{i+1,j+1}}{4}\,.
$$ 

Defining the vector $\bm{C} = [C,C^{\tau_1},C^{\tau_2}]^{\mathcal{T}}$, we construct the operator 
$\mathcal{D}^{\tau}_{\operatorname{DIFF}}$ as 
$$
\left[ \mathcal{D}^{\tau}_{\operatorname{DIFF}}(\bm{U}_{r,s},\bm{C}_{r,s}) \right]_{i,j} =
	\begin{bmatrix}
         0  \\[0.5em]
         \left[ \operatorname{DIFF}_{\tau}\left( u_{r,s}, C_{r,s}, C^{\tau_1}_{r,s}, C^{\tau_2}_{r,s}, \rho_{r,s} ; \tilde{\beta} \right) \right]_{i,j}  \\[1.5em]
         \left[ \operatorname{DIFF}_{\tau}\left( v_{r,s}, C_{r,s}, C^{\tau_1}_{r,s}, C^{\tau_2}_{r,s}, \rho_{r,s} ; \tilde{\beta} \right) \right]_{i,j}  \\[1.5em]
         0
        \end{bmatrix} \,,
$$
where the artificial viscosity coefficient $\tilde{\beta}$ is defined by \eqref{artificial-viscosity-Ctau}. 

The anisotropic Euler-$C^{\tau}$ system \eqref{EulerCtau-2D} may then be approximated by the semi-discrete
equations
$$
\begin{cases}
\frac{\mathrm{d}}{\mathrm{d}t} \bm{U}_{i,j} = \left[ \mathcal{F}_{\operatorname{WENO}}(\bm{U}_{r,s}) + \mathcal{D}^{\tau}_{\operatorname{DIFF}}(\bm{U}_{r,s},\bm{C}_{r,s}) \right]_{i,j} \,, \\
\frac{\mathrm{d}}{\mathrm{d}t} \bm{C}_{i,j} = \left[ \mathcal{L}(\bm{C}_{r,s};\varepsilon,\kappa) \right]_{i,j} \,, 
\end{cases}
$$
and time-integration is done using a $k^{\text{th}}$ order Runge-Kutta solver:
\begin{subequations}\label{WENO-Ctau-eqn}
\begin{align}
\bm{U}^{n+1}_{i,j} &= \mathrm{RK}\left(\bm{U}^{n}_{i,j},\left[ \mathcal{F}_{\mathrm{WENO}} (\bm{U}^n_{r,s}) + \mathcal{D}^{\tau}_{\operatorname{DIFF}}(\bm{U}^n_{r,s},\bm{C}^n_{r,s}) \right]_{i,j};k \right)\,, \\ 
\bm{C}^{n+1}_{i,j} &= \mathrm{RK} \left(\bm{C}^{n}_{i,j},\left[ \mathcal{L} (\bm{C}^n_{r,s} ; \varepsilon,\kappa) \right]_{i,j};k \right)\,.
\end{align}
\end{subequations}
We refer to the anisotropic scheme \eqref{WENO-Ctau-eqn} as the WENO-$C^{\tau}$ scheme or, when 
coupled with the noise detection and removal algorithm in \S\ref{sec:noise-implementation}, 
as the WENO-$C^{\tau}$-$N$ scheme. 

\subsection{Boundary conditions and ghost node values}\label{appendix:BCs}

In general, the boundary conditions for the various problems considered here are imposed through the 
assigning of values to the \emph{ghost nodes}. Generally, either an even extension or an odd 
extension of the variable in question is employed. For our (formally) fifth-order WENO reconstruction 
procedure, 7 nodes in each direction are used to reconstruct the flux at a cell-center, so that three ghost node 
values are required in each direction. By an even extension at the left and right boundaries 
of the variable $w$, we mean that 
$$
w_{1-i,j} = w_{1+i,j} \quad \text{ and } \quad w_{M+i,j} = w_{M-i,j} \quad \text{for } i =1,2,3\,,
$$
while an odd extension at the left and right boundaries means
$$
w_{1-i,j} = -w_{1+i,j} \quad \text{ and } \quad w_{M+i,j} = -w_{M-i,j} \quad \text{for } i =1,2,3\,. 
$$
An analogous identity holds for the top and bottom boundaries. 

\subsubsection{Boundary conditions for the RT problem}
The numerical solution to the Rayleigh-Taylor problem \ref{sec:RT} is computed on the half-domain 
$[0,1/6] \times [0 1]$ and then reflected appropriately to yield the solution on all of 
$\Omega = [-1/6,1/6] \times [0,1]$. Consequently, reflecting boundary conditions are used on the left boundary 
$x=0$ and right boundary $x=1/6$. These are imposed through the ghost-node values; an even extension
of $\rho$, $E$, $\rho v$, $C$, and  $C^{\tau_1}$ is imposed, while an odd extension of
 $\rho u $ and $C^{\tau_2}$  is enforced. 
 At the top and bottom boundaries, $\rho$, $\rho u$, $C$, $C^{\tau_1}$, and 
$C^{\tau_2}$ are extended in an even fashion, $\rho v$ is extended 
in an odd fashion, and the pressure $p$ is extended linearly so as to preserve hydrostatic equilibrium:
$$
p_{i,1-j} = 2p_{i,1-j+1} - p_{i,1-j+2}
\quad \text{ and } \quad 
p_{i,N+j} = 2p_{i,N+j-1} - p_{i,N+j-2}
$$
for $j=1,2,3$ and $i=1,\ldots,M$. 

\section*{Acknowledgements} 
Research reported in this publication was supported by the Office of Defense Nuclear Nonproliferation Research and Development and
by the Defense Threat Reduction Agency under Interagency Agreement number HDTRA1825370 (DTRA10027 -- 25370) as work for others.
SS was partially supported by DTRA HDTRA11810022.

We would like to express our gratitude to the anonymous referees for their numerous 
suggestions that have greatly improved the manuscript.


\end{document}